%% file: main.tex
\documentclass[10pt,journal,compsoc, cspaper, twoside]{IEEEtran}

\ifCLASSOPTIONcompsoc
  \usepackage[nocompress]{cite}
\else
  \usepackage{cite}
\fi
\ifCLASSINFOpdf
\else
\fi
\usepackage{cite}

\usepackage{multirow}
\usepackage{graphicx}
\usepackage[normalem]{ulem}
\useunder{\uline}{\ul}{}

\usepackage{url}
\usepackage{comment}

\usepackage{amsmath,amssymb,amsfonts}
\usepackage{graphicx}
\usepackage{textcomp}
\usepackage{xcolor}

\usepackage{listings}
\usepackage{capt-of}

\usepackage{booktabs} 
\usepackage{wrapfig}
\usepackage{algorithm}
\usepackage{algpseudocode}
\usepackage{multirow}
\usepackage{balance}
\usepackage{color}
\usepackage{xspace}
\usepackage{xtab}

\newcommand{\MT}{MT\xspace}

\newcommand{\NAZANIN}{}

\newcounter{cLINE}

\newcommand{\TEXTTT}[1]{$\mathtt{#1}$}
\newcommand{\MR}{MR\xspace}
\newcommand{\MRs}{MRs\xspace}

\newcommand{\SMRL}{SMRL\xspace}

\newcommand{\CHANGED}[1]{\textcolor{black}{#1}}
\newcommand{\CHANGEDLAST}[1]{\textcolor{black}{#1}}

\usepackage[backgroundcolor=white,bordercolor=blue,linecolor=blue]{todonotes}
\newcommand{\TSE}[2]{{#2}}

\newcommand{\TSEstart}[1]{}
\newcommand{\TSEstop}[1]{}

\newcommand{\manuallabel}[2]{\def\@currentlabel{#2}\label{#1}}

\usepackage{amsmath}
\usepackage{tikz}
\usepackage{makecell}

\definecolor{codegreen}{rgb}{0,0.6,0}
\definecolor{codegray}{rgb}{0.5,0.5,0.5}
\definecolor{codepurple}{rgb}{0.58,0,0.82}
\definecolor{backcolour}{rgb}{0.95,0.95,0.92}

\lstdefinestyle{mystyle}{
    commentstyle=\color{codegreen},
    keywordstyle=\color{magenta},
    numberstyle=\tiny\color{white},
    stringstyle=\color{codepurple},
    basicstyle=\ttfamily\footnotesize,
    breakatwhitespace=false,         
    breaklines=true,                 
    captionpos=b,                    
    keepspaces=true,                 
    numbers=left,                    
    numbersep=5pt,                  
    showspaces=false,                
    showstringspaces=false,
    showtabs=false,                  
    tabsize=2
}

\newcommand{\T}{\emph{MST-wi}\xspace}
\newcommand{\MST}{\T}

\begin{document}
%
\title{Metamorphic Testing for Web System Security}
%
%
%
%

\author{Nazanin~Bayati~Chaleshtari, 
        Fabrizio~Pastore,~\IEEEmembership{Member,~IEEE, }
        Arda~Goknil,
        and~Lionel~C.~Briand,~\IEEEmembership{Fellow,~IEEE}
\IEEEcompsocitemizethanks{
\IEEEcompsocthanksitem N.B. Chaleshtari is affiliated with the University of Ottawa, Canada. F. Pastore is with the SnT Centre for Security, Reliability and Trust, University of Luxembourg, Luxembourg. A. Goknil is affiliated with SINTEF Digital, Norway; the work of A. Goknil has been partially performed while affiliated with the SnT Centre. L.C. Briand is affiliated with both the University of Luxembourg and University of Ottawa.\protect \\
 E-mails: n.bayati@uottawa.ca  fabrizio.pastore@uni.lu arda.goknil@sintef.no lionel.briand@uni.lu lbriand@uottawa.ca
}

\thanks{Manuscript received August XX, 2022; revised Month XX, 2022.}}

%
%

\markboth{IEEE Transactions on Software Engineering,~Vol.~00, No.~0, MONTH~YEAR}%
{Bayati Chaleshtari \MakeLowercase{\textit{et al.}}: Metamorphic Testing for Web System Security}
%



\IEEEtitleabstractindextext{%
\begin{abstract}
Security testing aims at verifying that the software meets its security properties. In modern Web systems, however, this often entails the verification of the outputs generated when exercising the system with a very large set of inputs. Full automation is thus required to lower costs and increase the effectiveness of security testing.

Unfortunately, to achieve such automation, in addition to strategies for automatically deriving test inputs, we need to address the oracle problem, which refers to the challenge, given an input for a system, 
of distinguishing correct from incorrect behavior (e.g., the response to be received after a specific HTTP GET request). 

In this paper, we propose Metamorphic Security Testing for Web-interactions (\MST), a metamorphic testing approach that integrates test input generation strategies inspired by mutational fuzzing and alleviates the oracle problem in security testing. It enables engineers to specify 
metamorphic relations (\MRs) that capture many security properties of Web systems. To facilitate the specification of such \MRs, we provide a domain-specific language accompanied by an Eclipse editor. \MST automatically collects the input data and transforms the \MRs into executable Java code to automatically perform security testing. It automatically tests Web systems to detect vulnerabilities based on the relations and collected data.

We provide a catalog of 76 system-agnostic \MRs to automate security testing in Web systems. It covers 39\% of the OWASP security testing activities not automated by state-of-the-art techniques; further, our \MRs can automatically discover 102 different types of vulnerabilities, which correspond to 45\% of the vulnerabilities due to violations of security design principles according to the MITRE CWE database. We also define guidelines that enable test engineers to improve the testability of the system under test with respect to our approach. 

We evaluated \MST effectiveness and scalability with two well-known Web systems (i.e., Jenkins and Joomla). 
It automatically detected 85\% of their vulnerabilities and showed a high specificity (99.81\% of the generated inputs do not lead to a false positive); our findings include a new security vulnerability detected in Jenkins. Finally, our results demonstrate that the approach scale, thus enabling automated security testing overnight.

\end{abstract}

\begin{IEEEkeywords}
System Security Testing, Metamorphic Testing, Domain-specific Languages.
\end{IEEEkeywords}}

\maketitle

\IEEEdisplaynontitleabstractindextext

%
\IEEEpeerreviewmaketitle



\input{introduction}
\input{background}

\input{approach}

\input{dsl}

\input{transformation}

\input{data_framework}

\input{mt_framework}

\input{MRrelations}

\input{MRs_patterns}
\input{usage}

\input{evaluation}

\input{related}

\input{conclusion}


\ifCLASSOPTIONcompsoc
  \section*{Acknowledgments}
\else
  \section*{Acknowledgment}
\fi
This work has been carried out as part of the COSMOS Project, which has received funding from the European Union’s Horizon 2020 Research and Innovation Programme under grant agreement No. 957254.
This work was also supported by NSERC of Canada under the Discovery and CRC programs. The authors would like to thank Xuan Phu Mai for his preliminary investigation of CWE vulnerabilities and for setting up the Joomla case study subject.

\ifCLASSOPTIONcaptionsoff
  \newpage
\fi




\bibliographystyle{IEEEtran}
\bibliography{SMRL.bib}

%

%

%
%
%





\begin{IEEEbiography}[{\includegraphics[width=1in,height=1.25in,clip,keepaspectratio]{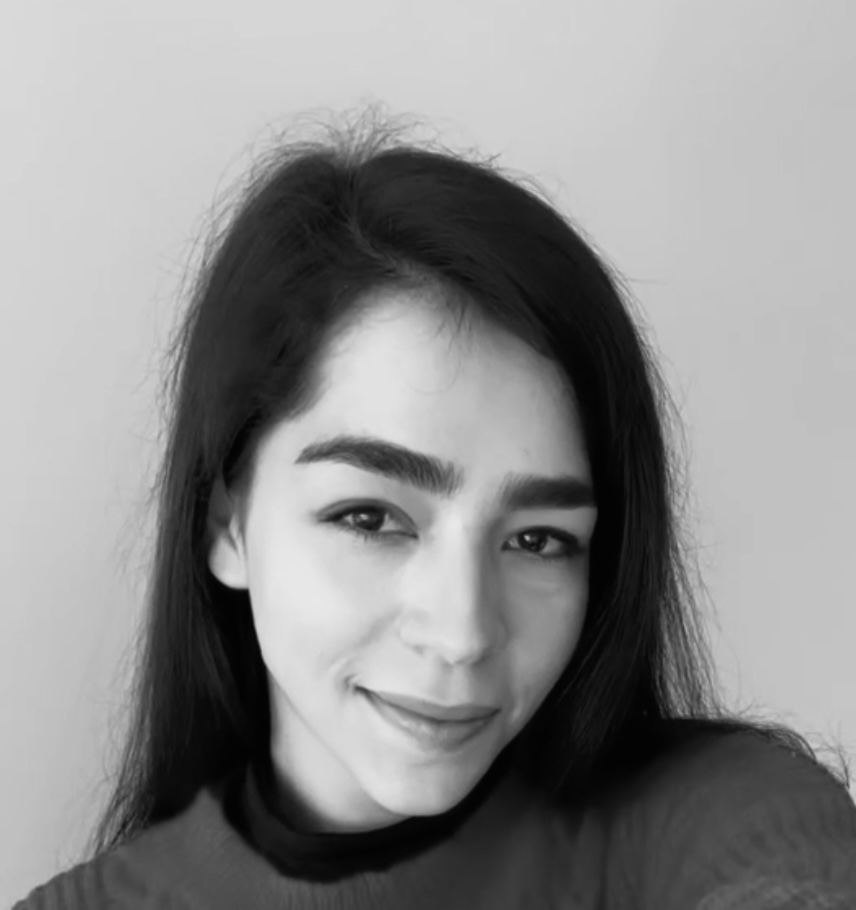}}]{Nazanin Bayati Chaleshtari} 
is a Ph.D. student at the School of EECS at the University of Ottawa and a member of the Nanda Lab. She received several academic awards, including a Ph.D. admission scholarship, an international doctoral scholarship from the University of Ottawa, and an honourable award for being an outstanding student during her master’s degree at the Iran University of Science and Technology. She was also ranked the best student among all computer engineering students at the Iran University of Science and Technology in 2019. Her research interests include automated software testing concerning security testing,  applied data science and empirical software engineering.
\end{IEEEbiography}

\begin{IEEEbiography}[{\includegraphics[width=1in,height=1.25in,clip,keepaspectratio]{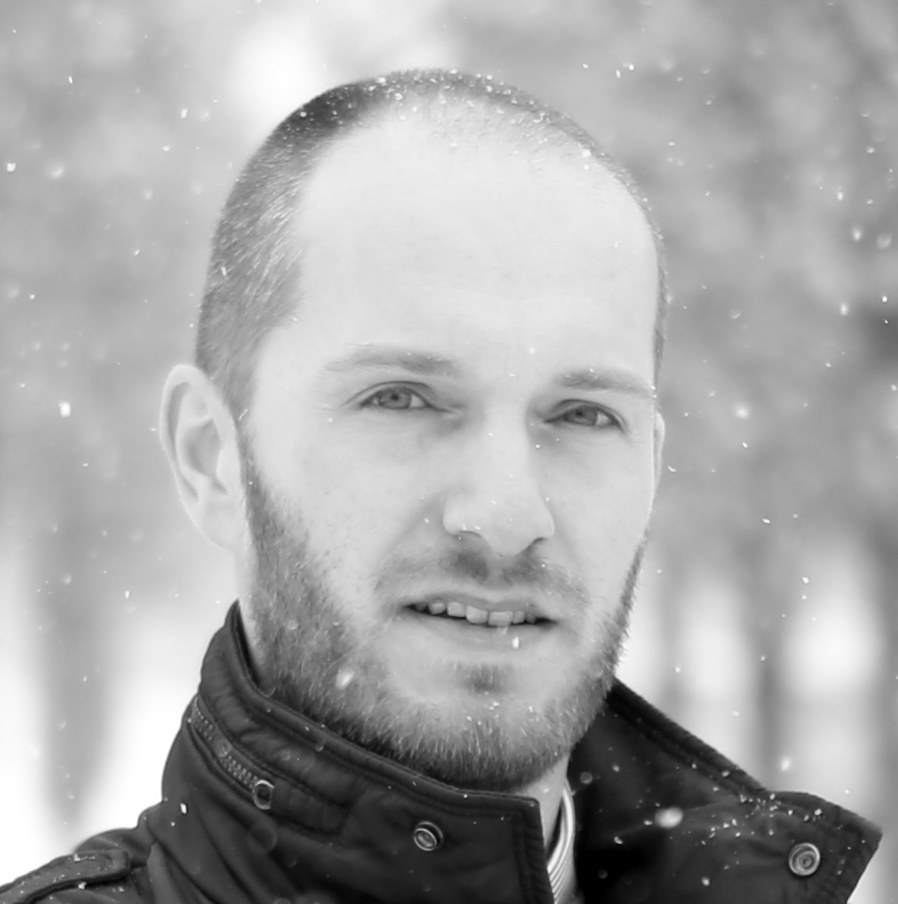}}]{Fabrizio Pastore}
is Chief Scientist II at the Interdisciplinary Centre for Security, Reliability and Trust (SnT), University of Luxembourg. He obtained his PhD in Computer Science in 2010 from the University of Milano - Bicocca.

His research interests concern automated software testing, including security testing and testing of AI-based systems; his work relies on the integrated analysis of different types of artefacts (e.g., requirements,  models, source code, and execution traces). He is active in several industry partnerships and national, ESA, and EU-funded research projects.
\end{IEEEbiography}

\begin{IEEEbiography}[{\includegraphics[width=1in,height=1.25in,clip,keepaspectratio]{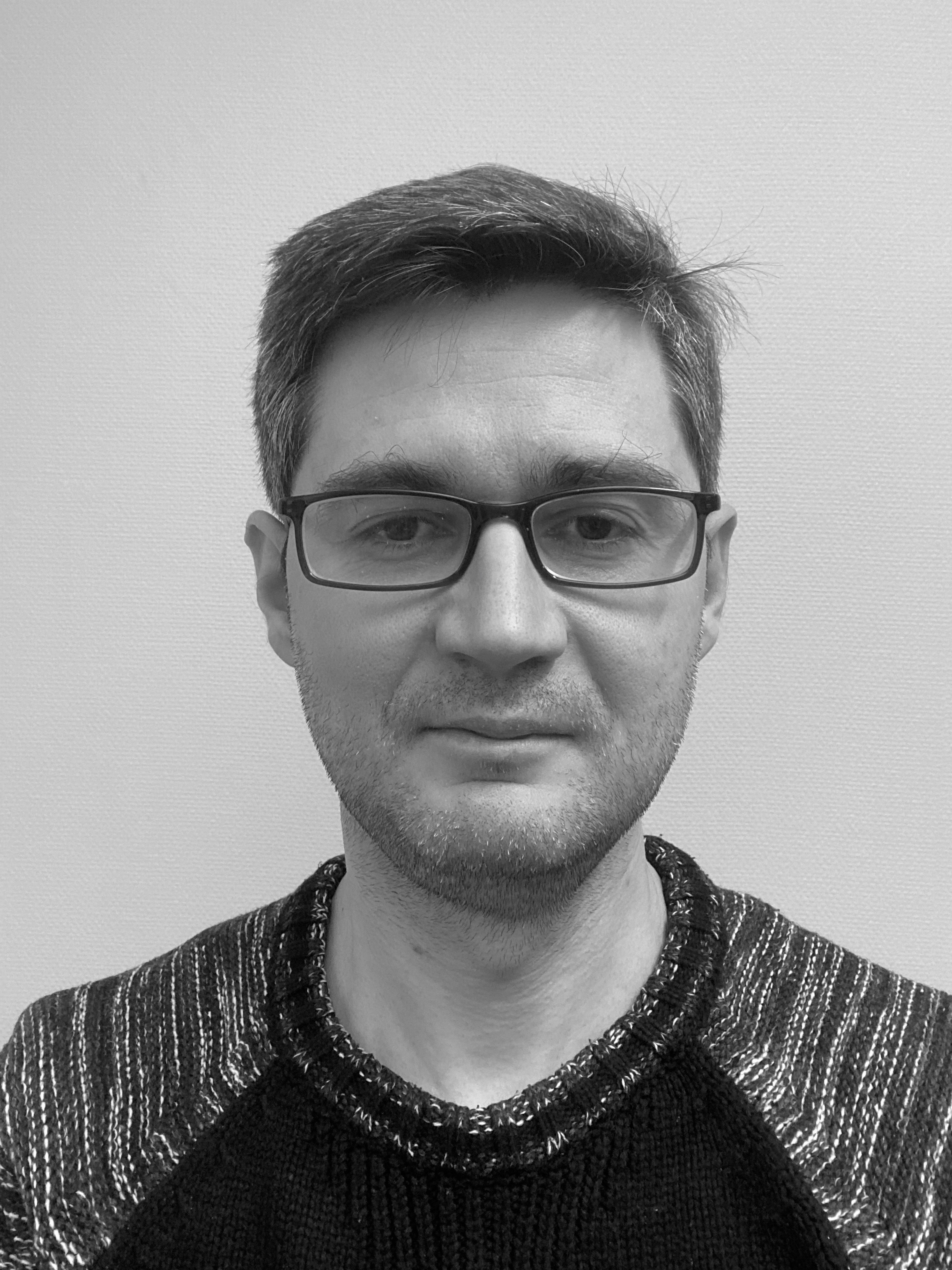}}]{Arda Goknil} received his Ph.D. degree in computer science from the University of Twente, in the Netherlands, in 2011. He is a senior research scientist at SINTEF, Norway. He was a research associate at the Interdisciplinary Centre for Security, Reliability, and Trust (SnT) at the University of Luxembourg. His research concerns AI Engineering, Model-Driven Engineering, Software Testing, Software Security, Intermittent Computing, Product Line Engineering, and Requirements Engineering. He is active on EU-funded and national research projects with several academic and industry partners.
\end{IEEEbiography}


\begin{IEEEbiography}[{\includegraphics[width=1in,height=1.25in,clip,keepaspectratio]{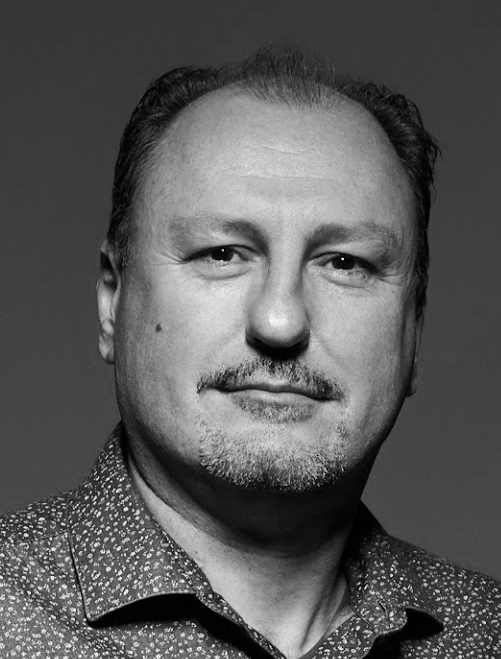}}]{Lionel C. Briand} is professor of software engineering and has shared appointments between (1) School of Electrical Engineering and Computer Science, University of Ottawa, Canada and (2) The SnT centre for Security, Reliability, and Trust, University of Luxembourg. He is the head of the SVV department at the SnT Centre and a Canada Research Chair in Intelligent Software Dependability and Compliance (Tier 1).

He has conducted applied research in collaboration with industry for more than 25 years, including projects in the automotive, aerospace, manufacturing, financial, and energy domains. In 2016, he received an ERC Advanced grant, the most prestigious European individual research award. He was elevated to the grades of IEEE and ACM fellow, granted the ACM SIGSOFT Outstanding Research Award (2022), the IEEE Computer Society Harlan Mills award (2012), and the IEEE Reliability Society Engineer-of-the-year award (2013) for his work on software verification and testing. 
His research interests include: Testing and verification, trustworthy AI, search-based software engineering, model-driven development, requirements engineering, and empirical software engineering.
\end{IEEEbiography}

\end{document}

%% file: introduction.tex
\section{Introduction}

Web systems (i.e., software systems providing services through Web pages consisting of HTML, Javascript, and CSS) are one of the main means to deliver online services, from e-commerce to online banking. These systems are business-critical, manage critical assets (e.g., card transactions), and often store sensitive information (e.g., customer data). Therefore, in Web systems,
discovering security vulnerabilities (i.e., faults preventing the system from fulfilling its security requirements) is essential~\cite{Haley2008,Felderer2016,Mai2018a,ISSRE,mai2019mcp}.
It is the objective of security testing.

Unfortunately, it is hard to discover vulnerabilities in Web systems during testing. Indeed, Web systems typically consist of several input interfaces (i.e., Web pages provided through URL requests). Each interface handles a large set of inputs (e.g., Web forms, cookies, URL parameters) that might be configured differently according to user roles. Considering that vulnerabilities might result from specific combinations of user roles, URL requests, and parameters, it is necessary to exercise the system with a large set of inputs, including inputs crafted to harm the system. Therefore, \emph{strategies to automatically generate security test inputs are necessary}.


However, automatically generating inputs is insufficient; indeed, an automated \textit{test oracle} (i.e., a mechanism for determining whether a test case has passed or failed) is needed. \emph{Security testing is known to suffer from the oracle problem}~\cite{Barr2015,Staats2011,Pezze2014}.
It is indeed generally infeasible to automatically determine the expected output for a given input or even manually specify expected outputs for a large number of test inputs. 
For instance, a security test case for a bypass authorization schema vulnerability should verify, for every user role, whether it is possible to access resources that should be available only to a user having a different role~\cite{OWASPtesting}.
We can discover this vulnerability by verifying 
access to various resources with different privileges and roles; however, to determine the test outcome, we need a mapping between roles and resources, which is not always available because of the complexity of the operations provided by modern Web systems.
Recent incidents involving corporate Web sites, such as Facebook, demonstrate that it is difficult to verify, at testing time, large sets of input sequences, including the ones that trigger vulnerabilities~\cite{FBviewAs,FBotherVuln,FBtakeover2022,FBunmaks2022}.

Although several security testing approaches exist in the literature, 
they do not address the oracle problem and assume the availability of an implicit test oracle~\cite{Barr2015}. Furthermore, most of them focus on a particular vulnerability type
(e.g., buffer overflows~\cite{Haller:2013:DOG,Ognawala-SymbExecutionLowLevelVuln-ASE-2016}) 
and only uncover vulnerabilities that prevent a system from providing outputs (e.g., system crashes because of buffer overflows). 

Metamorphic Testing (\MT) is a testing technique that has shown, in some contexts, to be very effective in alleviating the oracle problem~\cite{Chen1998,Liu2014}.
\emph{\MT is based on the idea that it may be simpler to reason about relations between outputs of multiple test executions, called metamorphic relations (\MRs), than to specify the system's input-output behavior}~\cite{Segura2016}. In \MT, system properties are captured as \MRs 
that are used to automatically transform an initial set of test inputs into follow-up test inputs. If the system outputs for the initial and follow-up test inputs violate the corresponding \MR, it is concluded that the system is faulty.

Considerable research has been devoted to developing \MT approaches for application domains such as computer graphics (e.g.,~\cite{Mayer2006,Guderlei2007,Just2009,Kuo2011}), 
Web services (e.g.,~\cite{Chan2007b,Sun2011,Zhou2012}), and embedded systems (e.g.,~\cite{Tse2004,Chan2007,Kuo2011b,Jiang2013}). 
Unfortunately, only a few approaches target security aspects~\cite{ChenMTSecurity2016}; also, their applicability is limited to the functional testing of security components (e.g., 
code obfuscators~\cite{ChenMTSecurity2016}) or the verification of specific security bugs (e.g., heartbleed~\cite{Heartbleed}).
They do not support the specification of general security properties by using \MRs.
Although \MT is automatable, few \MT approaches provide proper tool support~\cite{Segura2016}. 

Our goal in this paper is to use \MT to both automatically generate test inputs for security testing and address the test oracle problem in security testing. 
We propose Metamorphic Security Testing for Web-interactions (\MST), a technique that supports engineers in specifying \MRs to capture security properties of Web systems and automatically detect vulnerabilities (i.e., violations of security properties) based on those relations.

\MST automatically generates test inputs by altering valid inputs as an attacker would do; the strategies adopted to modify valid inputs are encoded in \MRs. It automatically collects valid inputs by relying on a Web crawler or reusing the scripts engineers implement for functional testing. For example, an \MR to spot bypass authorization schema vulnerabilities should ensure that 
\emph{a Web system returns different responses to two users when the first user (User-A) requests a URL that is provided to her by the GUI (e.g., in HTML links) while the second user (User-B) requests the same URL, which is not provided to her by the GUI.}
With \MST, it is possible to define an executable \MR that accesses, for User-B, all the URLs that can be reached by the GUI for User-A but not with the GUI for User-B. The \MR verifies that the output returned to User-B is different from the one for User-A; otherwise, a vulnerability is reported. If the output for the two users is the same, User-B can access the same page accessed by User-A instead of receiving an error message, which is not the functionality exposed by the user interface.

\MST is built on top of the following novel solutions:
\begin{itemize}
    \item Security Metamorphic Relation Language (SMRL), a Domain-Specific Language (DSL) for specifying \MRs for software security testing. SMRL is supported by a custom editor, implemented as a plug-in for the Eclipse IDE~\cite{EclipseIDE}, which facilitates the specification of \MRs.
    \item A catalog of system-agnostic \MRs targeting 102 security vulnerabilities of Web systems.
    \item A data collection framework that automatically collects the data required to perform \MT.
    \item A testing framework that automatically performs security testing based on the \MRs and the collected data.
\end{itemize}

Our DSL supports data representation functions and boolean operators to specify security properties in \MRs. It also provides a set of utility functions to express data properties that cannot be described with simple boolean or arithmetic operators. \MST automatically transforms \MRs written in our DSL into executable Java code. It extends the Crawljax Web crawler~\cite{crawljax:tweb12} to derive source inputs automatically from the system under test. It provides an \MT algorithm integrated into the JUnit framework~\cite{JUnit} as part of the testing framework. The algorithm performs \MT based on the executable \MRs in Java and the source inputs collected by the data collection framework.  
 
We evaluated our approach through the following analyses:

\begin{itemize} 

\item The capability of \MST to automate (including oracles) the security testing activities suggested by OWASP\footnote{OWASP (Open Web Application Security Project) is one of the best known organizations that focus on software security~\cite{OWASPWeb}.}. Our results show that \MST automates 39\% of the security testing activities not automated by other approaches relying on catalogs or implicit oracles (e.g., crashes) and addressing specific vulnerability types.  

\item The  applicability of \MST to discover common and important security vulnerability types described in the Common Weaknesses Enumeration Repository~\cite{CWE}. Our results show that \MST enables testing for 101 (45\%) vulnerability types due to errors in applying security design principles.

\item A study of testability guidelines that enable engineers to improve the testability of Web systems with \MST. We observe that controllability and an adequate test support environment are key for applying \MST.

\item An analysis of the effectiveness of \MST when applied to discover vulnerabilities in Jenkins, a leading open source automation server~\cite{Jenkins}, and Joomla, a content management system~\cite{Joomla}.
\MST has shown to be largely effective; indeed, it automatically detected 85\% of the targeted vulnerabilities affecting these two systems. Further, only a negligible fraction of follow-up test inputs leads to false alarms (0.19\% max).

\item An analysis of the \TSE{1.3}{efficiency} of \MST while testing Jenkins and Joomla. Our results show that, for most vulnerability types, \MST can be applied overnight; further, technical improvements in our framework implementation may enable overnight execution for all the \MRs in our catalog.

\end{itemize}


This paper largely extends our previous conference paper~\cite{Mai2020a} published at the 13th IEEE International Conference on Software Testing, Verification and Validation (ICST'20). An earlier version of our tool was demonstrated~\cite{Mai2020b} at the 42nd International Conference on Software Engineering (ICSE'20). This paper brings together, refines, and significantly extends the ideas from the above papers: 
\begin{itemize}
\item We extend our DSL with additional data representation and utility functions.
\item We extend our \MR catalog with \emph{54 new \MRs}.  
\item We present the results of an extensive study on the types of security vulnerabilities that can be addressed by our approach. Precisely, we study the security weaknesses organized in the CWE view for common security architectural tactics~\cite{CWE-ArchitectureView}, the ones belonging to the CWE Top-25 most dangerous software errors~\cite{CWE-Top25}, and the ones in the OWASP Top-10 Web security risks~\cite{OWASP-Top10}. 
\item We provide testability guidelines that assist engineers in designing and implementing their software to enable effective test automation with \MST. 
\item We provide substantial new empirical evidence to demonstrate the effectiveness and \TSE{1.3}{efficiency} of \MST by applying all the \MRs in our catalog to Jenkins and Joomla (the latter being entirely new). Our results include the discovery of a new security vulnerability in Jenkins that received a CVE identifier~\cite{NewCVE}.
\end{itemize}

Our toolset~\cite{WebMST,WebSMRLEditor}, our catalog of \MRs~\cite{WebMSTCWE}, and empirical data~\cite{Replicability} are publicly available.

This paper is structured as follows. Section~\ref{sec:background} provides the background information regarding \MT. In Section~\ref{sec:approach}, we present an overview of the approach. Sections~\ref{sec:dsl}~to~\ref{sec:mtframework} describe the core technical solutions. Section~\ref{sec:mrs} describes our catalog of \MRs. 
In Section~\ref{sec:usage}, we investigate the security vulnerability types that \MST can address and the testability guidelines for \MST to address these vulnerabilities. Section~\ref{sec:evaluation} reports on the results of the empirical validation conducted with two open-source case studies. Section~\ref{sec:empirical:validity} addresses threats to validity. Section~\ref{sec:related} discusses the related work. We conclude the paper in Section~\ref{sec:conclusion}.

%% file: background.tex
\section{Background: Metamorphic Testing}
\label{sec:background}

In this section, we present the basic concepts of MT.
The core of \MT is a set of \MRs, which are necessary properties of the program under test in relation to multiple inputs and their expected outputs~\cite{Chen2018}. 
\MRs \emph{resemble the traditional concept of program invariants, which are properties that hold at certain points in programs. However, the key difference is that an invariant has to hold for every possible program execution, whereas a metamorphic relation captures a property of inputs and outputs belonging to different executions}~\cite{Segura2016}.

In \MT, a single test case run requires multiple executions of the system under test with distinct inputs. 
The test outcome (pass or fail) results from verifying the outputs of different executions against the \MR. \TSE{1.1}{Below we provide the basic definitions underpinning MT.}

\vspace{0.10cm}
\TSEstart

\textit{Definition 1 (Metamorphic Relation - \MR).}
\label{definitionMR}
Let $f$ be a function under test. A function $f$ typically processes a set of arguments; we use the term \emph{input} to refer to the set of arguments processed by the function under test. 
An \MR is a relation concerning a set of inputs $\langle x_{1}, ..., x_{n} \rangle$, where $n \geq 2$, and a set of outputs generated with those inputs $\langle f(x_{1}), ..., f(x_{n}) \rangle$. 
\MRs are typically expressed as implications.

\vspace{0.10cm}
\textit{Definition 2 (Source Input and Follow-up Input).} 
An \MR defines how to generate a \emph{follow-up input} from a \emph{source input}. 
A source input is an input in the domain of $f$. A follow-up input 
satisfies the properties expressed by the \MR.

Follow-up inputs can be obtained by applying \emph{transformation functions} to the source inputs. The use of \emph{transformation functions} in \MRs simplifies the identification of follow-up inputs.

\vspace{0.10cm}
\textit{Definition 3 (Metamorphic Testing - MT).} MT consists of the following steps:

\begin{itemize}
\item[MT-1] Generate a number of source inputs (usually one) required by the \MR. 
\item[MT-2] Execute the \MR, which implies the following steps (they can be executed in any order, depending on the \MR):
\begin{itemize}
\item[MT-2.1] Execute the function under test with the source input(s).
\item[MT-2.2] Derive follow-up input(s) based on the \MR.
\item[MT-2.3] Execute the function under test with the follow-up input(s).
\end{itemize}
\item[MT-3] Check whether the results violate the \MR. If the \MR is violated, then the function under test is faulty.
\item[MT-4] Restart from (MT-1), up to a predefined number of iterations.
\end{itemize}

As an example, let us consider an algorithm $f$ that computes the shortest path for an undirected graph $G$. For any two nodes $a$ and $b$ in graph $G$, it may not be feasible to generate all possible paths from $a$ to $b$ and check whether the output path is really the shortest path. However, a property of the shortest path algorithm is that the length of the shortest path remains the same if nodes $a$ and $b$ are swapped.
By using this property, we can derive an \MR, i.e., $|f(G, a, b)| = |f(G, b, a)|$, in which we need two executions of the function under test, one with $(G, a, b)$ and another one with $(G, b, a)$. The results of the two executions are verified by the relation. If the relation is violated, $f$ is faulty.  

A source input in the example consists of a (random) graph $G$ to be generated and two vertices $a$ and $b$ in $G$ to be randomly selected. 
 Follow-up inputs can be generated by a \emph{transformation function}
that swaps the last two arguments of the source input. We call this function $\mathit{swapLastArguments}$ and apply it to $(G, a, b)$.


 Based on the above, our \MR can thus be defined as follows:\\ 
{\footnotesize{$x_1=(G, a, b) \land x_2=\mathit{swapLastArguments}(x_1)  \rightarrow  |f(x_1)| = |f(x_2)|$}} 

We first generate
a (random) graph $G$ and randomly select two vertices $a$ and $b$ in $G$ for the source input (see MT-1). We then execute the \MR (MT-2). We execute the shortest path function twice: first with $(G, a, b)$ (MT-2.1) and then with $(G, b, a)$ (MT-2.3). We
derive the follow-up input by applying function $\mathit{swapLastArguments}$ to $(G, a, b)$ (MT-2.2). Finally, in MT-3, we verify that the absolute value of the two retrieved outputs is the same; otherwise, we report a failure. 

\TSEstop










%% file: approach.tex
\section{Overview of the Approach}
\label{sec:approach}

\begin{figure}[h]
\hspace{3mm}
\includegraphics[width=8.25cm]{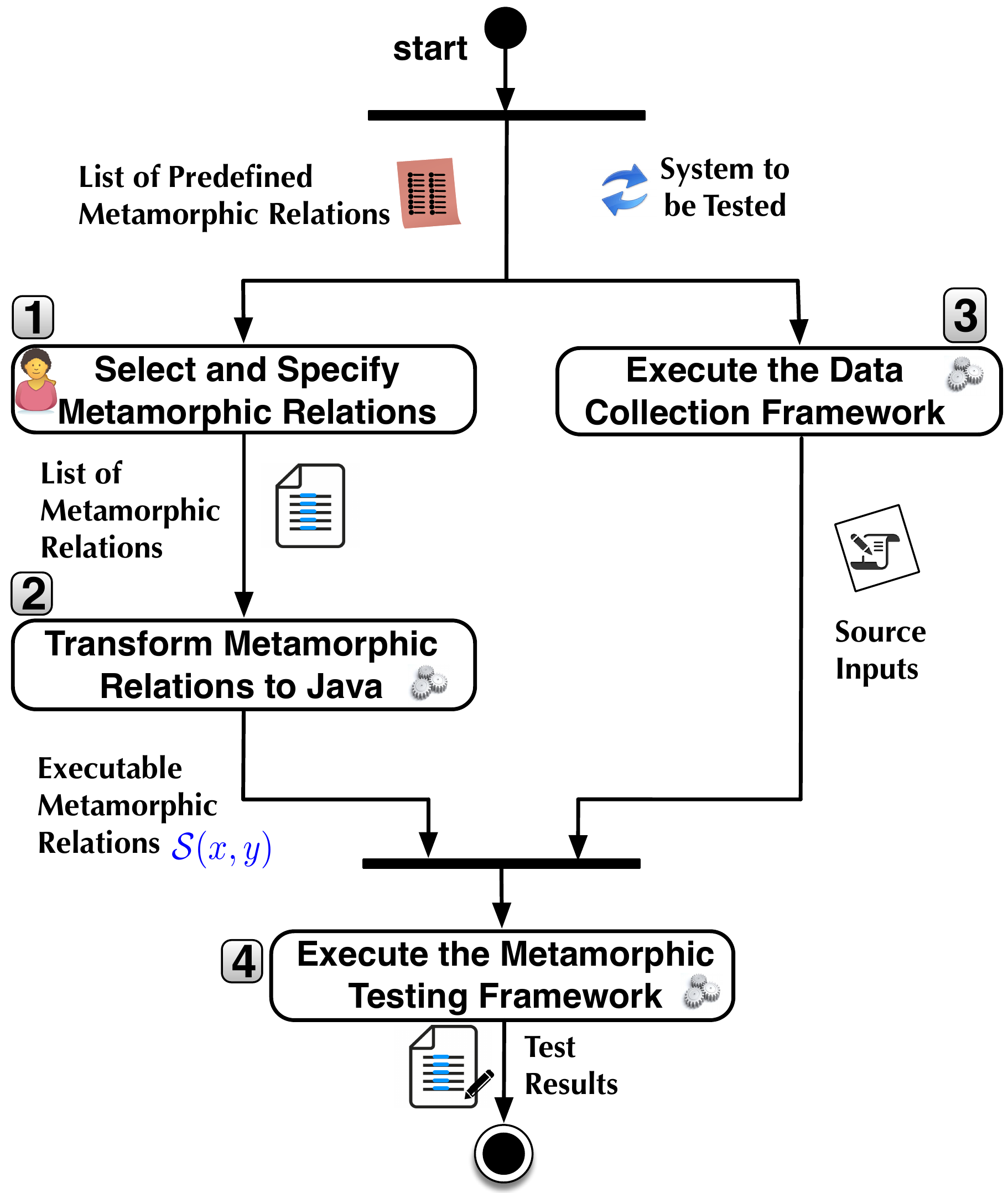}
\caption{Overview of the approach.}
\label{fig:approach}
\end{figure}

The process in Fig.~\ref{fig:approach} presents an overview of \MST. 
In Step 1, the engineer selects, from a catalog of predefined \MRs, the relations for the system under test. In general, we expect the engineer to choose the whole set of \MRs in our catalog. 
In addition, the engineer can also specify new relations by using our DSL. Step 1 is manual. 

In Step 2, \MST automatically transforms the selected \MRs into executable Java code. 



In Step 3, the engineer executes a Web crawler to automatically collect information 
about the Web system under test (hereafter, SUT).  
The crawler reveals the SUT structure (e.g., URLs that can be visited by an anonymous user) and the actions that trigger the 
generation of new content on a page (e.g., clicking on an anchor or a textual label).
The collected data is used to derive source inputs for \MT. 
The engineer can process manually implemented test scripts, if available, to obtain additional information.
Step 3 does not depend on other steps. 

In Step 4, \MST automatically loads source inputs and generates follow-up inputs as described by the relations. After executing the source and follow-up inputs, the outputs are checked according to the \MRs. 

Our DSL and the data collection framework can be extended to support new language constructs and data collection methods.
The \MT framework can also be improved to deal with input interfaces not supported yet (e.g., Silverlight plug-ins~\cite{Silverlight}) and to load data collected by new data collection methods.

Sections~\ref{sec:dsl} to~\ref{sec:mtframework} explain the details of each step in Fig.~\ref{fig:approach}, with a focus on how we achieved our automation objectives.

%% file: dsl.tex

\section{Step 1: Select and Specify \MRs}
\label{sec:dsl}

Step 1 in Fig.~\ref{fig:approach} concerns selecting and specifying \MRs. To enable specifying new \MRs, we provide a DSL called Security Metamorphic Relation Language (\SMRL). 

The most common usage scenario for \MST is the selection of \MRs for the SUT from our \MRs catalog (see Section~\ref{sec:mrs}) without specifying additional ones. Since our catalog covers generic vulnerabilities that may affect any Web system, we expect all the \MRs to be chosen in most cases. 

In our implementation, the catalog of \MRs is released as an Eclipse project folder, including our \MRs. The project can be copied and reconfigured for every SUT. The selection of \MRs is performed within the Eclipse environment by defining a JUnit test suite containing the \MRs to test (see Section~\ref{sec:mtframework}).

Since the selection of \MRs is straightforward, we focus on the \SMRL DSL in the following paragraphs. We introduce the \SMRL grammar, the boolean operators, the data representation functions, and the Web-utility functions.

\subsection{SMRL Grammar}
\label{subsec:grammar}

SMRL is an extension of Xbase~\cite{Efftinge2012xbase}, an expression language provided by Xtext~\cite{Xtext}.
Xbase specifications can be translated to Java programs and compiled into executable Java bytecode.

We rely on Xbase since DSLs extending Xbase inherit the syntax of a Java-like expression language and language infrastructure components, including a parser, 
a linker, a compiler, and an interpreter~\cite{Efftinge2012xbase}.
Further, it provides features common in modern programming languages, including lambda expressions, type inference, and simple operator overloading. 
These features will facilitate the adoption of \SMRL.


%
%

\begin{figure*}[tb]
\begin{center}
\includegraphics[width=12.7cm]{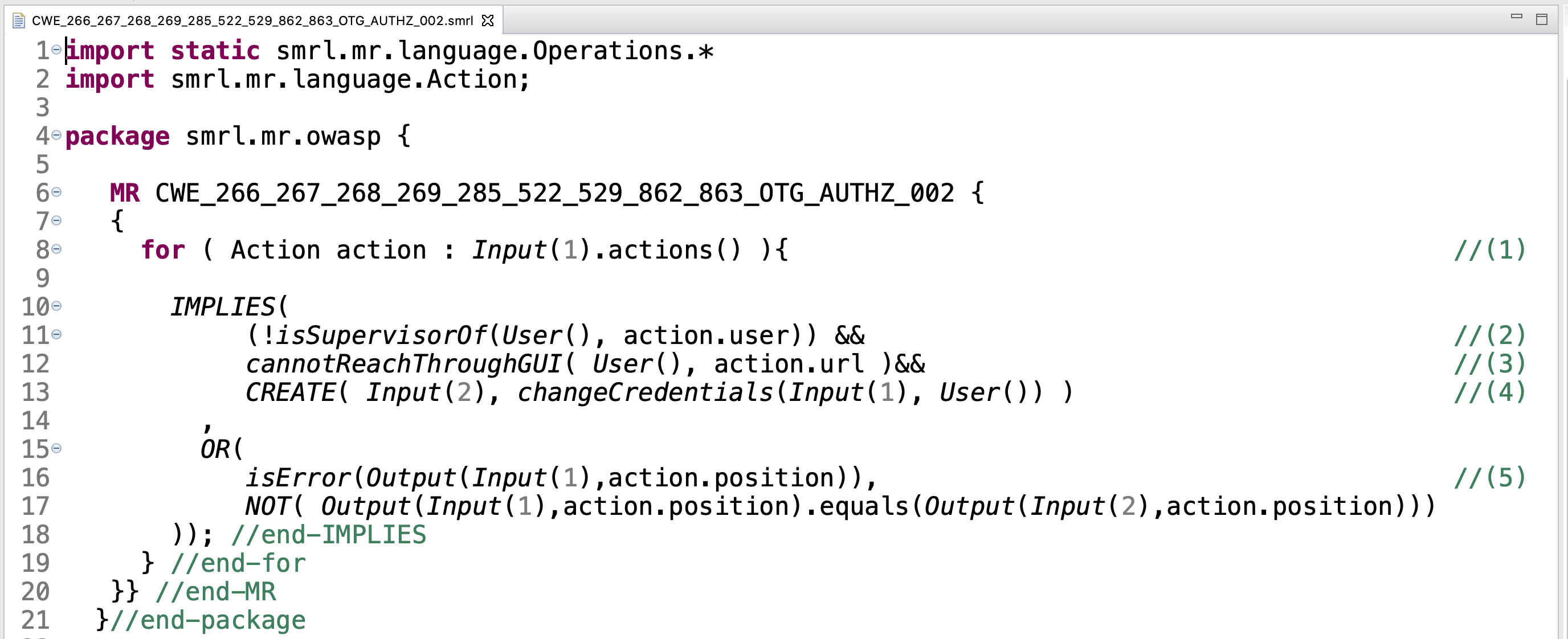}

 \raggedright{\footnotesize{

\emph{Description of the annotated \MR statements:} (1) For loop iterates over all actions of the Input. (2) Checks whether the user in User() is not a supervisor of the user performing the current action. (3) Verifies that the user cannot retrieve the URL of the action through the GUI (based on the data collected by the crawler). (4) Defines a follow-up input that matches the source input except that the credentials of User() are used in this case. (5) Verifies that the follow-up input leads to an error page or the output generated by the action containing the URL indicated above leads to two different outputs in the two cases.}}
\caption{An \MR for the Bypass Authorization Schema vulnerability.}
\label{fig:mrExample}
\end{center}
\end{figure*}

\SMRL extends Xbase by introducing (i) data representation functions, 
(ii) boolean operators to specify security properties, and (iii) Web-utility 
functions to express data properties and transform data. We can extend these functions by defining new Java APIs invoked in \MRs.

Fig.~\ref{fig:mrExample} presents an \MR written in our \SMRL editor. 
The relation checks whether other users can access the URLs dedicated to specific users through a direct request. We use it as a running example in the rest of the paper.


The SMRL grammar extends the Xbase grammar, which extends the Java grammar.
Each \SMRL specification can rely on external classes and APIs; in practice, it can have have an arbitrary number of import declarations indicating the APIs used in \MRs (Line 1 in Fig.~\ref{fig:mrExample}). 

A package declaration resembles the Java package structure and can contain one or more \MRs.
Line 4 in Fig.~\ref{fig:mrExample} declares the package \emph{smrl.mr.owasp}, which is the package for our \MRs automating the testing activities in the OWASP testing guidelines~\cite{OWASPtesting}.
Like in Java, \MRs defined in different \SMRL specification files can belong to the same package.

An \MR can contain an arbitrary number of \TEXTTT{XBlock}- \TEXTTT{Expressions}, which are nonterminal symbols defined in the Xbase grammar. An \TEXTTT{XBlockExpression} can have loops, function calls, operators, and other \TEXTTT{XBlockExpression}s.

\input{tables/tableDataOperators.tex}

\subsection{Data Representation Functions}
\label{subsec:datarep}

\SMRL provides 46 functions to represent different data types used to refer to SUT inputs or outputs in \MRs. 
At a high level, we distinguish between four main categories of data: 
\begin{itemize}
    \item \emph{Interaction data} characterize the interactions with the SUT and is collected by the \MST Web crawler.
    \item \emph{SUT data} is specific to the SUT and needs to be specified by the end-user in \MST configuration files (e.g., the path of log files generated by the SUT).
    \item \MST \emph{data} is provided with the MST framework and does not need to be modified by the end-user (e.g., attack vectors for SQL injection attacks).
    \item \emph{Output data} is generated by the SUT in response to an input  (e.g., Web pages).
\end{itemize}

In \SMRL, data is represented by a keyword followed by an index number used to identify different data items. To keep \SMRL simple, 
we refer to data by using functions (hereafter, \emph{data functions}) with capitalized names (e.g., \TEXTTT{Input(1)}). 
Table~\ref{table:dataMethods} presents the data functions in \SMRL, grouped by category. 
Each data function returns a data class instance.

\begin{figure*}[tb]
\begin{center}
\includegraphics[width=19cm]{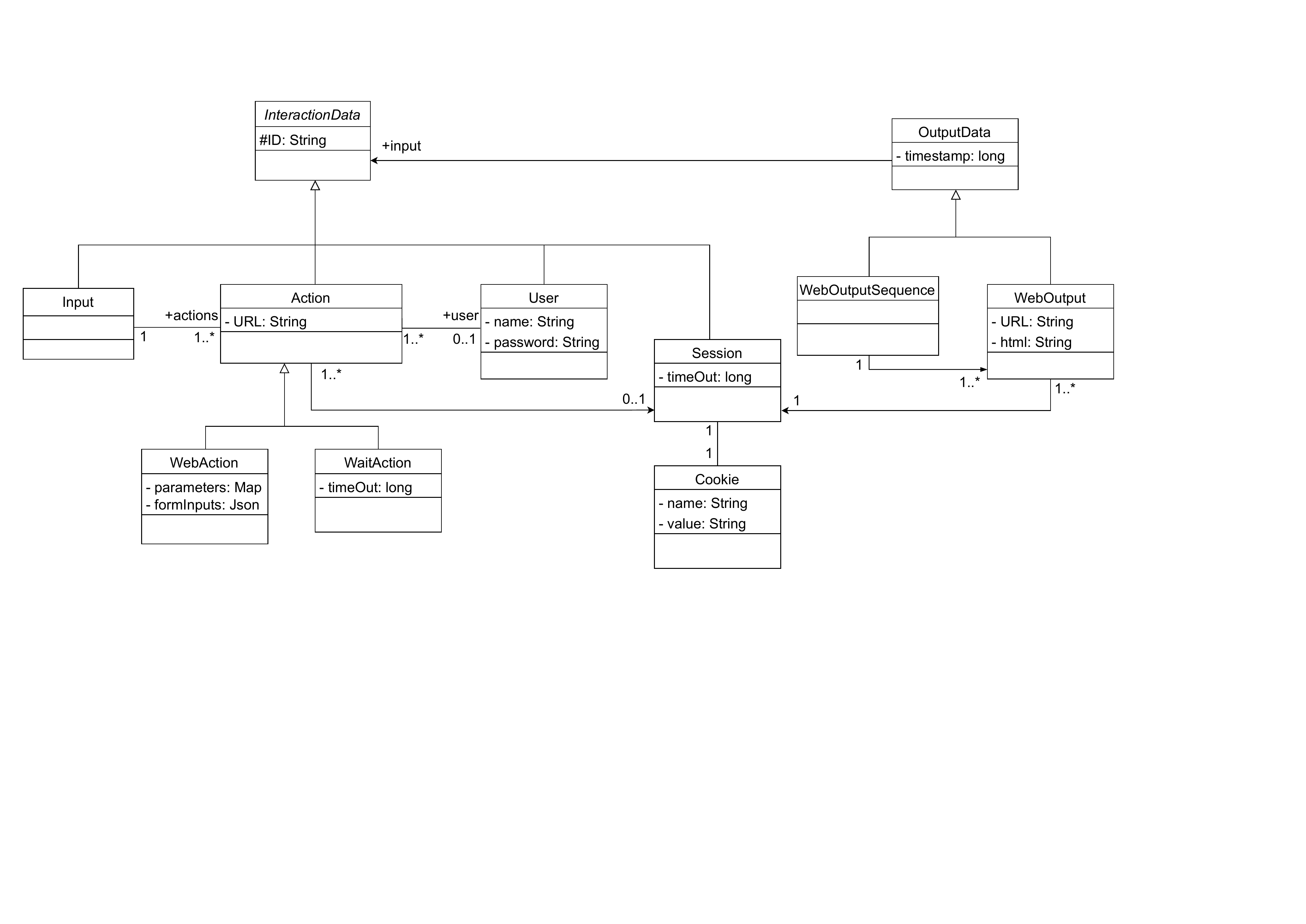}
\vspace*{-15.1em}
\caption{Metamorphic data classes in \SMRL.}
\label{fig:dataClasses}
\end{center}
\end{figure*}

Fig.~\ref{fig:dataClasses} presents the data model for Interaction and Output data. We focus on these two categories because they include the input and output types for the SUT.  The other data categories concern data that is represented using Strings and used to assign values to   Interaction data attributes. 
\TEXTTT{Input} refers to a sequence of interactions between a user and the SUT; such interaction sequences are the only way to give inputs to the SUT. It is consequently associated with \TEXTTT{Action},
which represents an activity performed by a user. 
\SMRL provides two types of actions: \TEXTTT{WebAction} and \TEXTTT{WaitAction}.
\TSE{1.2, 2.5}{\TEXTTT{WebAction} captures a browser activity (e.g., sending an HTML form).
It carries information about actions (e.g., the \emph{URL} where the form is submitted and the \emph{form inputs} provided to the URL with a POST request or the URL \emph{parameters} for GET requests).} 
\TEXTTT{WaitAction} simulates time passing by changing the system time. 
\TEXTTT{User} represents a system user. 

In \SMRL, source and follow-up inputs are always instances of the Input class, with the follow-up input derived through a Web-utility function (see Section~\ref{subsec:dsl_op}).
A follow-up input may differ from the source input because it includes a set of actions that differ from those in the source input or because it performs actions as a different user. In the rest of the paper, to simplify the writing, we use the terms \emph{follow-up action} and \emph{follow-up user} to indicate an Action or a User 
derived by modifying an action or a user in the source input. 

Instances of \TEXTTT{OutputData} capture outputs generated by the system during a system-user interaction; each instance of \TEXTTT{OutputData} is associated with an instance of \TEXTTT{InteractionData}.
\TSE{1.2, 2.5}{Finally, both Actions and WebOutputs can be associated with a Session; indeed, \MRs can modify the  Session Cookies set before executing an Action and retrieve the Session Cookie returned by a Web page.}


\subsection{\MR Operators}
\label{subsec:boolean}

\SMRL provides seven operators, i.e., \TEXTTT{IMPLIES}, \TEXTTT{AND}, \TEXTTT{OR}, \TEXTTT{TRUE}, \TEXTTT{FALSE}, \TEXTTT{NOT}, \TEXTTT{CREATE}, and \TEXTTT{EQUAL}. 
They enable the definition of \emph{metamorphic expressions}, which are boolean expressions that should hold for an \MR to be true.
A \emph{metamorphic expression} is a specific kind of \TEXTTT{XBlockExpression} in XText.
We use metamorphic expressions to decompose an \MR into simple properties.
They are defined in a declarative manner, which is standard practice in \MT.

The \MR in Fig.~\ref{fig:mrExample} includes a metamorphic expression using the operator \TEXTTT{IMPLIES}.
Since the expression is within a loop body, 
the relation holds only if the expression evaluates to true in all the iterations over the input actions.

The semantics of operators \TEXTTT{IMPLIES}, \TEXTTT{AND}, \TEXTTT{OR}, \TEXTTT{TRUE}, \TEXTTT{FALSE}, and \TEXTTT{NOT} is straightforward.
Operator \TEXTTT{CREATE} defines a follow-up input by creating a copy of the source input passed as a second parameter. The follow-up input is identified by the keyword provided as the first parameter. 
In Fig.~\ref{fig:mrExample}, operator \TEXTTT{CREATE} defines the follow-up input \TEXTTT{Input(2)} as a modified copy of \TEXTTT{Input(1)}.
Operator \TEXTTT{CREATE} returns true if the follow-up input is successfully created.

Depending on the context, operator \TEXTTT{EQUAL} either evaluates the equality of two arguments or defines a follow-up input similarly to \TEXTTT{CREATE}. The construct \TEXTTT{EQUAL( Input(2), Input(1) )} enables writing an \MR in a declarative manner without caring if \TEXTTT{Input(2)} should be generated as a copy of Input(1) or if \TEXTTT{Input(2)} already exists and it is equal to \TEXTTT{Input(1)}.
Operator \TEXTTT{EQUAL} is evaluated to false when its first parameter refers to an input that has
already been defined and used previously, in addition to not being equal to the second parameter.
Operator \TEXTTT{CREATE} in Fig.~\ref{fig:mrExample} can be replaced with \TEXTTT{EQUAL} to obtain an equivalent \MR (i.e., an \MR that passes and fails with the same source inputs). The main difference between \TEXTTT{CREATE} and \TEXTTT{EQUAL} is that operator \TEXTTT{CREATE} returns false if the identifier provided for the follow-up input already exists. Based on our experience, using operator CREATE simplifies the understanding of \MRs for external readers.

\subsection{Web-Utility Functions}
\label{subsec:dsl_op}

\MRs for security testing often capture complex properties of Web systems that we cannot express with simple boolean or arithmetic operators. 
Therefore, \SMRL provides some functions that capture standard Web system properties and alter Web data. 
Table~\ref{table:dataOperators} describes a portion of the {\NAZANIN{55}} Web-specific functions in \SMRL~\cite{WebSMRL}. 
Each one is provided as a method of the \SMRL API. Engineers can specify additional functions as Java methods. 
The new functions can be used in \SMRL thanks to the underlying Xtext framework.

\input{tables/dsOperators.tex}

The \MR in Fig.~\ref{fig:mrExample} uses the Web-specific functions \TEXTTT{cannotReachThroughGUI}, \TEXTTT{isSupervisorOf}, \TEXTTT{isError}, and \TEXTTT{changeCredentials}.
The relation indicates that 
the same sequence of actions should provide different outputs when performed by two users under a condition: the two users cannot access one of the requested URLs by browsing the GUI of the system.
In other words, if the system does not provide a URL to a user through its GUI, then she should not access the URL.
Also, to avoid false alarms, the user who cannot access the URL from the GUI (indicated as \TEXTTT{User(2)} in Fig.~\ref{fig:mrExample}) should not be a supervisor with access to all the resources of the other user (\TEXTTT{User(1)}).
Finally, we avoid source inputs that return an error message to \TEXTTT{User(1)} because it is impossible with these inputs to characterize the output that should be observed for \TEXTTT{User(2)}, which may be the same error, a different error, or an empty page. 

Function \TEXTTT{cannotReachThroughGUI} in Fig.~\ref{fig:mrExample} checks if the URL of the current action cannot be reached from the GUI (Line 9).
Function \TEXTTT{isSupervisorOf} checks if \TEXTTT{User(2)} is not a supervisor of \TEXTTT{User(1)} (Line 10).
Function \TEXTTT{isError} returns true based on a configurable regular expression (Line 11) which checks if an output page contains an error message.
Function \TEXTTT{changeCredentials} creates a copy of a provided input sequence using different credentials. 
It is invoked to define the follow-up input (Line 12).
Data function \TEXTTT{Output} executes the sequence of actions in an input sequence (e.g., requests a sequence of URLs) and returns the output of the $i^{th}$ action.

%% file: tables/tableDataOperators.tex
\begin{table*}[tb]
\scriptsize
\caption{Data functions in \SMRL.}
\begin{tabular}{|@{\hspace{0.05cm}}p{1cm}| @{\hspace{0.05cm}}p{3.2cm} | @{\hspace{0.05cm}}p{13.2cm} |}
\hline
\textbf{Category}&\textbf{Data function}&\textbf{Description}\\
\hline
\multirow{5}{*}{\emph{Interaction}}&Input(int i) & Returns the i\textsuperscript{\emph{th}} input sequence.\\
\cline{2-3}
&Action(int i) & Returns the i\textsuperscript{\emph{th}} input action.\\
\cline{2-3}
&ActionAvailable\-WithoutLogin(int i)& Returns the i\textsuperscript{\emph{th}} input action that can be performed without logging into the system.\\
\cline{2-3}
&User(int i) & Returns the i\textsuperscript{\emph{th}} user of the system.\\
\cline{2-3}
&ParameterValueUsedBy\-OtherUsers(Action i, par i) & Returns, for the same action and parameter position, the i\textsuperscript{\emph{th}} parameter value used by a user different than the one executing the action.\\

\hline
\multirow{4}{*}{\emph{SUT}}&RandomFilePath(int i) & Returns the i-th file system path. We select paths of files in the Web system subfolder, ignoring images, and replacing symbolic links (e.g., `plugins' is mapped to `plugin' in Jenkins).\\
\cline{2-3}
&RandomAdmin\-FilePath(int i) & Returns the i-th path to configuration file for the SUT. \\
\cline{2-3}
&Log(int i) & Returns the i-th path to a log file generated by the SUT. \\
\hline
\multirow{14}{*}{\MST}&HttpMethod(int i) & Returns the i-th name of an HTTP method (e.g., DELETE).\\
\cline{2-3}
&SQLInjectionString(int i) & Returns the i-th attack string to be used to perform an SQL injection (e.g., \texttt{'or '1' = '1}).\\
\cline{2-3}
&CodeInjectionString(int i) & Returns the i-th attack string to be used to perform an code injection, e.g.,
\texttt{/\%3C?php\%20system(\%22/bin/ls} \texttt{\%20-l\%22);?\%3E}.\\
\cline{2-3}
&XSSInjectionString(int i) & Returns the i-th attack string to be used to perform an XSS injection, e.g., \texttt{<SCRIPT>alert('XSS');</SCRIPT>}.\\
\cline{2-3}
&StaticInjectionString(int i) & Returns the i-th attack string to be used to perform an static code injection.\\
\cline{2-3}
&LDAPInjectionString(int i) & Returns the i-th attack string to be used to perform an LDAP  injection.\\
\cline{2-3}
&XQueryInjection(int i) & Returns the i-th attack string to be used to perform an XQuery  injection.\\
\cline{2-3}
&CommandInjection(int i) & Returns the i-th attack string to be used to perform a command  injection.\\
\cline{2-3}
&CRLFAttackString(int i) & Returns the i-th CRLF attack string, e.g., `\texttt{);die(2}'.\\
\cline{2-3}
&WeakPassword(int i) & Returns the i-th weak password to be tested.\\
\cline{2-3}
&SpecialCharacters(int i) & Returns the i-th special character to be tested.\\
\cline{2-3}
&FileWithInvalidType(int i) & Returns the path to the i-th file with invalid type to be tested.\\
\cline{2-3}
&XMLInjectedFile(int i) & Returns the path to the i-th XML file containing an XML injection.\\
\cline{2-3}
&RandomValue(Type~t) & Returns a random value of the given type.\\
\hline

\multirow{2}{*}{\emph{Output}}&Output(Input i) & Returns the sequence of outputs generated by Input \emph{i}.\\
&Output(Input i, int n) & Returns the output generated by the n\textsuperscript{\emph{th}} action of Input \emph{i}.\\



\hline
\end{tabular}
Note: for each data function in the table, except for the \emph{Output} category, we provide a convenience method that does not include the parameter \emph{int i} and simply returns the first data item for that type; this leads to 46 data functions.
\label{table:dataMethods}
\end{table*}%

%% file: tables/dsOperators.tex
\begin{table}[tb]
\scriptsize
\caption{Excerpt of the Web-specific functions in \SMRL.}
\begin{tabular}{|@{\hspace{0.1cm}}p{3.6cm} | @{\hspace{0.05cm}}p{4.6cm} |}
\hline
\textbf{Function}&\textbf{Description}\\
\hline
changeCredentials(Input i, User u) & Creates a copy of the provided input sequence where the credentials of the specified user are used (e.g., within login actions).\\
\hline
copyActionTo(Input i, int from, int to) & Creates a new input sequence where an action is duplicated in the specified position and the remaining actions are shifted by one.\\
\hline
cannotReachThroughGUI( User u, String URL) & Returns true if a URL cannot be reached by the given user by exploring the user interface of the system (e.g., by traversing anchors).\\
\hline
isLogin(Action a) & Returns true if the action performs a login.\\
\hline
isSupervisorOf(User a,User b) & Returns true if `a' can access the URLs of `b'.\\
\hline
afterLogin(Action a) & Returns true if the action follows a login.\\
\hline
isSignup(Action a) & Returns true if the action registers a new user on the system.\\
\hline
isError(Output page) & Returns true if the page contains an error message.\\
\hline
userCanRetrieveContent(User u, Object out) & Returns true if the output data (i.e., the argument `out') has ever been received in response to any of the input sequences executed by the given user during data collection. \\
\hline
{\NAZANIN
isResetPassword(Action action)}  &{\NAZANIN Returns true if the action is resetting the password. }\\

\hline
{\NAZANIN
EncodeUrl(String url)}  &{\NAZANIN Returns the  {\texttt{UTF\_8}} encoded version of the requested URL. }\\

\hline
{\NAZANIN
setChannel(String string)}  &{\NAZANIN Modifies the transfer protocol according to the string value (e.g. Http). }\\

\hline
{\NAZANIN
setSession(Session newSession)}  &{\NAZANIN Sets the session cookie value to match on the passed one.}\\

\hline
\end{tabular}
\label{table:dataOperators}
\end{table}%

%% file: transformation.tex
\section{Step 2: Transform \MRs to Java}
\label{sec:transformation}

In Step 2, \SMRL specifications are automatically transformed into Java code.
To this end, we extended the Xbase compiler (hereafter, \SMRL compiler).
Each \MR is transformed into a Java class with the relation name and package.
The generated classes extend class \TEXTTT{MR} and implement its method \TEXTTT{mr}. 

\begin{figure*}[tb]
\begin{center}
\includegraphics[width=18.0cm]{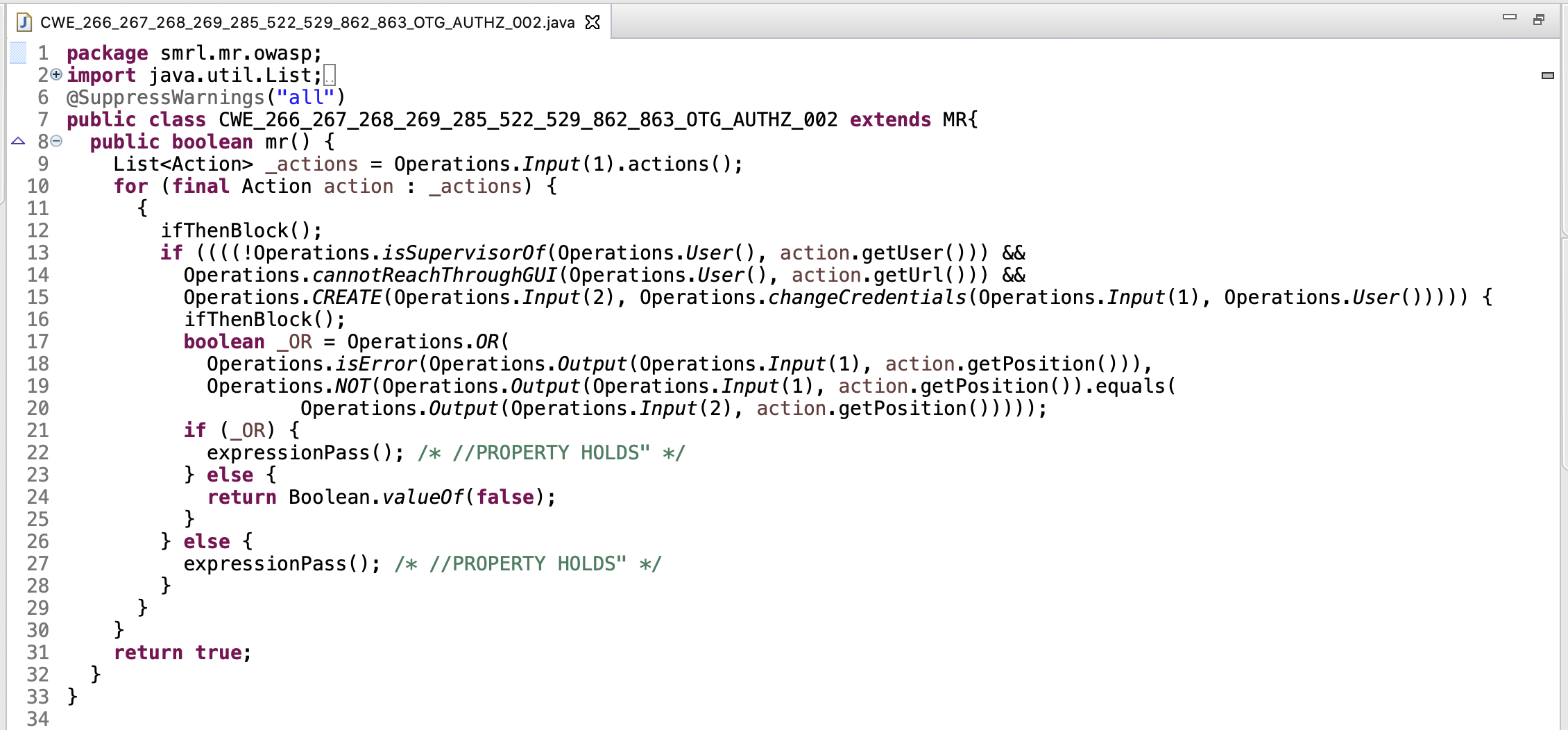}
\caption{Java code generated from the \MR in Fig.~\ref{fig:mrExample}.}
\label{fig:generated002}
\end{center}
\end{figure*}

\begin{figure*}[tb]
\begin{center}
\includegraphics[width=18.2cm] {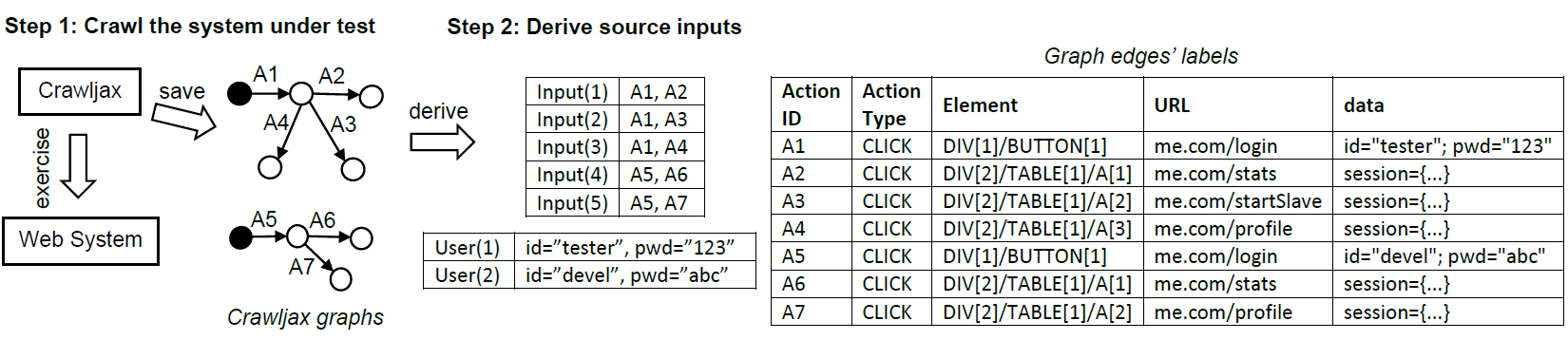}
\caption{Data collection with a simplified example.}
\label{fig:dataCollectionExample}
\end{center}
\end{figure*}

Method \TEXTTT{mr} executes the metamorphic expressions in the \MR. 
It returns \TEXTTT{true} if the relation holds and \TEXTTT{false} otherwise.
To do so, the \SMRL compiler transforms each boolean operator into a set of nested \TEXTTT{IF} conditions.  
For example, for operator \TEXTTT{IMPLIES}, the generated code returns \TEXTTT{false} when the first parameter is true and the second one is false. 
For the case in which the \MR holds, the SMRL compiler generates a statement that returns \TEXTTT{true} at the end of method \TEXTTT{mr}.

Fig.~\ref{fig:generated002} shows the Java code generated from the \MR in Fig.~\ref{fig:mrExample}. 
A loop control structure is derived from the loop instruction in the relation (Line 10). The loop body contains the Java code generated from the metamorphic expression using operator \TEXTTT{IMPLIES} (Lines 13-28). 
The first \TEXTTT{if} condition checks whether the first parameter of operator \TEXTTT{IMPLIES} holds (Lines 13-15). The nested \TEXTTT{IF} block examines whether the second parameter of \TEXTTT{IMPLIES} holds (Line 21). 
If the expression does not hold, \TEXTTT{mr} returns \TEXTTT{false} (Line~24).
The relation holds only if all the expressions in the loop hold. Therefore, the \SMRL compiler generates a \TEXTTT{return~true} statement after the loop body (Line 25). 
Calls to the methods \TEXTTT{ifThenBlock} and \TEXTTT{expressionPass} erase the generated follow-up inputs at each iteration and keep track of the last output observed. Function \TEXTTT{ifThenBlock} determines if
we are within the first \TEXTTT{ifThenBlock} of the \MR (it happens when \TEXTTT{ifThenBlock} is invoked before the first metamorphic expression of the \MR) and, in the affirmative case, erases the follow-up inputs generated so far. Indeed, the first \TEXTTT{ifThenBlock} function is executed at the beginning of each iteration. Function \TEXTTT{expressionPass} empties the list of inputs processed by the last metamorphic expression. Function \TEXTTT{Output} tracks these inputs to provide engineers with contextual information in the case of a failure.

%% file: data_framework.tex

\begin{figure*}[tb]
\begin{center}
\includegraphics[width=14cm]{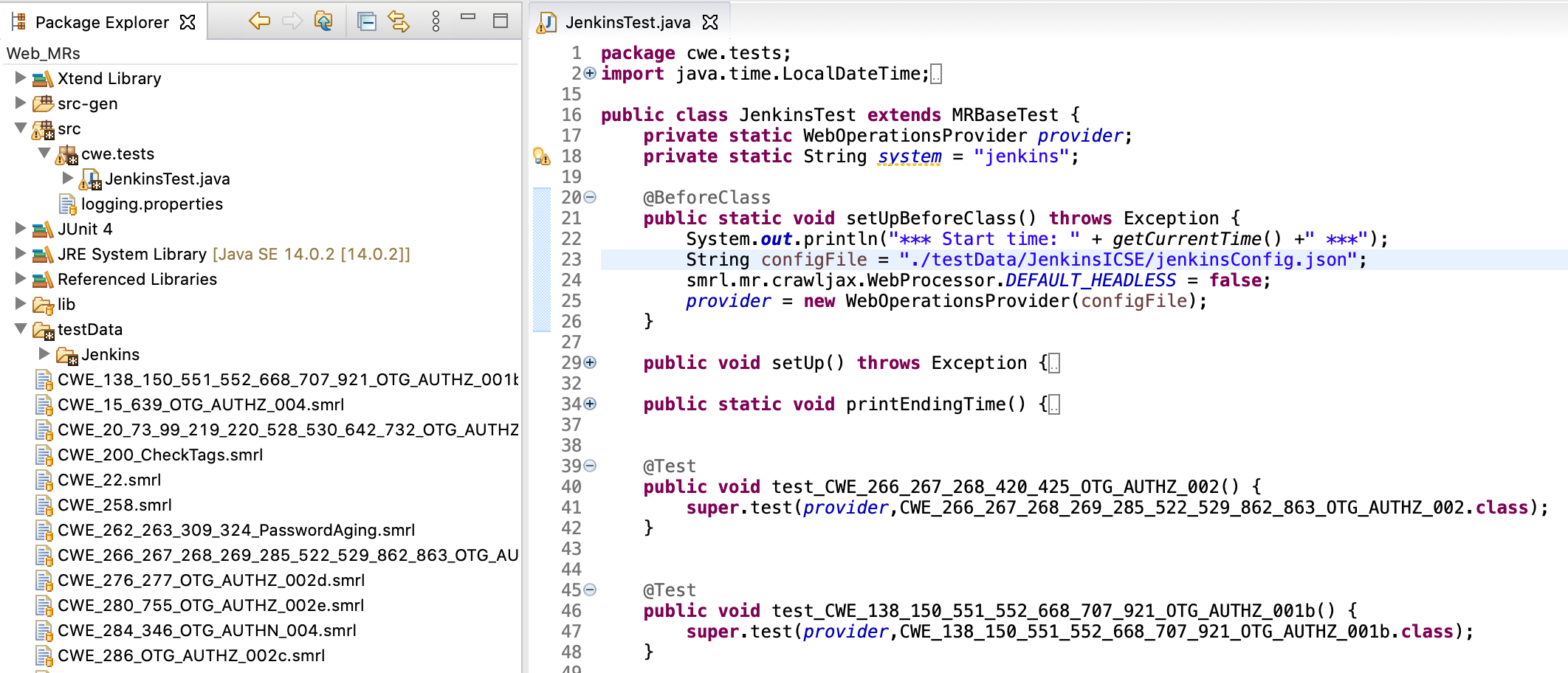}
\caption{An example JUnit test case to select and execute \MRs.}
\label{fig:selection}
\end{center}
\end{figure*}

\section{Step 3: Data Collection}
\label{sec:dataframework}

In Step 3, we rely on an extended version of the Crawljax Web crawler to automatically derive source inputs~\cite{Mesbah2012,Mesbah2008}. Crawljax explores the user interface of a Web system (e.g., by requesting URLs in HTML anchors or by entering text in HTML forms). It generates a graph whose nodes represent the system states reached through the user interface and whose edges capture the action performed to reach a given state (e.g., clicking on a button). Crawljax detects the system states based on the content of the displayed page. 
Our extension relies on the edit distance to distinguish the system states~\cite{Levenshtein}. 
We keep a cache of the HTML page associated with each state detected by Crawljax. 
When a new page is loaded, our extension computes the edit distance between the loaded page and all the pages associated with different system states. When the distance is below a given threshold (5\% of the page length), we assume that two pages belong to the same state. 
If a page does not belong to any state, 
Crawljax adds a new state to the graph.
Crawling stops when no more states are encountered, or when a timeout is reached.

Our Crawljax extensions enable replicating and modifying portions of a crawling session, which is necessary to simulate attacks (e.g., by changing the URL accessed within an input sequence).
To this end, in addition to (i) the Crawljax actions and (ii) the XPath of the elements targeted by the actions (e.g., a button clicked on), our extension records (iii) the URLs requested by the actions, (iv) the data in the HTML forms, and (v) the background URL requests.
This additional data enables, for example, replicating modified portions of crawling sessions that request URLs not appearing on the last Web page returned by the system.
To crawl the SUT, we require only 
its URL and a list of credentials.

Fig.~\ref{fig:dataCollectionExample} exemplifies the data collection steps. 
First, Crawljax generates the graphs of the system under test. 
Second, 
our toolset automatically derives
source inputs from the graphs. 
A source input is a path from the root to a leaf of a Crawljax graph in a depth-first traversal.
Third, the \SMRL functions query the source inputs (see Section~\ref{sec:mtframework}).
For example, \TEXTTT{Input(i)} returns the $i^{th}$ input sequence; \TEXTTT{User(i)} returns the $i^{th}$ unique login credentials in the input sequences. 

In addition to Crawljax, \MST processes manually implemented test scripts to generate additional source inputs.
It processes scripts based on the Selenium framework~\cite{web:selenium} and derives a source input from each script.
We rely on test scripts to exercise complex interaction sequences not triggered by Crawljax (see Section~\ref{sec:evaluation}).
Crawljax performs an almost exhaustive exploration of the Web interface, which is typically not achieved by test scripts. 
Engineers can reuse scripts developed for functional testing or define new ones.



%% file: mt_framework.tex

\section{Step 4: Execute the \MT Framework}
\label{sec:mtframework}


We automatically perform testing based on the executable \MRs in Java and the data collected by the data collection framework (Step 4 in Fig.~\ref{fig:approach}). 
Our testing framework relies on the JUnit framework~\cite{JUnit} to integrate \MT into traditional testing environments. To choose \MRs to be executed, the engineer writes a JUnit test case. Fig.~\ref{fig:selection} presents a JUnit test case selecting multiple \MRs (i.e., CWE\_266..\_OTG\_AUTHZ\_002, and CWE\_138..\_OTG\_AUTHZ\_001b) through function \texttt{test}. 
\MRs need to be copied into the workspace (see the files with extension \texttt{.smrl} in the project explorer window in Fig.~\ref{fig:selection}) and referred to in the JUnit test case (see JenkinsTest.java in Fig.~\ref{fig:selection}). 
\MST provides a Java class, \texttt{MRBaseTest} (Line 16 in Fig.~\ref{fig:selection}), which extends the JUnit framework with utility functions to facilitate the selection of \MRs. Method \texttt{setUpBeforeClass} is used to specify the configuration for \texttt{WebOperationsProvider} (Lines 21-26), i.e., the component in charge of loading source inputs. 

\MRs are executed as standard JUnit test classes through the Eclipse user interface or the console wrapper.
Each \MR is treated as a distinct JUnit test case (Lines 40-48 in Fig.~\ref{fig:selection}).
For each \MR, class \texttt{MRBaseTest} automatically invokes our \MT algorithm that executes the \MR.

Fig.~\ref{alg:executeMR} presents our \MT algorithm. 
The algorithm takes as input an \MR and a data provider exposing the collected data (source inputs).
We first process the bytecode of the \MR to identify the types of source inputs 
referenced by the relation (e.g., \emph{Input} and \emph{User}).
To do so, \TEXTTT{extractSourceInputTy\-pes} (Line~\ref{alg:executeMR:extract}) identifies the calls to the \emph{data representation functions} using the ASM static analysis framework~\cite{ASM}.
Function \TEXTTT{iterate\-Over\-InputTypes} (Line~\ref{alg:executeIterate:OverInputTypes}) ensures that each source input is used in at least one execution and that all possible source input combinations are stressed during the execution of the relation (e.g., all available URLs with all configured users).
It iterates on all available items for a given input type (e.g., all available users). It is invoked recursively for each input type in the \MR.

\input{algos/executeMRs.tex}

Function \TEXTTT{iterate\-Over\-InputTypes} is driven by the methods exposed by the data provider (Lines~\ref{alg:executeMR:hasMore}~and~\ref{alg:executeMR:nextView}). 
The data provider works as a circular array that provides, in each iteration of \TEXTTT{iterate\-Over\-InputTypes}, 
a different view of the collected data. 
For N input items of a given type (e.g., User), function \TEXTTT{nextView} (Line~\ref{alg:executeMR:nextView}) generates N different views with items shifted by one position. 

The \MR is executed (Line~\ref{alg:executeMR:execute}) after obtaining the views. 
The algorithm generates follow-up inputs from the source inputs at each invocation of operator \texttt{EQUAL}. For example, operator \TEXTTT{EQUAL} in Fig.~\ref{fig:mrExample} makes \TEXTTT{Input(2)} refer to a copy of the input sequence returned by function \TEXTTT{changeCredentials}.

Function \TEXTTT{addFailure} stores the failure context information (i.e., source inputs, follow-up inputs, and system outputs) when the \MR does not hold (Lines \ref{alg:executeMR:resultFalse} and \ref{alg:executeMR:addFailure}). \MST reports only failures that perform HTTP requests (e.g., accessing a URL) not generated by input sequences that led to previously reported failures. Therefore, it reduces the time spent to manually analyze failures triggered by distinct follow-up inputs exercising the same vulnerability.

To guarantee that all input item combinations are used, function \TEXTTT{nextView} is iteratively invoked until all the items of a given input type are processed (Line~\ref{alg:executeMR:hasMore}). 
The \MST data representation function \emph{RandomValue} (see Table~\ref{table:dataMethods}) returns a random data value. For scalability reasons, it is configured to generate up to 100 different values, thus leading to 100 views. 

\begin{figure*}[tb]
\begin{center}
\includegraphics[width=12.7cm]{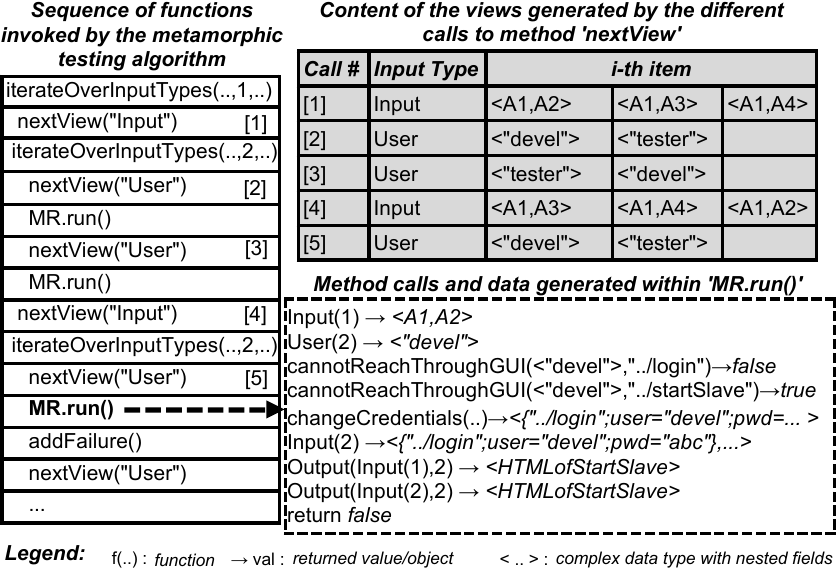}
\caption{Example data processing for the relation in Fig.~\ref{fig:mrExample}.} 
\label{fig:data_processing}
\end{center}
\end{figure*}

\CHANGED{Fig.~\ref{fig:data_processing} illustrates the execution of the relation in Fig.~\ref{fig:mrExample}.}
The table on the left represents the sequence of functions invoked by our algorithm.
In this example, two views for \TEXTTT{User} are inspected for each view of \TEXTTT{Input}.
The first two invocations of \TEXTTT{MR.run} return true (not shown in Fig.~\ref{fig:data_processing}) because the \emph{login} and \emph{stats} pages have been accessed by both users \emph{devel} and \emph{tester},
and, thus, the implication holds. 
The third invocation of \TEXTTT{MR.run} returns false because the output page for the \emph{startSlave} URL is the same for the two input sequences and, thus, the relation does not hold. We rely on edit distance to determine if Web pages are equal.

%% file: algos/executeMRs.tex

\begin{figure}[tb]

\begin{algorithmic}[1]

\scriptsize
\Require \emph{MR}, the bytecode of the metamorphic relation to be executed
\Require \emph{dataProvider}, an object that exposes the data collected by the crawlers
\Ensure \emph{Failures}, a list of failing executions with contextual information

\State srcTypes $\gets$ extractSourceInputTypes(MR) \label{alg:executeMR:extract}
\State iterateOverInputTypes(MR, dataProvider, 0, dataTypes) \label{alg:executeIterate:OverInputTypes}
\State \Return $\mathit{Failures}$
\vspace{2mm}
\Function{iterateOverInputTypes}{MR, dataProvider, i, dataTypes} 
\While {dataProvider.hasMoreViews(dataTypes[i])}  \label{alg:executeMR:hasMore}
\State dataProvider.nextView(dataTypes[i])  \label{alg:executeMR:nextView}
\If{(i $<$ dataTypes.lenght)} \hspace{2mm}\textbf{\textcolor{gray}{//need to iterate over other types}} \label{alg:executeMR:checkAll}
	\State iterateOverInputTypes(MR,dataProvider, i+1,srcTypes)   \label{alg:executeMR:iterateType}
\Else  \hspace{2mm}\textbf{\textcolor{gray}{//we have set a view for every input type in the relation}} \label{alg:executeMR:allViewsSet}
	\State result = MR.run() \textbf{\textcolor{gray}{//execute the metamorphic relation}} \label{alg:executeMR:execute}
	\State \textbf{if} ( result == false)	\textbf{\textcolor{gray}{//the MR does not hold}} \label{alg:executeMR:resultFalse}
	\State  \hspace{5mm}  addFailure(Failures,dataProvider) \textbf{\textcolor{gray}{//trace the failure}} \label{alg:executeMR:addFailure}
\EndIf	
	
\EndWhile
\EndFunction



\end{algorithmic}
\vspace{-3mm}
\caption{Metamorphic testing algorithm.}
\label{alg:executeMR}
\end{figure}

%% file: MRrelations.tex
\section{Catalog of Metamorphic Relations}
\label{sec:mrs}

We derived a catalog of \MRs 
from the activities described in the OWASP book on security testing~\cite{OWASPtesting} and from a set of security vulnerability types related to violations of security design principles and reported in the Common Weakness Enumeration (CWE) database~\cite{CWE-DB}.

The OWASP book on security testing presents detailed descriptions of 90 testing activities (hereafter \textit{OWASP security testing activities}) for Web systems; each OWASP security testing activity targets a specific vulnerability type. 
The activities provide information for implementing metamorphic relations. For example, for the bypass authorization schema vulnerability, OWASP suggests collecting links in administrative interfaces and directly accessing the corresponding URLs by using the credentials of other users. Using this suggestion, we defined the \MR in Fig.~\ref{fig:mrExample}. 

The CWE vulnerability types considered for our catalog concern violations of security design principles. As further discussed in Section~\ref{sec:usage}, they are the ones we selected to address our research questions about the applicability of \MST as they correspond to system-level violations resulting from design mistakes.




We can perform security testing activities (e.g., simulate attacks) in multiple ways and may have more than one relation for each OWASP activity or CWE vulnerability type. Also, not all testing activities benefit from \MT. We discuss the capabilities of \MT in Section~\ref{sec:usage}.

Our catalog includes 76 \MRs in total.
22 \MRs automate 16 OWASP activities; 54 \MRs detect 101 CWE vulnerability types. 15 \MRs both automate OWASP activities and detect CWE vulnerability types, which is unsurprising since OWASP activities aim to verify that secure design principles (e.g., correctly implement and configure authorization mechanisms) are in place. Overall, our \MR catalog discovers 102 security vulnerability types, including 101 belonging to the CWE catalog (some of which are discovered by OWASP testing activities) and one vulnerability type being targeted by OWASP testing activities only.


We perform security testing using follow-up inputs that cannot be generated by interacting with the GUI of the system but conform with the input format of the system and match its configuration (e.g., the URLs requested by the unauthorized user refer to existing system resources). We inherit, from mutational fuzzing, the idea of generating follow-up inputs by altering valid source inputs~\cite{fuzzingbook2019}. However, we do not rely only on random values to obtain valid inputs that match the system configuration. Instead, we modify source inputs using the data provided by the \SMRL Web-utility functions, which return domain-specific information (e.g., protocol names), random values, and crawled data. Finally, by capturing properties of the output generated from the source and follow-up inputs, we identify vulnerabilities that cannot be detected with implicit test oracles~\cite{Barr2015} (e.g., crashes).

We present some of the \MRs in our catalog in Figs.~\ref{fig:simulation} to~\ref{fig:simulation10}. 
The entire catalog is available for download~\cite{WebSMRL}. All the \MRs in our catalog follow the template in Fig.~\ref{alg:MRtemplate} in Section~\ref{sec:template}. Also, the sequences of invocations of Web-utility functions in our \MRs can be grouped into a set of patterns given in Section~\ref{sec:patterns}. The identification of patterns facilitates the definition of solutions to a problem~\cite{patterns}; in our case, the \MR patterns facilitate the identification of \MRs to address the problem of automatically performing security testing based on the description of a vulnerability type and corresponding attacks. In other words, \MR patterns help understand how to derive \MRs that simulate certain attacks; in turn, they enable engineers to define new \MRs for attacks not considered yet (e.g., by looking at similarities with attacks already considered).

%% file: MRs_patterns.tex

\subsection{Metamorphic Relations Template}
\label{sec:template}

\begin{figure}[tb]
\newcommand{\IND}[0]{2mm}
\newcommand{\INDD}[0]{4mm}
\begin{algorithmic}[1]

\scriptsize

\State Iterate over all the Actions of Input \label{alg:MRtemplate:loop}

\State \hspace{\IND} (optional) Iterate over all the Actions of another Input \label{alg:MRtemplate:combineInputs}

\State \hspace{\IND} (optional) Iterates over all the parameters, form inputs, or session cookies of the selected Action
\label{alg:MRtemplate:iterateParameters}

\State \hspace{\IND} \textbf{IMPLIES} $\big($ \label{alg:MRtemplate:implies}

\State \hspace{\INDD} $[$a precondition holds $\big]_{\text{repeatable multiple times}}$ \label{alg:MRtemplate:precondition}

\State \hspace{\INDD} $[$ and a follow-up input is successfully created$\big]_{\text{repeatable multiple times}}$ \label{alg:MRtemplate:follow}

\State \hspace{\INDD} $\big[$ and the follow-up input is successfully modified $\big]_{\text{repeatable multiple times}}$ \label{alg:MRtemplate:modify}

\State \hspace{\IND} \textbf{,}

\State \hspace{\INDD} condition on the outputs generated for the source and follow-up inputs

\State \hspace{\IND} $\big)$

\end{algorithmic}
\vspace{-3mm}
\caption{Template common to all the \MRs in our catalog.}
\label{alg:MRtemplate}
\end{figure}

This section presents the template followed by all the \MRs in our catalog. This template was not defined up-front but is the result of deriving \MRs for more than 100 specifications (16 OWASP activities and 101 CWE vulnerabilities) and analyzing their commonalities.

All the \MRs include a loop (Line~\ref{alg:MRtemplate:loop} in Fig.~\ref{alg:MRtemplate}) to define
multiple follow-up inputs by iteratively modifying the actions of the source input. 
\MRs may have additional nested loops used either to combine multiple source inputs (Line~\ref{alg:MRtemplate:combineInputs}) or to iterate over all the parameters, form entries, or session cookies belonging to the Web page on which the action is performed (Line~\ref{alg:MRtemplate:iterateParameters}). For example, \texttt{CWE\_287a\_425\_OTG\_AUTHN\_001} (Fig.~\ref{fig:simulation}) works with all the login actions observed in the source input. 
Function \TEXTTT{isLogin()} returns true only if the current action performs a login; 
otherwise, the implication trivially holds, and no follow-up input is generated.

We express all our \MRs using an implication (operator \TEXTTT{IMPLIES} in Line~\ref{alg:MRtemplate:implies} of Fig.~\ref{alg:MRtemplate}). The left-hand side of the implication often starts with verifying if a precondition holds (Line~\ref{alg:MRtemplate:precondition}). Then we check if the follow-up input is successfully defined (Line~\ref{alg:MRtemplate:follow}). 
Finally, we ensure that the follow-up input can be further modified through multiple function calls appearing on the left-hand side of the implication (Line~\ref{alg:MRtemplate:modify}). 
For instance, the first and second follow-up inputs in \MR \texttt{CWE\_302\_471\_472\_784\_807} (Fig.~\ref{fig:simulation1}) are for performing the actions of the source inputs with another user (to ensure the follow-up user cannot perform them) and the attack, respectively.



The right-hand side of the implication usually captures the relation between the outputs of the source and follow-up inputs. For instance, according to \texttt{CWE\_287a\_425\_OTG\_AUTHN\_001} in Fig.~\ref{fig:simulation}, the output for the follow-up input (which performs a login on the unencrypted HTTP channel) should be different from the output for the source input because it should not be possible to login using the HTTP channel.

In the following, we describe the elements of our template: preconditions, generation of follow-up inputs, and output conditions (postconditions).
Table~\ref{MRsTemplateElements} provides an overview of possible template element instances.

\input{tables/tableMRtemplateElementInstances}


\subsubsection{Preconditions}

Preconditions generally concern the actions in the source input or the user performing these actions. 

User preconditions are applied to identify the users performing the follow-up inputs based on access level. For instance, we use the function \textit{isSupervisorOf()} to ensure that the follow-up user doesn't have access to a resource from another user. Also, we rely on the function \textit{notTried(User, actionURL)} to avoid testing the same URL multiple times with the same user; indeed, this function indicates if we have already exercised a URL with a follow-up input for the specified user.

Action preconditions avoid redundant follow-up inputs and false positives.
For example, the function \textit{notTried(actionURL)} is used not to test the same URL twice (regardless of the user performing the action) in \MRs that do not address authorization and authentication vulnerabilities.
To reduce false positives, we may avoid actions whose output leads to error or alert messages. For instance,
\texttt{!Output(Input(1),pos).hasAlert} for \texttt{CWE\_79\_a\_XSSreflected} (Fig.~\ref{fig:simulation2}) checks if the source input action does not lead to a popup message, i.e., the condition used to determine if the XSS injection attack is successful.

72 MRs (94\%) and 33 MRs (43\%) in our catalog include an action precondition and a user precondition, respectively.

\subsubsection{Generation of follow-up inputs}

Follow-up inputs can be generated via input generation strategies focusing on the sequence of actions in the source input and the user performing actions.

We have two strategies for the \emph{user}: (i) keeping the same user of the original source input sequence and (ii) selecting another user. The second one is adopted for \MRs looking for vulnerabilities related to confidentiality, authorization, or authentication (e.g., accessing a resource with different users). 
The first one is used in all the other \MRs. In total, eight \MRs in our catalog create follow-up inputs whose actions are performed with a user different than the one in the source input.




Concerning actions, the strategies employed to derive follow-up inputs include (i) relying on the same sequence of actions of the source input sequence, (ii) retaining only a subset of the actions, (iii) adding actions to the source input, and (iv) modifying the actions in the source input. 

\begin{figure*}[tb]
\begin{center}

\begin{lstlisting}[language=Java]
MR CWE_287a_425_OTG_AUTHN_001 { 
{ 
   for ( Action action : Input(1).actions() ) {                     //(1)
      var pos = action.getPosition();                               //(2)
      IMPLIES(
          isLogin(action) &&                                        //(3)
          notTried(action.url) &&                                   //(4)
          CREATE ( Input(2), Input(1) ) &&                          //(5)
          Input(2).actions.get(action.position).setChannel("http")  //(6)
          ,
          different ( Output(Input(1),pos),  Output(Input(2),pos) ) //(7)   	
    );//end-IMPLIES
   }//end-for
  } 
 }//end-MR
}//end-package
\end{lstlisting}
 
 \raggedright{\footnotesize{
 
 (1) For loop iterates over all actions of Input(1).
 (2) Stores the parameters of the current action in a variable.
  (3) It checks that the action is a login.
  (4) It verifies that this login URL was not already tested.
  (5) Creates the follow-up input by copying the input(1).
 (6) Sets the HTTP channel for the follow-up input.
  (7) A login operation should not succeed if performed on the HTTP channel.
 	   checks if the output generated by the login operation is different in the two cases.}}
\caption{CWE\_287a\_425\_OTG\_AUTHN\_001: Testing for credentials transported over an encrypted channel}
\label{fig:simulation}
\end{center}
\end{figure*}

\begin{figure*}[tb]
\begin{center}

\begin{lstlisting}[language=Java]
MR CWE_302_471_472_784_807{ 
{
   for (Action action: Input(1).actions()){                                         //(1)
      var pos = action.getPosition();                                               //(2)
      var session = Output( Input(1), pos).session as CookieSession;                //(3)
      CREATE( Input(2), changeCredentials(Input(1), User()) )                       //(4)
      var session2 = Output( Input(2), pos).session as CookieSession;               //(5)
      var notTried = notTried( action.url, Input(1).actions().get(pos).getElementURL()) //(6) 
      var mappings = session.keyValueMappings.entrySet;                             //(7)
      
      for ( Entry<String,String> cookie : mappings){                                //(8)
         var type = typeOf(cookie.value);                                           //(9)
         IMPLIES( 
             notTried && 
             type == Boolean &&                                                    //(10)
             cannotReachThroughGUI( User(), action.url ) &&                        //(11)
             different( Output( Input(1), pos) , Output(Input(2) , pos)) &&        //(12)
             CREATE ( Input(3), Input(2) )  &&                                     //(13)
             Input(3).actions.get(pos).setSession( session2 ) &&                   //(14)
             session.setCookie( 
                 new Cookie(cookie.key, String.valueOf(!Boolean.valueOf(cookie.value)))) //(15)
            ,
         OR(                                                                       //(16)
             different(Output( Input(1), pos) , Output(Input(3), pos)), 
             isError(Output(Input(3) , pos)))
     );//end-IMPLIES
    }//end-for
   }//end-for	
  }
 }//end-MR
}//end-package

\end{lstlisting}
 
 \raggedright{\footnotesize{
 
(1) For loop iterates over all actions of the Input(1).
 (2) Stores the parameters of the current action in a variable.
 (3) Saves the cookie value of the Input(1) in a variable.
 (4) Creates a follow-up input with different credentials.
  (5) Saves the cookie value of the follow-up user in a variable.
  (6) To speed up the process, verifies that the element URL has not been tried before. 
  (7) Stores the session key values in a variable.
  (8) For loop iterates over the session values.
  (9) Assigns the type of the cookie to a variable.
  (10) Filters all cookie types other than Boolean
  (11) Makes sure the URL is not accessible without login.
 (12) To avoid False positives, filters the actions with the same output for Input(1) and Input(2).
 (13) Creates the follow-up input, named Input(3).
 (14) Sets the session2 as the session value of the Input(3).
  (15) Flip the value of the Boolean cookie.
 (16) The system should give a different output or should show an error.}}
\caption{CWE\_302\_471\_472\_784\_807: Testing for assumed-immutable elements} 
\label{fig:simulation1}
\end{center}
\end{figure*}

\begin{figure*}[tb]
\begin{center}

\begin{lstlisting}[language=Java]

 MR CWE_79_a_XSSreflected { 
{ 	
  keepDialogsOpen = true;                                                            //(1)
  for ( Action action : Input(1).actions() ) {                                       //(2)
     for ( var x = 0; x < action.parameters.size; x++){                              //(3)	
        var pos = action.getPosition();                                              //(4)
        IMPLIES(
              notTried(x+action.url, Input(1).actions().get(pos).getElementURL()) && //(5)
              !Output(Input(1), pos).hasAlert &&                                     //(6)
              CREATE( Input(2), Input(1) ) &&		                                     //(7)
              Input(2).actions().get(pos).setParameterValue(x,XSSInjectionString())  //(8)
              ,
           OR(                                                                       //(9)
               Output(Input(2), pos).emptyFile,       
              !Output(Input(2), pos).hasAlert) 
     );//end-IMPLIES
    }//end-for 
   }//end-for 
  }
 }//end-MR
}//end-package

\end{lstlisting} 
 
 \raggedright{\footnotesize{
 
(1) It avoids clicking on OK; because dialogs are normally ignored by our framework by clicking on OK.
  (2) For loop iterates over all actions of the Input(1).
  (3) The second loop iterates over all parameters of the action.
  (4) Defines a variable to store the position of the Input(1)'s action.
  (5) Checks if the parameter of the action URL has not seen before, to speed up the process.
  (6) Verifies that the action does not originally contain any alert.
 (7) Creates the follow-up input by copying the Input(1).
  (8) Sets the injected XSS string as the parameter value of the follow-up input.
  (9) Checks either the attack was not performed or no effect shall be observed (the effect of the XSS is usually visualized when reaching the page where we inject the XSS).  }}
\caption{CWE\_79\_a\_XSSreflected: Testing for reflected cross-site scripting} 
\label{fig:simulation2}
\end{center}
\end{figure*}

\begin{figure*}[tb]
\begin{center}
 \begin{lstlisting}[language=Java]
  MR CWE_286_OTG_AUTHZ_002c {
  {
    for(var y = Input(1).actions().size()-1; ( y > 0 ); y--){                       //(1)
        IMPLIES(
            (!isSupervisorOf(User(), Input(1).actions().get(y).user)) &&            //(2)
            afterLogin(Input(1).actions().get(y)) &&                                //(3)
            cannotReachThroughGUI(User(), Input(1).actions().get(y).getUrl()) &&    //(4)
            CREATE(Input(2), Input(LoginAction(User()), Input(1).actions().get(y))) //(5)
            , 
         OR(                                                                        //(6)
            isError(Output(Input(1), y)),
            different(Output(Input(1), y),Output(Input(2), 1)))
    ); //end-IMPLIES
   } //end-for
  }
 } //end-MR
} //end-package

\end{lstlisting}
 
 \raggedright{\footnotesize{
 
(1) The for loop iterates over all the actions of an input sequence.
 (2) Checks whether the user in User() is not a supervisor of the user performing the y-th action.
 (3) Verifies that the y-th action is performed after a login.
 (4) Verifies that the follow-up user cannot retrieve the URL of the action through the GUI (based on the data collected by the crawler).
  (5) Defines a follow-up input that performs the login as the follow-up user
 (6) The system checks if the y-th action from the source input leads to an error page Or the output generated by the action containing the URL indicated above, lead to two different outputs in the two cases.
}}
\caption{CWE\_286\_OTG\_AUTHZ\_002c: Testing for incorrect user management} 
\label{fig:simulation3}
\end{center}
\end{figure*}

\begin{figure*}[tb]
\begin{center}
 \begin{lstlisting}[language=Java]
MR CWE_841 {
{
   for( var x = 0; x < Input(1).actions().size(); x++ ){                         //(1)
      for( var y = x+2; isSignup(Input(1).actions().get(x)) && 
           (y < Input(1).actions().size()); y++){                                //(2)
         IMPLIES (
             isLogin(Input(1).actions().get(y))&&                                //(3)
             CREATE ( Input(2) , Input ( Input(1).actions().subList(y,
                     Input(1).actions().size())))  &&                            //(4)
             Input(2).addAction(0,Input(1).actions().get(x))   &&                //(5)
             Input(2).addAction(0, new ResetSUTAction())                         //(6)
             , 
             AND( different ( Output(Input(2),1),  Output(Input(1),y) )          //(7) 
                , 
                  isError(Output(Input(2),1)))
     ); //end-IMPLIES
    } //end-for
   } //end-for
  }
 } //end-MR
} //end-package

\end{lstlisting}
 
 \raggedright{\footnotesize{
 
 (1) The first loop iterates over all the actions to find a signup action.
 (2) The second loop iterates over all the actions to find a login action performed after the signup.
  (3) Verifies that the action y is a login.
  (4) Creates the follow-up input with the sequence of actions after the login (action y).
   (5) Makes sure that we start with a signup action
  (6) Reset the actions for next iteration
  (7) Verifies that the system gives two different outputs or will give an error by executing the follow-up input.  }}
\caption{CWE\_841: Testing for improper enforcement of behavioral workflow} 
\label{fig:simulation4}
\end{center}
\end{figure*}

\begin{figure*}[tb]
\begin{center}

 \begin{lstlisting}[language=Java]
MR OTG_SESS_003 {
{
   for( Action signup : Input(1).actions() ){                                 //(1)
     for ( var i=0;isSignup(signup) && i < Input (2).actions().size; i++ ) {  //(2)
        var f = Input(2).actions().get(i);
        var pos = f.getPosition();
        IMPLIES (
             afterLogin( f ) &&                                               //(3)
             CREATE(Input(3), addAction( Input(2), pos+1, signup ))           //(4)
             ,
             different(                                                       //(5)
                    Output(Input(3), pos).getSession(),
                    Output(Input(3), pos+1).getSession())
     );//end-IMPLIES
    }//end-for
   }//end-for
  }
 }//end-MR
}//end-package
\end{lstlisting}
 
 \raggedright{\footnotesize{
 
(1) The first loop iterates over the inputs to find a sign up action.
 (2) The second loop iterates over the actions that follow the sign up. The second loop is necessary to check that a sign up action repeated at any point of the action sequence leads to a new session ID.
 (3) Checks if the current action has been performed after a login.
 (4) Defines a follow-up input with the sign up action being duplicated in a certain position.
 (5) Checks if the session ID of the response page sent after the two successive login actions is different.  }}
\caption{OTG\_SESS\_003: Testing for session fixation} 
\label{fig:simulation5}
\end{center}
\end{figure*}

\begin{figure*}[tb]
\begin{center}
 \begin{lstlisting}[language=Java]
MR CWE_262_263_309_324 {
{
   for ( var x=0; x < Input(1).actions().size() ; x++ ){                    //(1)
       IMPLIES ( 
           isLogin( Input(1).actions().get(x) ) &&                          //(2)
           !isError ( Output(Input(1),x+1) )&&                              //(3)
           CREATE ( Input(2) , Input(1) ) &&                                //(4)
           Input(2).addAction ( x, Wait( 60*60*24*30*12*1000) )             //(5)    
           ,
           different ( Output(Input(2),x+2) ,  Output(Input(1),x+1))        //(6)
					
    ); //end-IMPLIES
   } //end-for
  }
 } //end-MR
} //end-package

\end{lstlisting}
 
 \raggedright{\footnotesize{
 
  (1) For loop iterates over the actions of Input(1).
 (2) Checks if the current action x is doing "log in".
  (3) Checks that the action-x of the Input(1) does not include any errors.
  (4) Creates the follow-up Input by copying the Input(1), named Input(2).
  (5) Add a wait action to Input(2) that moves time forward for a year i.e. the expected time for resetting the password.
  (6) Based on the above description the results should be different; it shall prompt a password update request.}}
\caption{CWE\_262\_263\_309\_324: Testing for password aging}
\label{fig:simulation6}
\end{center}
\end{figure*}

\begin{figure*}[tb]
\begin{center}

 \begin{lstlisting}[language=Java]
MR CWE_20_73_99_219_220_528_530_642_732_OTG_AUTHZ_001a {
{
   for ( Action action : Input(1).actions() ){                                   //(1)
      for ( var par=0; par < action.getParameters().size(); par++ ){             //(2)
        var pos = action.getPosition();                                          //(3)
        IMPLIES( 
            notTried( Input(1).actions().get(pos).getUrl() ) &&                  //(4)
            CREATE( Input(2), Input(1) )	 &&                                    //(5)
            Input(2).actions().get(pos).setParameterValue(par, RandomFilePath()) //(6)
            ,
        OR( 	                                                                   //(7)
            isError(Output(Input(2),pos)) 	
            ,
            userCanRetrieveContent(action.getUser(), Output(Input(2),pos)) )
     );//end-IMPLIES
    }//end-for
   }//end-for 
  }
 }//end-MR
}//end-package

\end{lstlisting}
 
 \raggedright{\footnotesize{
 
 (1) Iterates over the actions of the Input(1).
  (2) The second for loop iterates over the parameters of the action.
 (3) Stores the parameters of the current action in a variable.
 (4) To speed up the process, verifies that the URL has not been tried before.
 (5) Creates the follow-up input, named Input(2).
 (6) Sets the value of a parameter to a random file path.
 (7) Verifies that the system shows an error page or the returned content is something that the user has the right to access.}}
\caption{CWE\_20\_73\_99\_219\_220\_528\_530\_642\_732\_OTG\_AUTHZ\_001a: Testing for directory traversal} 
\label{fig:simulation7}
\end{center}
\end{figure*}

We \emph{keep the same sequence of actions} of the source input for \MRs that focus on authorization and authentication vulnerabilities; indeed, to test the authorization mechanisms of a system, we can perform the same actions of the source input but with a different access level (e.g., \texttt{CWE\_266\_..\_OTG\_AUTHZ\_002} in Fig.~\ref{fig:mrExample}). 
Eight \MRs in our catalog (10\%) perform the same actions of the source input either with a different user (e.g., \texttt{CWE\_266\_..\_OTG\_AUTHZ\_002}) or in an alternative channel (e.g., \texttt{CWE\_287a\_425\_OTG\_AUTHN\_001} in Fig.~\ref{fig:simulation}).

\begin{figure*}[tb]
\begin{center}

 \begin{lstlisting}[language=Java]
MR CWE_15_639_OTG_AUTHZ_004 {
{
   for ( Action action : Input(1).actions() ){                                //(1)
      for ( var par=0; par < action.getParameters().size() 
            && notTried( action.getUser(), action.url); par++ ){              //(2)
        for ( usedValue : parameterValuesUsedByOtherUsers(action, par) ){     //(3)
          var pos = action.getPosition();                                     //(4)
          IMPLIES( 
             CREATE ( Input(2), Input(1) ) &&	                                //(5)
             Input(2).actions().get(pos).setParameterValue(par, usedValue)    //(6)
             , 
          OR(                                                                 //(7)
             Output(Input(2),pos).isError()                                       
             , 
             userCanRetrieveContent( action.user, Output(Input(2),pos) ))
      );//end-IMPLIES
     }//end-for
    }//end-for
   }//end-for
  }
 }//end-MR
}//end-package

\end{lstlisting}
 
 \raggedright{\footnotesize{
 
(1) The first loop iterates over all the actions of the input sequence. 
  (2) The second iterates over all the parameters of the action.
  (3) The third loop iterates over all used values by other users.
  (4) Stores the parameters of the current action in a variable.
  (5) Defines the follow-up input. 
  (6) Sets a parameter value to a value that is used by other users.
  (7) Checks if the content of the output is either an error message Or some content that can be retrieved from the GUI.}}
\caption{CWE\_15\_639\_OTG\_AUTHZ\_004: Testing for externally controlled elements} 
\label{fig:simulation8}
\end{center}
\end{figure*}

\begin{figure*}[tb]
\begin{center}

 \begin{lstlisting}[language=Java]
 MR CWE_94_96_B {
{
  keepDialogsOpen = true;                                                          //(1)
  for (Action action : Input(1).actions()){                                        //(2)
     var pos = action.getPosition();                                               //(3) 
     for (var x=0; action.containFormInput() && x < action.formInputs.size; x++){  //(4)
        IMPLIES(
            action.isClickOnButton &&                                              //(5)
            notTried(x+action.url, Input(1).actions().get(pos).getElementURL()) && //(6)
            userCanRetrieveContent(action.getUser(), Output(Input(1),pos)) &&      //(7)
            CREATE(Input(2), Input(1) ) &&                                         //(8)
            Input(2).actions.get(pos).setFormInput(x, StaticInjectionString())     //(9)
            ,
        OR(                                                                        //(10)
            isError(Output(Input(2), pos)),       
            userCanRetrieveContent(action.getUser(), Output(Input(2),pos)))
     );//end-IMPLIES
    }//end-for parameter
   }//end-for action
  }
 }//end-MR
}//end-package

\end{lstlisting}
 
 \raggedright{\footnotesize{
 
(1) It avoids clicking on OK; because dialogs are normally ignored by our framework by clicking on OK.
  (2) For loop iterate over all actions of the Input(1).
  (3) Define a variable to store the position of the Input(1)'s action.
 (4) The second loop iterates over all parameters of the action if the action contains a form input.
  (5) makes sure the user will submit the form.
 (6) Verifies the current parameter was not tested before.
  (7) Filters out the web pages with dynamic content.
 (8) Creates the follow-up input by copying the Input(1).
 (9) Injects some Static injection strings to the follow-up input.
 (10) The system may show an error page to the user or the user is retrieving the content that has right to access it. }}
\caption{CWE\_94\_96\_B: Testing for static code injection} 
\label{fig:simulation9}
\end{center}
\end{figure*}

\begin{figure*}[tb]
\begin{center}

 \begin{lstlisting}[language=Java]
 MR CWE_792_793_794_795_796_797_A{ 
{
  keepDialogsOpen = true;                                                   //(1)
  for ( Action action : Input(1).actions() ) {                              //(2)
    for ( var par=0; par < action.getParameters().size(); par++ ){          //(3)  	
       var pos = action.getPosition();                                      //(4)
       var value = Input(1).actions().get(pos).getParameterValue(par);      //(5)
       IMPLIES(
           notTried( Input(1).actions().get(pos).getUrl() ) &&              //(6)
           value != Boolean &&                                              //(7)
           CREATE(Input(2), Input(1))&&                                     //(8)
           Input(2).actions().get(pos).setParameterValue(par,
                SCInjection_beginning(value,SpecialCharacters()))           //(9) 
           ,
        OR(                                                                 //(10)
           isError(Output(Input(2), pos)), 
           EQUAL(Output(Input(1), pos),Output(Input(2), pos))) 
     );//end-IMPLIES
    }//end-for
   }//end-for
  } 
 }//end-MR
}//end-package

\end{lstlisting}
 
 \raggedright{\footnotesize{
 
(1) It avoids clicking on OK; because dialogues are normally ignored by our framework by clicking on OK.
  (2) For loop iterates over all actions of the Input(1).
  (3) Iterates over all parameters of each action.
  (4) Stores the parameters of the action in the variable. 
  (5) Reads the user's input value and keeps it in a variable.
  (6) Checks if the URL was not tried before, to speed up the process.
 (7) Filters out the boolean input values.
  (8) Creates the follow-up input by copying the Input(1).
  (9) Sets the new input value which contains special character for the follow-up input.
 (10) Verifies that the system should show an error page or it would neutralize the input.   }}
\caption{CWE\_792\_793\_794\_795\_796\_797\_A: Testing for incomplete filtering of one or more instances of special elements} 
\label{fig:simulation10}
\end{center}
\end{figure*}

We take a \emph{subset} of the actions in the source input to speed testing up or ensure that the system enforces the precedence relation between the actions (if there are any). An example \MR of the first case is \texttt{CWE\_286\_OTG\_AUTHZ\_002c} (Fig.~\ref{fig:simulation3}). This \MR tests the condition that if a user navigating the GUI cannot access a URL, the same URL should not be available to the same user when she directly requests it from the server. It relies on a subset of the source input actions since, to test the condition above, it is sufficient to create a follow-up input including only the action accessing the reserved URL rather than performing all the actions. The system is vulnerable if the follow-up input leads to a successful retrieval of the resource pointed by the URL (i.e., the same output as the source input action).
\TSE{3.15.1}{An example of the second case} (i.e., ensuring that precedence relations are enforced) is that of a system that should validate a user session to ensure 
that she has confirmed her email address after signing up.
The system is vulnerable if the user can log in without confirming her email address. The source input sequence includes the registration to the service, the email confirmation, and a successful log-in. The follow-up input should cover only the registration action and the log-in without confirming the email address. If the log-in is successful, the system is vulnerable as it should ensure that the user is registered with a valid email address. Note that \MST automatically generates such follow-up inputs by iteratively generating distinct subsets of actions from the source input sequence (see \MR \texttt{CWE\_841} in Fig.~\ref{fig:simulation4}). Five \MRs (7\%) in our catalog are obtained by selecting a subset of actions from follow-up inputs.


We \emph{add} one or more actions to a source input to test scenarios when we expect \TSE{3.15.2}{such a change} in action sequence to lead to different results. 
Usually, the actions added are user actions belonging to a source input (not necessarily the one used to define the follow-up input) or actions capturing environmental factors (e.g., the passing of time).
An example user action copied into the follow-up input is given in \MR \texttt{OTG\_SESS\_003} (Fig.~\ref{fig:simulation5}); \texttt{OTG\_SESS\_003} scans the source inputs to identify a sign-up action to be copied into the follow-up input, after the login. In \texttt{OTG\_SESS\_003}, the presence of a sign-up action after the login enables us to verify that the session ID is updated after every sign-up. 
An example environment action added to a source input sequence is given in \texttt{CWE\_262\_263\_309\_324} (Fig.~\ref{fig:simulation6}), which tests the system's password aging mechanism through a \emph{DelayAction}.
We consider a source input including login and add a \emph{DelayAction} that changes the system's date to next year, thus simulating the passing of time. When the user performs the next action in the sequence, the system should ask the user to change his credentials instead of completing the requested action and returning the same results as the source input.
We obtain the follow-up inputs in nine \MRs by adding actions to the source input.


We \emph{modify} actions by changing the assignments to parameters that are inputs of the SUT (i.e., URL parameters, form entries, session cookies, certificates) 
or that control the communication channel (i.e., encryption algorithms, communication protocol, HTTP method). These changes may be applied to follow-up inputs that match the source inputs or include other differences (e.g., subsets).
For example, to test a system for code injection vulnerabilities, we execute the same sequence of actions as in the source inputs and modify form entries by relying on known attack vectors (e.g., concatenating SQL injection commands to the original input values). An example \MR that involves the modification of a communication protocol is \texttt{CWE\_287a\_425\_OTG\_AUTHN\_001} in Fig.~\ref{fig:simulation}.
In 57 \MRs of our catalog, the follow-up inputs perform the same sequence of actions as the source input but with at least one parameter modification.


\subsubsection{Output conditions}

Output conditions (postconditions) in our \MRs check if the outputs of the source and follow-up inputs (or a subset of their actions) are equal, different, or satisfy a predicate (e.g., the output is erroneous or includes information that has already been observed by the user) implemented by an SMRL function. 


SMRL \MRs capture properties that should hold if the system is not vulnerable. Therefore, we expect the \emph{same output} for the source and follow-up inputs when the attack captured by the follow-up input should not alter the system behavior, that is, when the system is supposed to detect an attack vector and ignore its effects. For example, the SUT should sanitize the received inputs by removing the code injection attack vector added to the original input and return the same output as the source input.
In 20 of our 76 \MRs (26\%), the follow-up input is supposed to lead to the same output as the source input.

In contrast, for 30 \MRs in our catalog (39\%), the outputs generated by the follow-up inputs are expected to be \emph{different} than those of the source inputs.
For instance, in \MR \texttt{CWE\_266\_..\_OTG\_AUTHZ\_002} (Fig.~\ref{fig:mrExample}), accessing a resource dedicated to the source input's user should lead to a different output (e.g., the home page or an error page).


Last, we can also verify \emph{predicates} on the generated outputs. 
Predicates are used, for example, to verify that the returned output should be accessible by the user in the \MRs \texttt{CWE\_20\_..OTG\_AUTHZ\_001a} and \texttt{CWE\_15\_639\_OTG\_AUTHZ\_004}.
\MR \texttt{CWE\_20\_..OTG\_AUTHZ\_001a} (Fig.~\ref{fig:simulation7}) replaces URL parameters (e.g., an ID) with paths of files on the SUT. Similarly, \texttt{CWE\_15\_639\_OTG\_AUTHZ\_004} (Fig.~\ref{fig:simulation8}) replaces URL parameter values with values observed only with other users. Though we cannot predict the effect of altering URL parameters --- we cannot know in advance if the value is legal for the user performing the action --- 
the user should be able to access the output when browsing the GUI. 
Since the crawler browses the GUI with the different users, function \texttt{userCanRetrieveContent} checks if the output has already been observed with any source input collected by the crawler. In our catalog, 61 \MRs include at least one predicate on the generated outputs.


\subsection{Metamorphic Relation Patterns}
\label{sec:patterns}


This section presents how the instances of the template elements are combined to form the \MR patterns. 
A pattern is captured by a set of element instances in one or more \MRs.
\TSE{2.7}{Our focus on the common set of element instances across \MRs aims to enable the systematic identification of patterns while organizing them in a taxonomy. Further, with patterns, the reader can more easily compare MR characteristics. For example, at the end of this section, we discuss why some patterns are less frequent than others. Covered element instances provide us with a systematic approach to identify patterns.  
The following conditions determine our patterns: (1) the user in the follow-up inputs should either match or differ from the one in the source input, (2) in the follow-up inputs, we should observe a sequence of action that contain either the same sequence of actions as the source inputs, or additional actions, or a subset of source inputs' actions, or modified actions. 
} 

To identify \MR patterns, we first filled in a mapping table tracing \MRs to the template elements in the previous section. 
Then, we grouped the \MRs covering the same combination of elements. By definition, an \MR can implement only one pattern. 
The resulting \MR patterns are reported in Table~\ref{tab:PatternsOfMRs}.

 
In total, we identified 23 patterns. Six patterns are implemented by at least five \MRs, and twelve patterns are implemented by only one \MR. In the following, we discuss the three most frequent patterns.
 
 \input{tables/tableMRsPatterns}
 



Two patterns occur with the same frequency (i.e., \emph{P1} and \emph{P2} in Table~\ref{tab:PatternsOfMRs}); we present them in the order they appear in Table~\ref{tab:PatternsOfMRs}. The first pattern (i.e., \emph{P1}) includes the following template elements: \emph{user  precondition(s), action precondition(s), same user, modified action(s), and verify other predicates}. It is implemented by 13 \MRs and used to ensure that a 
user cannot retrieve protected resources by providing crafted data as input.

Indeed, the same user performs the same actions as in the source input sequence after modifying some of them. User and action preconditions enable the selection of cases where the \MR should hold. All these 13 \MRs include an action precondition verifying that the URL is tested only once (to speed testing up). Further, 12 of these 13 \MRs include a user precondition checking if the user has already retrieved the content of the source input (to avoid testing with non-deterministic sequences); one \MR verifies that the user performing the follow-up actions is not an administrator since she might have \TSE{3.15.3}{the permission} to perform these actions. 
The predicates ensure that the generated output is either an error page or content that the user is supposed to be able to access (i.e., it has been accessed during crawling or functional testing as reported by functions \emph{userCanAccess} or \emph{userCanRetrieveContent}).
After providing a code injection string, \texttt{CWE\_94\_96\_B} (Fig.~\ref{fig:simulation9}), for instance, verifies that a user can retrieve only an error page or a resource that she can access.


The other most frequent pattern (i.e., \emph{P2}) is similar to the first one, except it does not verify user preconditions and verifies output equality. It includes the following template elements: \emph{action precondition, same user, modified action(s), verify equality, and verify other predicates}. It is used to simulate attacks where the user provides a crafted input (e.g., invalid character) that should be either sanitized (the same output is returned) or lead to an error page (this last condition is verified with a dedicated predicate on the output). An example \MR of this pattern is \MR CWE\_792\_793\_794\_795\_796\_797\_A (Fig.~\ref{fig:simulation10}).




The third pattern (i.e., \emph{P3}) 
includes the following template elements: \emph{action precondition(s), same user, added action(s), and verify the difference}. It is implemented by seven \MRs where a follow-up input with a different number of steps should lead to different outputs (e.g., \emph{OTG\_SESS\_003} in Fig.~\ref{fig:simulation5}, which duplicates a signup action and verifies that the action leads to a session cookie different from the cookie of the previous signup action).

Twelve patterns are implemented by only one \MR. Six of these patterns result from \MRs creating multiple follow-up inputs, which leads to combinations of template elements normally not appearing together (e.g., equal and subset). 
Another pattern (i.e., \emph{P12}) concerns an \MR that does not verify any precondition (it generates follow-up inputs from all the available source inputs) because discovering the vulnerability likely depends on the system state, the sequence of actions previously executed, or the user performing the action; it differs from all the other patterns because they all include at least one precondition. The presence of state-dependent vulnerabilities is however rare (e.g., the successful exploitation of a code injection vulnerability is unlikely to depend on the actions performed before), which is why all our patterns, except one, have at least one precondition.

%% file: tables/tableMRtemplateElementInstances.tex
\begin{table}[]


\centering
\caption{MRs template element instances}
\label{tab:my-table}
\resizebox{\columnwidth}{!}{%
\begin{tabular}{|c|ll|}
\hline
\textbf{Template element} & \multicolumn{2}{c|}{\textbf{Element instance}}\\
\hline
\multirow{2}{*}{Preconditions} & \multicolumn{2}{l|}{User precondition}                                \\ \cline{2-3} 
                               & \multicolumn{2}{l|}{Action precondition}                              \\ \hline
\multirow{6}{*}{Generation of follow-up inputs}          & \multicolumn{2}{l|}{Same user}\\ \cline{2-3} 
                               & \multicolumn{2}{l|}{Different user}                          \\ \cline{2-3} 
                               & \multicolumn{2}{l|}{Same actions}\\ \cline{2-3} 
                               & \multicolumn{2}{l|}{Actions subset }                            \\ \cline{2-3} 
                               & \multicolumn{2}{l|}{Added action(s)}                         \\ \cline{2-3} 
                               & \multicolumn{2}{l|}{Modified Action(s)}\\ \hline
\multicolumn{1}{|l|}{\multirow{3}{*}{Output conditions (postconditions)}} & \multicolumn{2}{l|}{Verify equality}                         \\ \cline{2-3} 
\multicolumn{1}{|l|}{}         & \multicolumn{2}{l|}{Verify difference}                           \\ \cline{2-3} 
\multicolumn{1}{|l|}{}         & \multicolumn{2}{l|}{Verify other predicate}                   \\ \hline
\end{tabular}%
}\label{MRsTemplateElements}
\end{table}

%% file: tables/tableMRsPatterns.tex
\begin{table*}[!htbp]
\centering
\setlength\tabcolsep{.2\tabcolsep}%
\caption{MR Patterns}
\label{tab:PatternsOfMRs}

{
\begin{tabular}{|c|c|c|c|c|c|c|c|c|c|c|c|c|}
\hline
\textbf
{Pattern ID}&
\multicolumn{2}{c|}{\textbf
{Preconditions}} &
  \multicolumn{6}{c|}{\textbf{Generation of follow-up inputs}} &
  \multicolumn{3}{c|}{\textbf{Output condition}} &
  \multirow{2}{*}{\textbf{Number} \textbf{of} \textbf{MRs}} \\ \cline{2-12}
 & \makecell{\textbf{User} \\\textbf{precondition}} &
 \makecell{\textbf{Action} \\\textbf{precondition}} &
\makecell{\textbf{Same} \\\textbf{user}} &
\makecell{\textbf{Different} \\\textbf{user}} &
\makecell{\textbf{Same} \\\textbf{actions}} &
\makecell{\textbf{Actions} \\\textbf{subset}} &
\makecell{\textbf{Added} \\\textbf{ action(s)}} &
\makecell{\textbf{Modified} \\\textbf{ action(s)}} &
\makecell{\textbf{Verify} \\\textbf{equality}} &
\makecell{\textbf{Verify} \\\textbf{difference}} &
\makecell{\textbf{Verify} \\\textbf{other} \\ \textbf{Predicates}} &
   \\ \hline
P1&
\multicolumn{1}{c|}{1} &
  1 &
  \multicolumn{1}{c|}{1} &
  \multicolumn{1}{c|}{0} &
  \multicolumn{1}{c|}{0} &
  \multicolumn{1}{c|}{0} &
  \multicolumn{1}{c|}{0} &
  1 &
  \multicolumn{1}{c|}{0} &
  \multicolumn{1}{c|}{0} &
  1 &
  13 \\ \hline
P2& \multicolumn{1}{c|}{0} &
  1 &
  \multicolumn{1}{c|}{1} &
  \multicolumn{1}{c|}{0} &
  \multicolumn{1}{c|}{0} &
  \multicolumn{1}{c|}{0} &
  \multicolumn{1}{c|}{0} &
  1 &
  \multicolumn{1}{c|}{1} &
  \multicolumn{1}{c|}{0} &
  1 &
  13 \\ \hline
P3& \multicolumn{1}{c|}{0} &
  1 &
  \multicolumn{1}{c|}{1} &
  \multicolumn{1}{c|}{0} &
  \multicolumn{1}{c|}{0} &
  \multicolumn{1}{c|}{0} &
  \multicolumn{1}{c|}{1} &
  0 &
  \multicolumn{1}{c|}{0} &
  \multicolumn{1}{c|}{1} &
  0 &
  7 \\ \hline
P4&\multicolumn{1}{c|}{0} &
  1 &
  \multicolumn{1}{c|}{1} &
  \multicolumn{1}{c|}{0} &
  \multicolumn{1}{c|}{0} &
  \multicolumn{1}{c|}{0} &
  \multicolumn{1}{c|}{0} &
  1 &
  \multicolumn{1}{c|}{0} &
  \multicolumn{1}{c|}{0} &
  1 &
  6 \\ \hline
P5&\multicolumn{1}{c|}{0} &
  1 &
  \multicolumn{1}{c|}{1} &
  \multicolumn{1}{c|}{0} &
  \multicolumn{1}{c|}{0} &
  \multicolumn{1}{c|}{0} &
  \multicolumn{1}{c|}{0} &
  1 &
  \multicolumn{1}{c|}{0} &
  \multicolumn{1}{c|}{1} &
  1 &
  5 \\ \hline
P6&\multicolumn{1}{c|}{1} &
  1 &
  \multicolumn{1}{c|}{1} &
  \multicolumn{1}{c|}{0} &
  \multicolumn{1}{c|}{0} &
  \multicolumn{1}{c|}{0} &
  \multicolumn{1}{c|}{0} &
  1 &
  \multicolumn{1}{c|}{0} &
  \multicolumn{1}{c|}{1} &
  1 &
  5 \\ \hline
P7&\multicolumn{1}{c|}{1} &
  1 &
  \multicolumn{1}{c|}{1} &
  \multicolumn{1}{c|}{0} &
  \multicolumn{1}{c|}{0} &
  \multicolumn{1}{c|}{0} &
  \multicolumn{1}{c|}{0} &
  1 &
  \multicolumn{1}{c|}{1} &
  \multicolumn{1}{c|}{0} &
  1 &
  4 \\ \hline
P8&\multicolumn{1}{c|}{0} &
  1 &
  \multicolumn{1}{c|}{1} &
  \multicolumn{1}{c|}{0} &
  \multicolumn{1}{c|}{0} &
  \multicolumn{1}{c|}{0} &
  \multicolumn{1}{c|}{0} &
  1 &
  \multicolumn{1}{c|}{0} &
  \multicolumn{1}{c|}{1} &
  0 &
  4 \\ \hline
P9&\multicolumn{1}{c|}{1} &
  0 &
  \multicolumn{1}{c|}{1} &
  \multicolumn{1}{c|}{0} &
  \multicolumn{1}{c|}{0} &
  \multicolumn{1}{c|}{0} &
  \multicolumn{1}{c|}{0} &
  1 &
  \multicolumn{1}{c|}{0} &
  \multicolumn{1}{c|}{0} &
  1 &
  3 \\ \hline
P10&\multicolumn{1}{c|}{1} &
  1 &
  \multicolumn{1}{c|}{0} &
  \multicolumn{1}{c|}{1} &
  \multicolumn{1}{c|}{1} &
  \multicolumn{1}{c|}{0} &
  \multicolumn{1}{c|}{0} &
  0 &
  \multicolumn{1}{c|}{0} &
  \multicolumn{1}{c|}{1} &
  1 &
  2 \\ \hline
P11&\multicolumn{1}{c|}{0} &
  1 &
  \multicolumn{1}{c|}{1} &
  \multicolumn{1}{c|}{0} &
  \multicolumn{1}{c|}{1} &
  \multicolumn{1}{c|}{0} &
  \multicolumn{1}{c|}{0} &
  0 &
  \multicolumn{1}{c|}{0} &
  \multicolumn{1}{c|}{0} &
  1 &
  2 \\ \hline
P12&\multicolumn{1}{c|}{0} &
  0 &
  \multicolumn{1}{c|}{1} &
  \multicolumn{1}{c|}{0} &
  \multicolumn{1}{c|}{0} &
  \multicolumn{1}{c|}{0} &
  \multicolumn{1}{c|}{1} &
  0 &
  \multicolumn{1}{c|}{1} &
  \multicolumn{1}{c|}{0} &
  0 &
  1 \\ \hline
P13&\multicolumn{1}{c|}{1} &
  1 &
  \multicolumn{1}{c|}{0} &
  \multicolumn{1}{c|}{1} &
  \multicolumn{1}{c|}{0} &
  \multicolumn{1}{c|}{1} &
  \multicolumn{1}{c|}{0} &
  0 &
  \multicolumn{1}{c|}{0} &
  \multicolumn{1}{c|}{1} &
  1 &
  1 \\ \hline
P14&\multicolumn{1}{c|}{0} &
  1 &
  \multicolumn{1}{c|}{0} &
  \multicolumn{1}{c|}{1} &
  \multicolumn{1}{c|}{1} &
  \multicolumn{1}{c|}{0} &
  \multicolumn{1}{c|}{0} &
  1 &
  \multicolumn{1}{c|}{0} &
  \multicolumn{1}{c|}{1} &
  0 &
  1 \\ \hline
P15&\multicolumn{1}{c|}{0} &
  1 &
  \multicolumn{1}{c|}{1} &
  \multicolumn{1}{c|}{0} &
  \multicolumn{1}{c|}{0} &
  \multicolumn{1}{c|}{1} &
  \multicolumn{1}{c|}{0} &
  0 &
  \multicolumn{1}{c|}{0} &
  \multicolumn{1}{c|}{1} &
  1 &
  1 \\ \hline
P16&\multicolumn{1}{c|}{0} &
  1 &
  \multicolumn{1}{c|}{1} &
  \multicolumn{1}{c|}{0} &
  \multicolumn{1}{c|}{0} &
  \multicolumn{1}{c|}{1} &
  \multicolumn{1}{c|}{0} &
  1 &
  \multicolumn{1}{c|}{0} &
  \multicolumn{1}{c|}{1} &
  1 &
  1 \\ \hline
P17&\multicolumn{1}{c|}{0} &
  1 &
  \multicolumn{1}{c|}{1} &
  \multicolumn{1}{c|}{0} &
  \multicolumn{1}{c|}{1} &
  \multicolumn{1}{c|}{0} &
  \multicolumn{1}{c|}{0} &
  0 &
  \multicolumn{1}{c|}{0} &
  \multicolumn{1}{c|}{1} &
  0 &
  1 \\ \hline
P18&\multicolumn{1}{c|}{0} &
  1 &
  \multicolumn{1}{c|}{1} &
  \multicolumn{1}{c|}{0} &
  \multicolumn{1}{c|}{1} &
  \multicolumn{1}{c|}{0} &
  \multicolumn{1}{c|}{0} &
  1 &
  \multicolumn{1}{c|}{0} &
  \multicolumn{1}{c|}{0} &
  1 &
  1 \\ \hline
P19&\multicolumn{1}{c|}{1} &
  1 &
  \multicolumn{1}{c|}{0} &
  \multicolumn{1}{c|}{1} &
  \multicolumn{1}{c|}{0} &
  \multicolumn{1}{c|}{0} &
  \multicolumn{1}{c|}{0} &
  1 &
  \multicolumn{1}{c|}{0} &
  \multicolumn{1}{c|}{1} &
  1 &
  1 \\ \hline
P20&\multicolumn{1}{c|}{1} &
  1 &
  \multicolumn{1}{c|}{0} &
  \multicolumn{1}{c|}{1} &
  \multicolumn{1}{c|}{0} &
  \multicolumn{1}{c|}{0} &
  \multicolumn{1}{c|}{1} &
  0 &
  \multicolumn{1}{c|}{0} &
  \multicolumn{1}{c|}{0} &
  1 &
  1 \\ \hline
P21&\multicolumn{1}{c|}{1} &
  1 &
  \multicolumn{1}{c|}{0} &
  \multicolumn{1}{c|}{1} &
  \multicolumn{1}{c|}{0} &
  \multicolumn{1}{c|}{1} &
  \multicolumn{1}{c|}{0} &
  0 &
  \multicolumn{1}{c|}{1} &
  \multicolumn{1}{c|}{0} &
  0 &
  1 \\ \hline
P22&\multicolumn{1}{c|}{1} &
  1 &
  \multicolumn{1}{c|}{1} &
  \multicolumn{1}{c|}{0} &
  \multicolumn{1}{c|}{0} &
  \multicolumn{1}{c|}{1} &
  \multicolumn{1}{c|}{0} &
  0 &
  \multicolumn{1}{c|}{0} &
  \multicolumn{1}{c|}{1} &
  1 &
  1 \\ \hline
P23&  \multicolumn{1}{c|}{{1}} &
 { 1} &
  \multicolumn{1}{c|}{{0}} &
  \multicolumn{1}{c|}{{1}} &
  \multicolumn{1}{c|}{{1}} &
  \multicolumn{1}{c|}{{0}} &
  \multicolumn{1}{c|}{{0}} &
 { 0} &
  \multicolumn{1}{c|}{{1}} &
  \multicolumn{1}{c|}{{0}} &
 { 1 }&
  {1} \\ \hline
\end{tabular}%

}
\end{table*}

%% file: usage.tex
\section{Analysis of \MST's Applicability and Testability Guidelines}
\label{sec:usage}

In this section, we investigate (i) the types of security testing activities presenting an oracle problem that can only be addressed by \MST, (ii) the types of security vulnerabilities that can be identified by \MST, and (iii) the guidelines (hereafter testability guidelines) that engineers should follow to make \MST as effective as possible. We investigate the following Research Questions (RQs):

\begin{itemize}

\item \textit{\textbf{RQ1.}} \textit{\textbf{To what extent can \MST address the oracle problem in the context of security testing?}} This research question aims to identify the security testing activities that, due to the oracle problem, can only be automated using \MST.

\item \textit{\textbf{RQ2.}} \textit{\textbf{What vulnerability types can \MST detect?}} 
\MST has been designed and implemented to perform security testing by reasoning on outputs of multiple interactions with the system under test. Not every type of vulnerability can be discovered through relationships between outputs of multiple user-system interactions (e.g., some may require program analysis).
This research question aims to determine, in a systematic way, the types of security vulnerabilities that \MST can discover \TSE{3.4}{and compare \MST with state-of-the-art (SOTA) security testing tools}.

\item \textit{\textbf{RQ3.}} \textit{\textbf{
\TSE{2.8}{What testability guidelines can we define to enable effective test automation with \MST?}}} 
Software testability is the degree to which a software artifact (i.e., a software system, module, requirements, or design document) supports its testing~\cite{voas1995software}. A higher degree of testability results in decreased test effort, increased quality of test activities, a higher probability of finding software defects, and, as a result, higher-quality software. 
This question investigates if it is possible to identify testability guidelines that assist engineers in designing, implementing, and configuring their software to enable effective test automation with \MST.

\end{itemize}

\input{setup}

\input{RQ1.tex}

\input{RQ2.tex}

\input{RQ3.tex}

%% file: setup.tex
\subsection{Targeted Vulnerabilities and Testing Activities}
\label{subsec:setup}

To address our first research question, we study the security testing activities recommended by OWASP~\cite{OWASPWeb}. These activities have been described as part of the OWASP testing guidelines Version 4.0~\cite{OWASPtesting} to help engineers understand the \textit{what}, \textit{why}, \textit{when}, \textit{where}, and \textit{how} \TSE{3.15.4}{of testing the security of Web applications.} The project provides these activities as a complete testing framework, not merely a simple checklist or prescription of issues that should be addressed. There are, in total, 90 testing activities organized into 11 categories. For instance, in testing activity \textit{Testing Session Timeout}, engineers check that the system under test automatically logs out a user when that user has been idle for a certain amount of time, ensuring that it is not possible to “reuse” the same session and that no sensitive data remains stored in the browser cache~\cite{WSTG-TestingSessionTimeout}. In our analysis, we include the security testing activities recommended by OWASP because they provide a comprehensive list of security testing methods that we can analyze to identify whether they suffer from the oracle problem.

To address our second and third research questions in a systematic way, we study the list of weaknesses reported in the Common Weakness Enumeration (CWE) database~\cite{CWE-DB}. We provide the following definitions of \textit{vulnerability} and \textit{weakness} since their definitions in the CWE framework~\cite{CWE_FAQ} lack clarity\footnote{The definitions provided by CWE are unclear since they rely on \emph{synonyms} to distinguish vulnerability and weakness, as follows: `Weaknesses are \emph{flaws}, \emph{faults}, \emph{bugs}, and other \emph{errors} in system design, architecture, code, or implementation that if left unaddressed could result in systems and networks, and hardware being vulnerable to attack. Weaknesses can lead to vulnerabilities. A vulnerability is a \emph{mistake} in software or hardware that can be used by a malicious user to gain access to a system or network~\cite{CWE_FAQ}'.}.
A \emph{vulnerability} is a specific fault of the system under test that causes the system to breach its security requirements.
A \emph{weakness} represents a fault type (i.e., the type of a vulnerability). It describes a human error made in the analysis, design, or implementation of the system that may affect the degree to which the system meets its security requirements.

The CWE database is organized into distinct views, each view grouping weaknesses according to a different set of categories, 
which are  
\emph{common security architectural tactics}~\cite{CWE-ArchitectureView},
\emph{software development concepts}~\cite{CWE-SoftwareView}, \emph{research concepts}~\cite{CWE-ResearchView}, 
\emph{software fault patterns}~\cite{CWEPatterns},
\emph{most dangerous errors}~\cite{CWE-Top25}, and \emph{hardware design}~\cite{CWE-HardwareView}. 
Other views map the weaknesses to some security-related catalogs (e.g., OWASP Top 10~\cite{OWASP-Top10} and SERT CEI C Coding standards~\cite{CWECoding}).

The CWE view for \emph{common security architectural tactics} organizes weaknesses according to \emph{security design principles}. This view has twelve categories
representing the individual security design principles that are part of
a secure-by-design approach to software development.
It covers, in total, 223 weaknesses. 
The security design principles assist engineers in identifying potential mistakes that can be made when designing software~\cite{santos2017understanding,santos2017catalog}.
A weakness is thus the result of a design principle not being followed.
For instance, the weaknesses \TSE{3.15.5}{in the design principle \textit{Authenticate Actors}} are related to authentication-based components in the system. These components deal with verifying that the actor interacting with the system is who she claims to be. The weaknesses in this category lead to a degradation of the quality of authentication if they are not addressed when designing and implementing the system under test~\cite{CWE-ArchitectureView}.
The views for \emph{software development concepts} and \emph{hardware design} organize weaknesses based on
the types of errors
that affect the software implementation (e.g., illegal pointer dereferences) and hardware design (e.g., faults in semiconductor logic), respectively.
The views for \emph{software fault patterns} and \emph{research concepts} group implementation errors into categories capturing fault patterns~\cite{Calloni2011} or \TSE{3.15.6}{high-level descriptions} of the faulty software behaviour (e.g., incorrect comparison and improper access control).


In our analysis, we focus on the weaknesses in the CWE view for common security architectural tactics~\cite{CWE-ArchitectureView}, the weaknesses in the view for the CWE Top 25 most dangerous
software errors (CWE Top 25)~\cite{CWE-Top25}, and the weaknesses in the view for the OWASP Top 10 Web security
risks (OWASP Top 10)~\cite{CWE_OWASPTOP10}. 
We select the common security architectural tactics view because it enables us to determine
the security design principles that \MST can verify. 
We do not consider the view for
\emph{software development concepts}
because \MST is a black-box testing approach,
that does not aim to discover specific implementation errors (e.g., type errors). 
We also ignore the views for \emph{software fault patterns}, \emph{research concepts}, and mappings to \emph{coding standards}~\cite{CWECoding} since they mainly focus on software implementation.
We ignore the CWE view for \emph{hardware design}
since \MST does not address hardware vulnerabilities. 

In our analysis, we include the CWE Top 25 and OWASP Top 10 views to assess to what extent \MST can address the most widespread and critical security vulnerabilities. The CWE Top 25 view lists twenty-five most widespread weaknesses which are often easy to find and exploit. These weaknesses are considered dangerous because they typically \textit{allow attackers to completely take over the control of software, steal data, or prevent software from working~\cite{CWE-Top25}.} 
The OWASP Top 10~\cite{OWASP-Top10} is the list of the ten most common web application security risks edited by the 
Open Web Application Security Project~\cite{OWASP}, i.e., an online community producing freely-available articles, methodologies, documentation, tools, and technologies in the field of web application security. It is updated every three to four years. The most up-to-date version includes 43 weaknesses grouped into 10 categories~\cite{CWE_OWASPTOP10}. 

\input{tables/tableSecurityTestingActivities}

%% file: tables/tableSecurityTestingActivities.tex
\begin{table*}[t]

\scriptsize
\caption{Subset of the security testing activities recommended by OWASP.}
\label{table:ST-Activities}
\begin{center}
\begin{tabular}{|p{6,3cm}|p{7.0cm}|p{3.3cm}|}
\hline
\textbf{Security Testing Category} &
\multicolumn{1}{|c|}{\textbf{Security Testing Activity}} &
  \textbf{Oracle Automation Strategy}  \\ \hline
\multirow{2}{*}{\textbf{Configuration and Deployment Management Testing}} &
Test Application Platform Configuration             & Manual Oracle \\ \cline{2-3}
&Test HTTP Methods & No Oracle Needed \\ \hline

\multirow{2}{*}{\textbf{Authentication Testing}}&
Testing for Credentials Transported over an Encrypted Channel                                  & Metamorphic Testing  \\ \cline{2-3}
&Testing for Default Credentials                                  & Catalog-based  \\ \cline{2-3}
&Testing for Weak Password Policy                & Catalog-based \\ \cline{2-3}

&Testing for Weak Lock Out Mechanism                & Implicit Oracle \\ \hline

\multirow{2}{*}{\textbf{Authorization Testing}}&Testing Directory Traversal File Include                         & Metamorphic Testing  \\ \cline{2-3}

&Testing for Bypassing Authorization Schema                         & Metamorphic Testing  \\ \cline{2-3}

&Testing for Privilege Escalation                                       & Metamorphic Testing  \\ \cline{2-3}

&Testing for Insecure Direct Object References                                       & Metamorphic Testing  \\ \hline

\multirow{2}{*}{\textbf{Session Management Testing}}  &
Testing for Cookies Attributes                     & Manual Oracle  \\ \cline{2-3}

&Testing for Session Fixation                     & Metamorphic Testing  \\ \cline{2-3}

&Testing for Exposed Session Variables                     & No Oracle Needed  \\ \cline{2-3}

&Testing for Cross Site Request Forgery             & Vulnerability Specific  \\ \hline

\multirow{2}{*}{\textbf{Input Validation Testing}} &
Testing for SQL Injection   & Vulnerability Specific \\ \cline{2-3}

&Testing for HTTP Verb Tampering   & Metamorphic Testing \\ \cline{2-3}

&Testing for Buffer Overflow              & Implicit Oracle  \\ \hline

\multirow{2}{*}{\textbf{Business Logic Testing}} &
Test for Process Timing   & No Oracle Needed \\ \cline{2-3}

&Test Number of Times a Function Can Be Used Limits   & Metamorphic Testing \\ \cline{2-3}

&Testing for the Circumvention of Work Flows              & Manual Oracle  \\ \hline

\end{tabular}
\end{center}
\end{table*}

%% file: RQ1.tex
\subsection{RQ1: Oracle Problem}
\label{subsec:RQ1}


\subsubsection{Analysis Procedure}
 
To respond to RQ1, we study the security testing activities recommended by OWASP~\cite{OWASPtesting}. 
We systematically analyzed them to identify applicable, SOTA oracle automation strategies. For each activity, we first inspect its description, objective, methods, and tools, if available. 

Based on the information collected from our inspection, we identify oracle automation strategies, including metamorphic testing, that can be applied to address the oracle problem if the activity is subject to it. For instance, testing activity \textit{Testing for Bypass Authorization Schema} focuses on verifying how the authorization schema has been implemented for each role or privilege to get access to reserved functions and resources~\cite{WSTG:ATHZ:02}. \TSE{3.15.7}{One of its testing methods} is to \TSE{3.15.8}{try to access resources} assigned to a different role; another one is to \TSE{3.15.8}{try to access administrative functions.} In the first method, it is not always feasible to verify the access to resources with different privileges and roles when the expected outputs need to be identified for a large set of inputs (i.e., the oracle problem). To address the oracle problem in the activity, we specify the \MR in Fig.~\ref{fig:mrExample}. It compares the outputs of the executions of the system under test for the same resource and different credentials.       
We discuss the proportion of software testing activities that cannot be automated with SOTA oracle automation strategies but can be automated with \MST. 
 
\subsubsection{Results}

In our analysis, we identified four oracle automation strategies for security testing: \textit{implicit oracle}, \textit{catalog-based},  \textit{vulnerability-specific}, and \textit{metamorphic testing}. In the following, we discuss the details of each strategy; in addition, we discuss why in certain cases \textit{no oracle is needed} or only  a \textit{manual oracle} is feasible.
To exemplify our descriptions, we provide in Table~\ref{table:ST-Activities}, for each OWASP testing category, some of its security testing activities along with oracle automation strategies. 

\emph{No oracle needed.} 
Some activities collect data to reverse engineer the system under test. They do not verify the security properties of the system. They aim to retrieve information which might be useful to identify potential weaknesses (e.g., the use of vulnerable versions of a Web framework). Therefore, these activities do not have an oracle problem. 
For instance, in security testing activity \emph{Map Application Architecture}~\cite{OTG:INFO:010}, engineers identify the components of a Web system, e.g., reverse proxy, type of front-end Web server, and version of the LDAP server. Commonly, hundreds of Web applications are hosted on an interconnected Web server infrastructure. A single vulnerability in one application may risk the security of the entire infrastructure. Even small risks may evolve into severe ones for other applications on the same server. Therefore, it is important to perform an in-depth review of known security issues for each application. Before the review, engineers need to map the application architecture through some tests to determine which application components are used~\cite{OTG:INFO:010}.

\emph{Manual oracle.} Some activities require humans to determine vulnerabilities based on system specifications.
For instance, testing activity \emph{Testing for the Circumvention of Work Flows} concerns vulnerabilities for the misuse of a system in a way that allows malicious users to circumvent the intended workflow~\cite{OTG:BUSLOGIC:006}. 
Vulnerabilities related to the circumvention of workflows are very system-specific.  
In short, the business process of the system under test must ensure that transactions and actions proceed in the right order. 
For example, when testing a pay-per-view system (i.e., a pay television service by which a viewer can purchase events to view via private telecast), it is necessary to determine transactions that should not enable service access.
Only a human can decide, based on system specifications, whether pending transactions should grant service access.
\TSE{2.9}{Although oracles that present similarities with the exemplifying case described above might be automated in some contexts (e.g., by encoding the logic to decide if a transaction can be performed), their implementation requires substantial manual effort and is unlikely to be generalizable and integrated with test input generation approaches.}

\emph{Implicit oracle.} We can automate some of the security testing activities by following randomized test input generation strategies relying on implicit oracles. An implicit oracle refers to the detection of ``obvious'' faults such as a program crash~\cite{Barr2015}. For instance, testing activity \emph{Testing for Buffer Overflow} in Table~\ref{table:ST-Activities}
is automated by looking for system crashes in response to lengthy inputs\cite{OTG:INPVAL:014}. One of the testing methods in the activity is testing for the format string~\cite{FormatString}. 
A format string is a null-terminated character sequence with conversion specifiers interpreted or converted at run-time. For systems concatenating user input with a format string, we add additional conversion specifiers to cause a buffer overflow and eventually a system crash. 

\emph{Catalog-based.} We can automate some activities based on a predefined catalog in which we specify test inputs and oracles. 
For instance, testing activity \emph{Testing for Default Credentials} focuses on systems installed on servers with minimal configuration or customization by the server administrator~\cite{OTG:AUTHN:002}. Often these systems are not properly configured, and the default credentials for initial authentication and configuration are never changed. These default credentials are well-known by malicious users, who use them to access various types of systems. As part of the security testing activity, we use a catalog of default credentials to test whether easy-to-guess pairs of usernames and passwords (e.g., $\langle$admin, admin$\rangle$) can be used to log into the system under test.






\input{tables/tableDistributionTestingActivity}

\emph{Vulnerability-specific.} 
Some activities can get automated by SOTA tools such as Burp Suite~\cite{BURP} and thus may not necessarily benefit from \MT.
These include OWASP testing activities that detect cross-site scripting and code injection vulnerabilities. 
Other activities are either not targeted or partially automated.
For example, Burp Suite does not automate oracles for activity \emph{Testing for Bypassing Authorization Schema}~\cite{BurpSuiteAC}. Though it enables engineers to compare the content of site maps~\cite{BurpSuiteScanner} recorded in different user sessions (e.g., with and without certain privileges), it requires engineers to manually identify privileged resources and inspect the differences in the observed system outputs, which is error-prone (e.g., overlooking resources) and expensive. Even Burp Suite plug-ins using Crawljax to build site maps do not address the oracle problem but generate JUnit tests that simply retrieve the mapped resources~\cite{BurpSuiteCSJ}.
With \MST, engineers, instead, can focus on the specifications of system-level properties without performing such manual testing activities. Testing activities, including oracles, are automated by the \MST framework.



\MST. 
In general, \MST can automate the testing of  activities that verify if a resource of the system under test can be accessed under circumstances that should prevent it (e.g., an unauthenticated user or unencrypted channel).
These activities benefit from \MST since they entail the verification of numerous system resources and specific security properties (e.g., each Web page might be accessed by a different set of users). 
For instance, testing activity \textit{Testing Directory Traversal File Include} focuses on reading directories or files which normally cannot be read, accessing data outside the web document root, and including scripts and other kinds of files from external websites~\cite{WSTG:ATHZ:01}. With a large set of test inputs, it is not feasible to list all the directories and files which a user normally cannot reach. To address the oracle problem in this testing activity, we specify an \MR which verifies that a file path should never enable a user to access data that is not provided by the user interface (see relation \texttt{CWE\_20\_..\_OTG\_AUTHZ\_001a} in  Fig.~\ref{fig:simulation7}).

Table~\ref{table:distributionTestingActivity} presents a summary of the OWASP security testing categories and the oracle automation strategies. The first column lists the security testing categories. The rest of the columns present the numbers of testing activities in the categories automated by each strategy.  

In total, 19 out of 90 activities (21\%) do not require a test oracle while 71 activities (79\%) do. 
Also, only 30 out of these 71 activities (42\%) can benefit from existing oracle automation solutions (i.e., implicit oracle, catalog-based, and vulnerability-specific), thus highlighting the severe impact of the oracle problem on security testing automation.

More than half of the activities not requiring a test oracle (53\%) are \TSE{3.15.9}{associated with the testing category \textit{Information Gathering}} because it covers all the activities which require reverse engineering or manual information retrieval to collect data about the system under test.

Among traditional oracle automation solutions,
vulnerability-specific approaches are the ones covering the largest proportion of OWASP activities (i.e., 22 out of 90, 24\%). Half of these activities \TSE{3.15.10}{are organized into the category \textit{Input Validation Testing}} and concern cross-site scripting and injection vulnerabilities; however, as reported in Section~\ref{subsec:RQ2}, \MST can still be used to test for these vulnerabilities thus enabling engineers to rely on a single testing framework (i.e., \MST) rather than several ones. \textit{Catalog-based} and \textit{Implicit} oracles address a low percentage of activities (8\% and 2\%, respectively). \TSE{3.15.11}{The limited applicability of implicit oracles shows} that most of the solutions relying on them (e.g., fuzz testing tools~\cite{manes2019art}), though useful, partially address the needs of security testing engineers.

Among the activities that cannot benefit from  traditional oracle automation solutions, 16 (39\%) can be automated thanks to \MST, which highlights the relevance of the contribution of this paper. A majority of these activities (69\%) are in the categories \textit{Authentication Testing}, \textit{Authorization Testing}, and \textit{Session Management Testing}.
The testing activities in these three categories require multiple interactions between the system under test and one or more users (actors), two characteristics well supported by \MST. Among them,
\MST can automate all the testing activities \TSE{3.15.10}{in the category \textit{Authorization Testing}} (i.e., the activities \textit{Testing Directory Traversal File Include}, \textit{Testing for Bypassing Authorization Schema}, \textit{Testing for Privilege Escalation}, and \textit{Testing for Insecure Direct Object References} in Table~\ref{table:ST-Activities}). Further, it can automate half of the activities \TSE{3.15.10}{in the category \textit{Session Management Testing}}. An example is \textit{Testing for Session Fixation}, which is automated by \texttt{OTG\_SESS\_003} in Fig.~\ref{fig:simulation5}. A session fixation vulnerability occurs when the system under test does not renew its session cookie(s) after successful user authentication~\cite{WSTG:SESS:03}.  \texttt{OTG\_SESS\_003} verifies whether a signup action leads to a new session ID, even when the action is performed by a user already logged in.


\input{tables/tableArchitectureWeaknesses}


Based on the above, \textit{we conclude that \MST can play a key role in addressing the oracle problem in security testing}. The activities for which \MST can automate oracles are mostly those that (i) verify that resources can be accessed only by authorized users (\textit{Authorization Testing}), (ii) test the initial authentication and the transfer of the user’s authentication data  (\textit{Authentication Testing}), (iii) discover vulnerabilities associated with session management (\textit{Session Management Testing}), and (iv) test the system's response to HTTP methods and parameters (\textit{Input Validation Testing}). 


%% file: tables/tableDistributionTestingActivity.tex
\begin{table*}[tb]

\scriptsize
\caption{Oracle automation strategies for the OWASP security testing activities*.}
\label{table:distributionTestingActivity}
\begin{center}

  \begin{tabular}{|p{6.75cm}@{}|p{1.10cm}@{}|p{1.10cm}@{}|p{1.35cm}@{}|p{1.25cm}@{}|p{1.75cm}@{}|p{1.65cm}@{}|}
\hline
\multirow{2}{3.55cm}{\textbf{Security Testing Category}}&\multicolumn{6}{c|}{\textbf{Oracle Automation Strategy}}\\
\cline{2-7}
&
  \textbf{Implicit Oracle} &
  \textbf{Catalog- based} &
  \textbf{No\,Oracle Needed} &
  \textbf{Manual Oracle} &
  \textbf{Vulnerability- specific} &
  \textbf{\MST} \\ \hline

Information Gathering  		& - & - & 10 & - & - & -   \\ \hline
Configuration\,and\,Deployment Management Testing			        & - & 1 & 4 & 3 & - & 1   \\ \hline
Identity Management Testing			        & - & 2 & - & 3 & - & -   \\ \hline
Authentication Testing			& 1 & 3 & - & 3 & - & 3   \\ \hline
Authorization Testing		    & - & - & - & - & - & 4   \\ \hline
Session Management Testing	        & - & - & 1 & 2 & 1 & 4 \\ \hline
Input Validation Testing  	& 1 & - & 1 & 2 & 11 & 2   \\ \hline
Testing for Error Handling     	        & - & - & 2 & - & - & -   \\ \hline
Testing for Weak Cryptography     	    & - & - & - & 3 & - & 1   \\ \hline

Business Logic Testing     	    & - & - & 1 & 6 & 1 & 1   \\ \hline

Client Side Testing     	    & - & - & - & 3 & 9 & -   \\ \hline

\textbf{Total} 	                & 2 & 6 & 19 & 25 & 22 & 16   \\ \hline
\textbf{\% of testing activities} 	& 2\tiny{\%} & 8\tiny{\%} & 21\tiny{\%} & 28\tiny{\%} & 24\tiny{\%} & 18\tiny{\%}   \\ 
\hline
\end{tabular}
\end{center}

\footnotesize{*Details are available online~\cite{WebSMRL}. For readability, the symbol '-' stands for zero. The \emph{\% of testing activities} (last row) is computed with respect to the 90 activities in the OWASP book~\cite{OWASPtesting}.}\\
\end{table*}

%% file: tables/tableArchitectureWeaknesses.tex
\begin{table*}[tb]

\scriptsize
\caption{Subset of the security weaknesses in the CWE view for common security design principles.}
\label{table:AC-Weaknesses}
\begin{center}
\begin{tabular}{|p{3cm}|p{5cm}|p{1.1cm}|p{1.05cm}|p{1.05cm}|p{1.1cm}|p{3.2cm}|}
\hline
\textbf{Design principle} &
\multicolumn{1}{|c|}{\textbf{Weakness}} &
  \textbf{\begin{tabular}[c]{@{}l@{}}Belongs to\\ Top 25\end{tabular}} &
    \textbf{\begin{tabular}[c]{@{}l@{}}Belongs to\\ Top 10\end{tabular}} &
  \textbf{\begin{tabular}[c]{@{}l@{}}Generic \\Weakness\end{tabular}} &
  \textbf{\begin{tabular}[c]{@{}l@{}}Addressed \\by \MST\end{tabular}} &
  \textbf{\begin{tabular}[c]{@{}l@{}} Testability Feature (TF) /\\ Reason \MST cannot be \\applied (R) \end{tabular}} \\ \hline
\multirow{3}{*}{\textbf{Audit}} &
Omission of Security-relevant Information             & \color{black}No & \color{black}Yes & \color{black}Yes & \color{black}No &  \color{black}R3 \\ \cline{2-7}
&Obscured Security-relevant Information by Alternate Name & \color{black} No & \color{black}No & \color{black}Yes &\color{black} No &  \color{black}R3 
 \\ \cline{2-7}
 \hline

\multirow{2}{*}{\textbf{Authenticate Actors}}&
Improper Authentication                                  & \color{black} Yes &  \color{black}Yes & \color{black}Yes & \color{black}Yes &\color{black} TF3 \\ \cline{2-7}
&Weak Password Recovery Mechanism for Forgotten Password                &\color{black} No & \color{black}Yes & \color{black}Yes & \color{black}No & \color{black}R2 \\ \hline

\multirow{2}{*}{\textbf{Authorize Actors}   }&
Improper Privilege Management                         & \color{black}No & \color{black}Yes & \color{black}Yes &\color{black} Yes & \color{black}TF3 \\ \cline{2-7}
&Process Control                                       & \color{black}No & \color{black}No & \color{black}Yes & \color{black}No & \color{black}R1 \\ \hline


\multirow{2}{*}{\textbf{Encrypt Data}}  &
Small Space of Random Values                     & \color{black}No & \color{black}No & \color{black}Yes & \color{black}No &  \color{black}R4 \\ \cline{2-7}
&Missing Encryption of Sensitive Data             & \color{black}No & \color{black}Yes & \color{black}Yes & \color{black}No &  \color{black}R3 \\ \hline

\multirow{2}{*}{\textbf{Identify Actors}}  &
Improper Validation of Certificate with Host Mismatch
& No & No & Yes & No & \color{black} R5 \\
\cline{2-7}

&Improper Validation of Certificate Expiration

& No & No & Yes & Yes &  TF10 \\

\hline

\multirow{2}{*}{\textbf{Limit Access}} &
Improper Restriction of XML External Entity Reference   & \color{black}Yes &\color{black} Yes & \color{black}Yes & \color{black}Yes & \color{black}TF2 \\ \cline{2-7}
&External Control of File Name or Path              & \color{black}No & \color{black}Yes & \color{black}Yes & \color{black}Yes &\color{black} TF1 \\ \hline

\multirow{4}{*}{\textbf{Manage User Sessions}}&
J2EE Bad Practices: Non-serializable Object Stored in Session           & \color{black} No &  \color{black}Yes &  \color{black}No &  \color{black}No &   \color{black}R1 \\ \cline{2-7}
& Insufficient Session Expiration                &  \color{black}No &  \color{black}Yes &  \color{black}No &  \color{black}Yes & \color{black}  TF2, TF8 \\  \cline{2-7}
 \hline

\multirow{3}{*}{\textbf{Validate Inputs} }&
Cross-site Scripting                          &\color{black} Yes &\color{black} Yes & \color{black}Yes &\color{black}Yes &  \color{black}TF2 \\ \cline{2-7}
&Deserialization of Untrusted Data               &\color{black} Yes & \color{black}Yes & \color{black}No & \color{black}No &\color{black} R2  \\ \cline{2-7}
\hline

\end{tabular}
\end{center}
\end{table*}

%% file: RQ2.tex
\subsection{RQ2: Vulnerability Types}
\label{subsec:RQ2}


\subsubsection{Analysis Procedure}

We aim to study which types of vulnerabilities can be discovered by \MST. To enable a discussion structured according to well-defined categories of vulnerabilities, we compute, for each category in the CWE views mentioned in Section~\ref{subsec:setup} (i.e., \emph{Common security architectural tactics}, \emph{CWE Top 25}, and \emph{OWASP Top 10}), the percentage of weaknesses that can be automatically discovered by \MST.

We systematically analyzed all the weaknesses with the objective of writing, for each one, one or more \MRs using \SMRL. 
For each weakness, we first inspect its description, its demonstrative examples, the description of concrete vulnerabilities (CVE) and common attack patterns (CAPEC)~\cite{CAPEC} associated with the weakness. Based on the information collected from our inspection, we implemented, using \SMRL, a new \MR that address the weakness
or reused, if possible, an \MR already available in the \MST catalog. 
Each time we could not do so, we kept track of the reasons preventing the writing of an \MR.
All the \MRs resulting from this analysis are part of the \MRs catalog provided online~\cite{Replicability}.

We report the percentage of weaknesses for which it has been possible to implement an \MR; in other words, the weaknesses that can be automatically tested with \MST.
Since some of the weaknesses in the CWE database are specific to certain types of systems (e.g., Java Enterprise~\cite{J2EE}), we distinguish between the results achieved with all the weaknesses (i.e., generic and specific), and the results achieved with the generic weaknesses only.

To better characterize the weaknesses that cannot be addressed by \MST, we analyze the distribution of the reasons preventing its application, across the categories of the views considered in our analysis. Finally, we discuss the percentage of the weaknesses belonging to the CWE Top 25 and OWASP Top 10 lists. \TSE{2.2}{To this end, we rely on the definitions available on the CWE Web site. More precisely, we rely on the list of CWE weakness IDs belonging to the views for CWE Top 25 and OWASP Top 10 provided on the CWE Web site. Within these lists, we identify the CWE IDs belonging to the CWE view for common security architectural tactics, our main target in this study, and track the CWE IDs tested by our \MRs.}

\input{tables/tableAC-SUM_MST}

To provide concrete examples of the weaknesses in our analysis, we report, in Table~\ref{table:AC-Weaknesses}, a subset of the weaknesses in the CWE view for common security architectural tactics. We refer to Table~\ref{table:AC-Weaknesses} in the rest of the section. Columns \emph{Design principle} and \emph{Weakness} report the security design principle affected by a weakness and its name, respectively. Columns \emph{Belongs to Top 25} and \emph{Belongs to Top 10} indicate whether a weakness also belongs to the \emph{CWE Top 25} or the \emph{OWASP Top 10} view, respectively. We also indicate if the weakness can be addressed by \MST. The remaining columns refer to concepts introduced later in this section.

\TSE{3.4}{Finally, to compare the capabilities of \MST with SOTA approaches, we refer to a recent empirical study from Elder et al.~\cite{Elder2022} comparing SOTA vulnerability detection tools based on dynamic and static program analysis. In the security context, static and dynamic program analysis tools are often referred to as Static application security testing (SAST) and Dynamic Application Security Testing (DAST). Elder et al. consider two open-source tools (i.e., OWASP Zap~\cite{ZapProxy} and Sonarqube~\cite{Sonar} for DAST and SAST, respectively) and another two proprietary tools whose name has not been disclosed due to licensing contracts (they are referred to as \emph{Dynamic Analysis 2 - DA2} and \emph{Static Analysis 2 - SA2)}. To the best of our knowledge, the study of Elder et al. is the only one providing the list of CWE IDs targeted by SOTA security tools (see Table 10 in their paper). Based on their list, we determine which CWE IDs belonging to the sets considered in our study are addressed by Zap, Sonarqube, SA2, and DA2. Using the collected data, we compare \MST with SOTA static and dynamic analysis regarding the overall number of targeted vulnerabilities. Also, 
we determine the number of vulnerability types that can be discovered
only by \MST and not by a single competing approach or by any of the four approaches. To be adopted in practice, \MST should complement SOTA tools (i.e., address weaknesses not discovered by these tools).}
\subsubsection{Results}

Table~\ref{table:ACSUM_MST} presents a summary of the CWE security design principles and related security weaknesses addressed by \MST. The first column in Table~\ref{table:ACSUM_MST} lists the security design principles appearing in the common security architectural tactics view. The second and third columns give, for each design principle, the overall number of weaknesses and the number of generic weaknesses, respectively. The fourth and fifth columns report the weaknesses that can be automatically discovered by \MST among all and generic weaknesses (we report the number, percentage, and ranking).

\input{tables/tableCWETop25Sum_MST}

\input{tables/tableOWASP-SUM_MST}

\input{tables/tableReason_MST}

In total, 101 out of all 223 weaknesses (45\%) and 98 out of 204 generic weaknesses (48\%) in the view can be addressed by \MST. These numbers show that our approach enables engineers to automatically discover a large subset of the weaknesses. Readers can download the details of our analysis for all 223 weaknesses from our replicability package~\cite{Replicability}.
If we sort security design principles based on the percentage of weaknesses addressed by \MST, we observe that the rankings for generic and all weaknesses match except for the first two security design principles in the rankings, which are swapped. These top ranked security design principles are
\textit{Manage User Sessions} ($1^{st}$ for generic weaknesses, $2^{nd}$ considering all the weaknesses) and \textit{Validate Inputs}   ($2^{nd}$ for generic weaknesses, $1^{st}$ considering all the weaknesses).
They are about 
malicious actors accessing resources because of session management faults (\textit{Manage User Sessions}), and
malicious actors providing malformed input data (e.g., code injection) to the system (\textit{Validate Inputs}). The main reason for the difference is that specific weaknesses for session management faults include two weakness (i.e., CWE-6 and CWE-579) related to the serialization of J2EE objects that cannot be discovered through \MT but only through code inspection. Indeed, CWE-6 is discovered by determining if the session includes a nonserializable object; CWE-579 can be discovered by checking if the session ID is stored in a field that is shorter than the one used in the J2EE session object. Since the rest of the ranking match, in the following discussion, we report only the percentages for generic weaknesses to simplify reading.

Other security design principles with a high percentage of weaknesses (above 40\%) being detected by \MST are  \textit{Authorize Actors} and \textit{Authenticate Actors} which concern malicious actors (external systems or users) accessing resources they are not authorized to access. These weaknesses are often discovered through interactions with the system and can be tested with \MST. \MST cannot cover cases in which program analysis is needed (e.g., \emph{Insufficient Compartmentalization} and \emph{Reliance on Security Through Obscurity}).

\MST addresses a low percentage (below 20\%) of the weaknesses related to the security design principles \textit{Audit}, \textit{Limit Exposure}, and \textit{Lock Computer} (i.e., 16\%, 0\%, and 0\% respectively). Such percentages are due to \MST relying, for source inputs, on sequences of user-system interactions. The weaknesses related to \textit{Audit}, \textit{Limit Exposure}, and \textit{Lock Computer} are, on the contrary, about quality of recorded logs, information that the system exposes, and restrictions of the lockout mechanism (e.g., lock an account after a predefined number of failed logins). They all require manual data inspection.

\input{tables/tableDistributionReason_MST}

Unsurprisingly, the design principles \textit{Authorize Actors} (34 weaknesses), \textit{Validate Inputs} (31),
and \textit{Authenticate Actors} (12) also have a high number of weaknesses addressed by \MST.
Indeed, they concern interactions between external actors and the system, which is the main focus of \MST.

Table~\ref{table:CWETop25Sum_MST} gives a summary of the CWE Top 25 weaknesses addressed by \MST. It can automatically discover 15 out of the 25 CWE top weaknesses (60\%) and 14 out of the 18 generic CWE top weaknesses (78\%), which shows that \MST is a key solution to identify widely spread weaknesses.

Among the CWE Top 25 weaknesses, \MST cannot address the ones that require program analysis or interactions with third parties (e.g., other system users or system administrators) to be detected:{\NAZANIN{ \emph{Out-of-bounds Write}, \emph{Out-of-bounds Read},
\emph{Improper Neutralization of Special Elements used in an OS Command}, \emph{Use After Free}, \emph{Integer Overflow or Wraparound}, \emph{Deserialization of Untrusted Data}, \emph{NULL Pointer Dereference}, \emph{Use of Hard-coded Credentials}, \emph{Improper Restriction of Operations within the Bounds of a Memory Buffer}, and \emph{Server-Side Request Forgery (SSRF)}.}}

Table~\ref{table:OWASP-SUM_MST} presents a summary of the security weaknesses related to the OWASP Top 10 security risks addressed by \MST.
It addresses {\NAZANIN{64}} out of the 126 weaknesses {\NAZANIN{(51\%)}} and {\NAZANIN{61}} out of the {\NAZANIN{117}} generic weaknesses {\NAZANIN{(52\%)}} in this view.
It addresses a high percentage {\NAZANIN{(above 55\%)}} of the weaknesses leading to security risks \textit{Broken Access Control}, \textit{Injection}, \textit{Insecure Design}, \textit{Security Misconfiguration}, and \textit{Identification and Authentication Failures}
{\NAZANIN{ (i.e., 75\%, 86\%, 55\%, 60\%, and 55\%, respectively).}} These risks are about unauthorized access to resources, injecting malicious client-side scripts into a website, leveraging the lack of security controls (e.g., an unprotected primary channel), gaining system information thanks to system misconfiguration, and bypassing
authentication. All involve malicious user-system interactions. 

{\NAZANIN \MST can address none of the weaknesses related to \textit{Vulnerable and Outdated Components}, which concern the use of outdated libraries affected by known vulnerabilities. Although it would be feasible to implement \MRs detecting failures of specific components, we cannot provide a catalog of \MRs that cover all the outdated libraries in the market. Therefore, we excluded them from our analysis.}



%

Table~\ref{table:Reason_MST} presents the reasons preventing the application of \MST to discover weaknesses. \MST has been designed to select source inputs by relying either on a Web crawler or on manually implemented test scripts that automate the
user-system interactions. Source inputs are automatically turned into follow-up inputs which are used to discover vulnerabilities. \MST can discover vulnerabilities that can be exercised through a sequence of interactions. Therefore, \MST cannot be applied 
when the weakness concerns a system that is not Web-based, or when it concerns components of a Web system that exchange data through dedicated protocols (see R1 in Table~\ref{table:Reason_MST}). 
Also, some weaknesses can be discovered only through program analysis, which \MST does not support (see R2 in Table~\ref{table:Reason_MST}). In that case, the analysis should be either performed without actually executing programs (e.g., through code inspection) or the output of program analysis should be reviewed manually to eliminate false positives, which typically come in large numbers~\cite{Austin2013}.
Example cases are reported in Table~\ref{table:AC-Weaknesses} and concern 
log files omitting or inappropriately reporting information about attempted attacks,
relying on a weak password recovery mechanism,
improper validation of certificates,
configuring a small space for random variables, 
lack of encryption,
storing objects in session tokens, and
deserializing untrusted data.

It is not possible to define an \MR in cases where a human is needed to inspect the system output (R3). For instance, to discover weakness \textit{Weak Password Recovery Mechanism for Forgotten Password} in Table~\ref{table:AC-Weaknesses}, a human needs to indicate that the system under test contains a mechanism for the recovery of passwords that is weak (e.g., it is based on a security question whose answer can be easily determined~\cite{CWE_WeakPasswordRecovery}).

\input{tables/tableDistributionReasonOWASP_MST}

\input{tables/tableDistributionReasonCWETop25_MST}

Due to the \MT foundations (based on \MRs), \MST cannot be applied for weaknesses that can be discovered only through data analysis (see R4 in Table~\ref{table:Reason_MST}). Some weaknesses can be determined only by analyzing large amounts of data (e.g., log files) based on statistics or machine learning; this analysis cannot be performed with an \MR. 


Finally, some weaknesses can be discovered only by controlling a third-party system
(see R5 in Table~\ref{table:Reason_MST}).
This is the case for \emph{Improper Validation of Certificate with Host Mismatch} (see Table~\ref{table:ST-Activities}), which requires setting up a malicious host with the same IP of the SUT.
Expressing such interactions in \MRs, in general, is infeasible.


Table~\ref{table:distributionReason_MST} presents the distribution of reasons preventing the application of \MST for the weaknesses regarding common security architectural tactics. 
In 60 out of 122 weaknesses (49\%) that cannot be discovered by \MST, program analysis is required (see R2). This is expected since program analysis complements software testing in software verification. 
R2 is particularly prevalent for the security design principle \textit{Encrypt Data}, where 21 out of 30 weaknesses (70\%) are not addressed due to R2. 
These 21
 weaknesses are associated with the protection of credentials or password (e.g., hard-coded cryptography key, password in configuration file), or the use of encryption/hash algorithm (e.g., hash without a salt). Therefore, in these cases, a static program analysis approach should be used; for example, to find hard-coded cryptography keys.

The second most frequent category is R3, which, in most of the cases, concerns the output generated in exceptional situation (e.g., messages provided in log files). R3 uniformly affects, with one to four cases, all the design principles except the two principles concerning the handling of input data (\emph{Validate Inputs} and \emph{Verify Message Integrity}). For these two cases, static program analysis is more effective than testing; this is the case for \emph{CWE-391} (\emph{Unchecked Error Condition}), where it is enough to examine source code for missing operations following the verification of function results.

The third most frequent category is R1; however, it is the prevalent reason for not applying \MST in the case of authorization weaknesses (15 out of 26 weaknesses) and has limited impact on all the other principles (i.e., one or no cases). Authorization is a generic security property that goes beyond Web-based systems; therefore, 15 out of these 26 weaknesses (58\%) concern functionalities that are not implemented by Web-systems. They concern the file system (e.g., preserving or managing the permissions related to files), the process control in an operating system, or the process communication in a mobile operating system. These weaknesses cannot be discovered by an automated test framework dedicated to Web-based interactions. 


\input{tables/tableApplicationCondition}

R4 has a limited impact on almost all the design principles except for \textit{Encrypt Data}, where it prevents \MST from detecting six weaknesses;
indeed, in several cases, the limitations of encryption algorithms (e.g., \emph{Insufficient entropy}) can be discovered only through a statistical test.

R5 is the least frequent reason for not applying \MST because most of the weaknesses are not due to  interactions among multiple components.

Tables~\ref{table:distributionReasonOWASP_MST} and~\ref{table:distributionReasonCWETop25_MST} report the reasons preventing the application of \MST to discover some highly critical vulnerabilities. In these cases as well, the main reason is the necessity to rely on static program analysis. This is expected since testing and program analysis are complementary quality assurance activities.

\input{tables/tableRQ2_design.tex}
\input{tables/tableRQ2_OWASP.tex}
\input{tables/tableRQ2_top25.tex}

\TSE{3.4}{Tables~\ref{table:RQ2:comparison:design} to~\ref{table:RQ2:comparison:top} provide the results of the comparison of \MST with state-of-the-art SAST and DAST tools. Concerning the security design principle verification (Table~\ref{table:RQ2:comparison:design}), we conclude that \MST targets most weaknesses and outperforms  the best competing approach (SA2). Also, the set of weaknesses targeted by \MST is larger than what can be targeted by applying all four competing approaches together. Indeed, all the other approaches can discover only 84 weaknesses (see Column \emph{Weaknesses addressed by any other} in Table~\ref{table:RQ2:comparison:design}), while \MST addresses 101 weaknesses (see Column \emph{Weaknesses addressed by \MST}). Concerning complementarity, we observe that \MST can detect 56 weaknesses that any other approach cannot address (see column \emph{Weaknesses addressed by MST but not by Any other}). Most of the weaknesses addressed by only \MST 
belong to the principles Authorize actors (25), Validate inputs (15), and Authenticate Actors (7), which are the principles with most of the weaknesses addressed by \MST (see Column \emph{Weaknesses addressed by \MST}). Together, all SOTA approaches can address only 39 weaknesses that \MST does not address (see column \emph{Weaknesses not addressed by MST but addressed by Any other}). The tool addressing most of the weaknesses not addressed by \MST is SA2, which fares the best in the study of Elder et al.~\cite{Elder2022}. 
The design principles in which SOTA tools provide greater benefits (i.e., the number of weaknesses addressed only by \MST is lower than the number of weaknesses addressed by other tools only) are (1) \emph{Encrypt data}, which we already reported above as one of the weakest for \MST, (2) \emph{Identify Actors} since SAST tools leverage code inspection to determine lack of certificate validation, (3) \emph{Audit}, which we already reported as benefiting from static analysis (e.g., it can determine lack of logging after exceptions or data-flows from sources of private data to log files), (4) \emph{Limit Access}, where SA2 and DA2 can detect information exposure through messages, (5) \emph{Limit Exposure}, where ZAP and likely SA2 can be used to detect cross-domain JavaScript file inclusion, and (6) \emph{Verify Message Integrity}, where SA2 and ZAP can be used to check the integrity of cookies and whether there are error conditions not being handled (the latter is enabled only by SA2).}

\TSEstart{}
Concerning the security design weaknesses in the OWASP Top 10  (see Table~\ref{table:RQ2:comparison:design}), we note that \MST targets most of the weaknesses. \MST discovers 64 weaknesses, while the best SOTA approach (i.e., SA2) finds 54 weaknesses (see column \emph{Weaknesses addressed by SA2}). We conclude that \MST complements the four other approaches by detecting 24 weaknesses they do not discover. 
The best-competing approach (SA2) can address 20 weaknesses not discovered by \MST, while \MST targets 30 weaknesses not addressed by SA2. However, all the SOTA approaches detect together 29 weaknesses not addressed by \MST, which indicates complementarity between \MST and other tools. SOTA approaches work better for cryptographic failures (13 weaknesses not discovered by \MST but discovered by others and two weaknesses detected only by \MST). Such a result is expected since most of those issues are related to outdated encryption algorithms or the wrong setup of security keys, which can be easily detected using program analysis (in particular static analysis). However, \MST outperforms the other approaches for \emph{Broken Access Control} (indeed, \MST includes several \MRs to determine if resources not available from the GUI can be accessed by modifying inputs), \emph{Code injection} (\MST includes several \MRs that modify requests by relying on catalogs of injections, including special characters, which are not covered by other approaches), and \emph{Insecure design}  (similarly to the case of \emph{Broken Access Control}, the detection of failures in the handling of privileges is enabled by \MST's \MRs verifying if resources not available from the GUI can be accessed by modifying inputs). 

As for the CWE Top 25 list (see Table~\ref{table:RQ2:comparison:top}), the number of weaknesses addressed by \MST and the best SOTA approach is similar (15 versus 14); however, they complement each other. \MST addresses five weaknesses not addressed by other approaches:
\emph{CWE-306 (Missing Authentication for Critical Function)},
\emph{CWE-276 (Incorrect Default Permissions)},
\emph{CWE-732 (Incorrect Permission Assignment for Critical Resource)},
\emph{CWE-77 (Improper Neutralization of Special Elements used in a Command - Command Injection)},
\emph{CWE-434 (Unrestricted Upload of File with Dangerous Type)}. In line with our discussion above, the first four weaknesses concern permission management or command injections not addressed by other approaches; the latter requires test automation not supported by the four SOTA approaches. Indeed, they do not generate test inputs automatically but rely on modifying inputs provided by the end user. \MST, instead, derives specific inputs for CWE-434. For the weaknesses not addressed by \MST but by SOTA approaches, we report that dynamic program analysis approaches (Zap and DA2) and SA2 discover a code injection vulnerability (\emph{CWE 78 - OS Command Injection}) that \MST does not target because it is not specific to Web systems. As discussed above, the other weaknesses not covered by \MST are due to \emph{them being only discoverable by means of program
analysis} (R2 in Table~\ref{table:Reason_MST}): \emph{CWE-190 (Integer Overflow or Wraparound)}, \emph{CWE-502 (Deserialization of Untrusted Data)}, \emph{CWE-476 NULL Pointer Dereference}, \emph{CWE-798  Use of Hard-coded Credentials}.
\TSEstop{}




\textbf{Summary.} As a result of our analysis, we conclude that \MST can address a large percentage (45\%) of the weaknesses organized in the CWE view for common security architectural tactics. Such percentage increases to 48\% if we consider generic weaknesses only. \MST can also address a majority of high-risk weaknesses (51\% of the weaknesses related to the OWASP Top 10 security risks and 60\% of the CWE Top 25 weaknesses). These results are promising as they demonstrate that \MST is relevant for a large subset of vulnerabilities occurring in practice. The weaknesses that \MST cannot address are mostly those (i) that can only be discovered  using program analysis, (ii) that are not related to  user-system
interactions, or (iii) that concern non-Web-based systems. \TSE{3.4}{Further, we demonstrated that, for all the category of weakness considered (i.e., security design principles, OWASP Top 10, or CWE Top 25), \MST complements state-of-the-art SAST and DAST approaches in terms of weaknesses addressed. Specifically, combining \MST with SA2 seems to be a particularly effective combination as it enables detecting 129 weaknesses (i.e., 101 + 28), that is 92\% of the 140 weaknesses that can be detected by any approach. Such complementarity is a key factor that justifies the adoption of metamorphic testing as a solution to improve software security. The weaknesses not discovered by any approach considered in our assessment (83 out of 223) can mainly be discovered through manual code inspection activities (e.g., to determine if \emph{a client/server product performs authentication within client code but not in server code}, CWE-603), activities which are difficult to automate through generic tools since they often require a specific understanding of the system under analysis.} 

%% file: tables/tableAC-SUM_MST.tex
\begin{table}
\begin{center}
\scriptsize
\caption{Summary of the CWE architectural security design principles and weaknesses addressed by \MST.}
\label{table:ACSUM_MST}

\begin{tabular}{|l|r|r|r|r|r|r|}
\hline
\multicolumn{1}{|c|}{\textbf{Security Design}} &
\multicolumn{2}{c|}{\textbf{weaknesses}} &
\multicolumn{2}{c|}{\textbf{addressed weaknesses}}
\\
 \multicolumn{1}{|c|}{\textbf{Principle}} &
  \textbf{all} &
  \textbf{generic} &
    \textbf{all} &
  \textbf{generic} 
 \\ 
 \hline
Audit                                                       & 6 & 6 &  1 (16\%) $10^{th}$&  1 (16\%) $10^{th}$   \\ \hline
Authenticate Actors                                         & 28 &  27 &  12 (43\%) $4^{th}$&   12 (44\%)   $4^{th}$\\ \hline
Authorize Actors                                            & 60 &   55 &   34 (57\%) $3^{rd}$&    34 (62\%)   $3^{rd}$\\ \hline
Cross Cutting                                               & 9  & 8 &  3 (33\%)   $6^{th}$&  3 (37\%) $6^{th}$\\ \hline
Encrypt Data                                                & 38 &  37 &  8 (21\%) $8^{th}$&   8 (22\%)   $8^{th}$\\ \hline
Identify Actors                                             & 12 &   12 &  3 (25\%)  $7^{th}$&  3 (25\%)  $7^{th}$\\ \hline
Limit Access                                                & 8  &   5 &  3 (38\%) $5^{th}$&  2 (40\%)   $5^{th}$\\ \hline
Limit Exposure                                              & 6  &   6 &  0 (0\%) $11^{th}$&  0 (0\%)   $11^{th}$\\ \hline
Lock Computer                                               & 1  & 1 &  0 (0\%) $11^{th}$&  0 (0\%)   $11^{th}$\\ \hline
Manage User Sessions                                        & 6  &   4 &  4 (67\%) $2^{nd}$&  4 (100\%)   $1^{st}$\\ \hline
Validate Inputs                                             & 39 & 33 &  31 (79\%)  $1^{st}$&   29 (88\%)  $2^{nd}$\\ \hline
Verify Message Integrity                                    & 10 & 10 &  2 (20\%) $9^{th}$&  2 (20\%)   $9^{th}$\\ \hline
\multicolumn{1}{|c|}{{\color[HTML]{343434} \textbf{Total}}} & 223&   204 &   101 (45\%) &   98 (48\%)   \\ \hline
\end{tabular}
\end{center}
\end{table}

%% file: tables/tableCWETop25Sum_MST.tex
\begin{table}[t]
\center
\scriptsize
\caption{Summary of the CWE Top 25 weaknesses addressed by \MST.}
\label{table:CWETop25Sum_MST}
\begin{tabular}{|r|r|r|r|}
\hline
\multicolumn{2}{|c|}{\textbf{Weaknesses}} &
\multicolumn{2}{c|}{\textbf{Addressed weaknesses}}
\\
  \textbf{all} &
  \textbf{generic} &
    \textbf{all} &
  \textbf{generic} 
 \\ 
 \hline
25&
   18&
  \NAZANIN{ 15 (60\%)}&
  \NAZANIN{ 14 (78\%)}\\
\hline
\end{tabular}
\end{table}

%% file: tables/tableOWASP-SUM_MST.tex

\begin{table}[t]

\scriptsize
\caption{Summary of the security weaknesses for OWASP Top 10 security risks addressed by \MST.}
  \vspace{-7pt}
\label{table:OWASP-SUM_MST}

\begin{tabular}{|p{3.4cm}|r|r|r|r|}
\hline
\multirow{2}{*}{\textbf{OWASP Security Risk}} &
\multicolumn{2}{c|}{\textbf{Weaknesses}} &
\multicolumn{2}{c|}{\textbf{Addressed weaknesses}}
\\
 &
  \textbf{all} &
  \textbf{generic} &
    \textbf{all} &
  \textbf{generic} 
 \\ 
 \hline

   \multirow{2}{*}{ Broken Access Control}                                         &    { 20} &    18 &    {15 (75\%)} &     {15 (83\%)}  \\[2ex] \hline
   \multirow{2}{*}{ Cryptographic Failures}                                                      &   24 &      24 &     {3 (13\%)} &     {3(13\%)}   \\[2ex] \hline
    \multirow{2}{*}{ Injection   }                                                    &   21 &   18 &    {18 (86\%)} &    {16 (89\%)}   \\[2ex] \hline
     \multirow{2}{*}{Insecure Design    }                                      &   22  &    20 &    { 12(55\%)} &     {12 (60\%)}   \\[2ex] \hline

     \multirow{2}{*}{Security Misconfiguration     }                                        &     5  &    4 &    {3 (60\%)} &    {2 (50\%)}  \\[2ex] \hline
   Vulnerable and Outdated Component                                              &   0 &    0&   {0 (0\%)} &     {0 (0\%)}    	 \\[2ex] \hline

   Identification and Authentication Failures                                                &    20 &    20  &    {11 (55\%)} &   {11 (55\%)}   \\ \hline

   Software and Data Integrity Failures                                               &   9  &    8 &    1 (11\%) &    1 (13\%)  \\ \hline

   Security Logging and Monitoring Failures                                        &     4 &     4 &    1 (25\%) &    1 (25\%)  \\    \hline

    Server-Side Request Forgery (SSRF)                                            &     1  &     1  &   { 0 (0\%)} &   0 (0\%)  \\ \hline

\multicolumn{1}{|c|}{{\color[HTML]{343434} \textbf{Total}}} &    126  &    117 &    {64 (51\%)} &     {61 (52\%)}  \\ \hline

\end{tabular}
\end{table}

%% file: tables/tableReason_MST.tex
\begin{table}[t]

\scriptsize
\caption{Reasons preventing the application of \MST.}
\label{table:Reason_MST}
\begin{center}
\begin{tabular}{|p{1cm}|p{7cm}|}
\hline
\textbf{ID} & \textbf{Reason}                             \\ \hline
R1          & The weakness concerns a system that is not Web-based or mobile-based. \\ \hline
R2          & The weakness can be discovered only by means of program analysis.  \\ \hline
R3          & It is not possible to distinguish valid and invalid behaviour based on system output; a human needs to inspect it. \\ \hline
R4          & The weakness can be discovered only by means of data analysis.  \\ \hline
R5          & The weakness can be discovered only by controlling a third-party component.  \\ \hline
\end{tabular}
\end{center}
\end{table}

%% file: tables/tableDistributionReason_MST.tex
\begin{table}[t]

\scriptsize
\caption{Distribution of reasons preventing the application of \MST to verify security design principles. The second column (\#) reports the number of weaknesses not discovered by \MST.}
\label{table:distributionReason_MST}
\begin{center}
\begin{tabular}{|l|r|r|r|r|r|r|r|}
\hline
\textbf{Security Design Principle}   & \textbf{\#} &  \textbf{R1} & \textbf{R2} & \textbf{R3} & \textbf{R4} &  \textbf{R5} & \textbf{Sum} \\ \hline
Audit                 &  5 &    &2  &  3     &  &  &   5     \\ \hline
Authenticate Actor       & 16   & &	11&	3&	1&	1&	16    \\ \hline
Authorize Actor          &26& 15&	6&	3&	1&	1&	26      \\ \hline
Cross Cutting            &  6  &1&	1&	4& &	&	6 \\ \hline
Encrypt Data             & 30  &1&	21&	2&	6&	&	30    \\ \hline
Identify Actors          &  9& 1&	2&	3&	&	3&	9  \\ \hline
Limit Access             & 5   &1&	1&	3&	&		&5\\ \hline
Limit Exposure           &  6  &1&	3&	2&	&	&	6\\ \hline
Lock Computer            & 1   &  &  & 1 &  &  &       1\\ \hline
Manage User Sessions     & 2    & 1   &    & 1 &  &  &   2\\ \hline
Validate Inputs          &  8  &    1&	7&	&	&	&	8\\ \hline
Verify Message Integrity &   8 &  &	6	& &	&	2&	8\\ \hline

\textbf{Total}           &    122 & 22&	\textbf{60}&	25&	8&	7	&122\\ \hline

\end{tabular}
\end{center}
\end{table}

%% file: tables/tableDistributionReasonOWASP_MST.tex
\begin{table}[tb]

\scriptsize
\caption{Distribution of reasons preventing the application of \MST to discover weaknesses associated with the OWASP Top 10 security risks.  The second column (\#) reports the number of weaknesses not discovered by \MST.}
\label{table:distributionReasonOWASP_MST}
\begin{center}
\begin{tabular}{|p{3cm}|r|r|r|r|r|r|r|}
\hline
\textbf{OWASP Security Risk}   & \textbf{\#} & \textbf{R1} & \textbf{R2} & \textbf{R3} & \textbf{R4} & \textbf{R5} & \textbf{Sum} \\ 
 \hline
   Broken Access Control                     &    5 &1&	3&	1&		& &	5 \\ \hline
   Cryptographic Failures       &   21 &  &	18&	1&	2&	&	21 \\ \hline
   Injection     &   3 & 1&	2&	&	&	&	3 \\ \hline
   Insecure Design        &   10 & 2&	6&	2&	&	&	10  \\ \hline
   Security Misconfiguration         &   2 & &	2&		& &	&	2   \\ \hline
   Vulnerable and Outdated Component   &   0 &   &   &     &   &   &    0  \\ \hline
   Identification and Authentication Failures       &    9 &1&	4&	1&	&	3&	9   \\ \hline
   Software and Data Integrity Failures     &    8 & 1&	6&	1&	&	&	8 \\ \hline
   Security Logging and Monitoring      &   3 &  &	3&		& &	&	3  \\ \hline
    Server-Side Request Forgery (SSRF)      &   1 &   &   &      &   &  1   &    1\\ \hline
\textbf{Total}              &    62 &   6&	\textbf{44}&	6&	2	&4&	62 \\ \hline

\end{tabular}
\end{center}
\end{table}

%% file: tables/tableDistributionReasonCWETop25_MST.tex

\begin{table}[tb]

\scriptsize
\caption{Distribution of reasons preventing the application of \MST to discover CWE Top 25 weaknesses.}
\label{table:distributionReasonCWETop25_MST}
\begin{center}
\begin{tabular}{|p{4.5cm}|r|r|r|r|r|}
\hline
\textbf{Weakness}    & \textbf{R1} & \textbf{R2} & \textbf{R3}  & \textbf{R4}  & \textbf{R5}\\ \hline
     Out-of-bounds Write    &   &     1  &   &   &      \\ \hline
      Out-of-bounds Read                  &   &     1 &   &   &      \\ \hline
   Improper Neutralization of Special Elements used in an OS Command ('OS Command Injection')                  &    1  &   &   &   &      \\ \hline
    Use After Free                          &   &1 &   &   &    \\ \hline

   Integer Overflow or Wraparound          &   &   &     1 &   &      \\ \hline
    Deserialization of Untrusted Data      &   &     1  &   &      &   \\ \hline
    NULL Pointer Dereference                    &   &     1 &   &   &      \\ \hline
   Use of Hard-coded Credentials               &  & &    1     &   &      \\ \hline
   Improper Restriction of Operations within the Bounds of a Memory Buffer     &   &       1  & &  &      \\ \hline

   Server-Side Request Forgery (SSRF)          &   &   & &  &    1    \\ \hline

\textbf{Total}                        &    1  &     \textbf{6} &    2  &       0  &     1 \\ \hline

\end{tabular}
\end{center}
\end{table}

%% file: tables/tableApplicationCondition.tex
\begin{table}[tb]

\scriptsize
\caption{Testability features and factors for \MST.}
\label{table:ApplicationCondition}
\begin{center}
\begin{tabular}{|l|p{4cm}|l|}
\hline
\textbf{ID} & \textbf{Testability feature}                                                           & \textbf{Testability factor} \\ \hline
TF1         & The feature under test is accessible via a URL/path                            & Controllability             \\ \hline
TF2         & The testing framework supports modifying parameter values                            & Test support environment             \\ \hline
TF3         & It is possible to log-in with a predefined list of credentials                       & Controllability             \\ \hline
TF4  & System settings or configuration elements can be controlled by 
the test engineer  & Controllability         \\ \hline
TF5 & The testing framework can control the Web-browser (e.g., click on back button) & Test support environment \\ \hline
TF6        & The type of the parameters of the request (in URL or post-data) is known or can be easily determined & Controllability             \\ \hline
TF7        & It is possible to access system artefacts (e.g., log files)                       & Observability               \\ \hline
TF8        & The system under test provides a feature to configure the system time                  & Controllability             \\ \hline
TF9 & The testing framework supports handling multiple user sessions in parallel             & Test support environment \\ \hline
TF10        & The testing framework has a feature to select certificates                   & Test support environment     \\ \hline
\end{tabular}
\end{center}
\end{table}

%% file: tables/tableRQ2_design.tex
\begin{table*}[]
\centering
\caption{Comparison of \MST with SOTA approaches for the verification of security design principles. Bold font is used to indicate the approach that performs best for a specific design principle. 
For the subtable \emph{Weaknesses addressed by}, we highlight the highest value on a row. 
Precisely, we highlight the highest value in the leftmost subtable, and we compare all columns across the two subtables to the right and highlight the highest value.
In the leftmost subtable, we underline cases where \MST performs better than the joint use of SOTA approaches. Also, in the middle subtable, we underline cases where the number of weaknesses addressed exclusively by \MST is higher than the number of weaknesses addressed exclusively by the combined use of other approaches.}
\label{table:RQ2:comparison:design}

\resizebox{\textwidth}{!}{%
\begin{tabular}{|l|cccccc||ccccc|ccccc|}
\hline
\multirow{2}{*}{\textbf{Security Design Principle}} &
  \multicolumn{6}{c||}{\textbf{Weaknesses addressed by}} &
  \multicolumn{5}{c|}{\textbf{Weaknesses addressed by MST but not by}} &
  \multicolumn{5}{c|}{\begin{tabular}[c]{@{}c@{}}\textbf{Weaknesses not addressed by MST}\\ \textbf{but addressed by}\end{tabular}} \\ \cline{2-17} 
 &
  \multicolumn{1}{c|}{MST} &
  \multicolumn{1}{c|}{Any other} &
  \multicolumn{1}{c|}{Zap} &
  \multicolumn{1}{c|}{DA2} &
  \multicolumn{1}{c|}{Sonar} &
  SA2 &
  \multicolumn{1}{c|}{Any other} &
  \multicolumn{1}{c|}{Zap} &
  \multicolumn{1}{c|}{DA2} &
  \multicolumn{1}{c|}{Sonar} &
  SA2 &
  \multicolumn{1}{c|}{Any other} &
  \multicolumn{1}{c|}{Zap} &
  \multicolumn{1}{c|}{DA2} &
  \multicolumn{1}{c|}{Sonar} &
  SA2 \\ \hline
Audit &
  \multicolumn{1}{c|}{1} &
  \multicolumn{1}{c|}{3} &
  \multicolumn{1}{c|}{0} &
  \multicolumn{1}{c|}{0} &
  \multicolumn{1}{c|}{0} &
  \textbf{3} &
  \multicolumn{1}{c|}{0} &
  \multicolumn{1}{c|}{{1}} &
  \multicolumn{1}{c|}{{1}} &
  \multicolumn{1}{c|}{{1}} &
  0 &
  \multicolumn{1}{c|}{\textbf{2}} &
  \multicolumn{1}{c|}{{0}} &
  \multicolumn{1}{c|}{{0}} &
  \multicolumn{1}{c|}{0} &
  \textbf{2} \\ \hline
Authenticate Actors &
  \multicolumn{1}{c|}{{\underline{\smash{\textbf{12}}}}} &
  \multicolumn{1}{c|}{10} &
  \multicolumn{1}{c|}{0} &
  \multicolumn{1}{c|}{2} &
  \multicolumn{1}{c|}{1} &
  9 &
  \multicolumn{1}{c|}{{\underline{\smash{{\textbf{7}}}}}} &
  \multicolumn{1}{c|}{{{12}}} &
  \multicolumn{1}{c|}{{11}} &
  \multicolumn{1}{c|}{{11}} &
  \textbf{7} &
  \multicolumn{1}{c|}{5} &
  \multicolumn{1}{c|}{0} &
  \multicolumn{1}{c|}{1} &
  \multicolumn{1}{c|}{0} &
  4 \\ \hline
Authorize Actors &
  \multicolumn{1}{c|}{{\underline{\smash{\textbf{34}}}}} &
  \multicolumn{1}{c|}{13} &
  \multicolumn{1}{c|}{2} &
  \multicolumn{1}{c|}{0} &
  \multicolumn{1}{c|}{1} &
  13 &
  \multicolumn{1}{c|}{{{25}}} &
  \multicolumn{1}{c|}{{{32}}} &
  \multicolumn{1}{c|}{\textbf{34}} &
  \multicolumn{1}{c|}{\textbf{34}} &
  {25} &
  \multicolumn{1}{c|}{{4}} &
  \multicolumn{1}{c|}{0} &
  \multicolumn{1}{c|}{0} &
  \multicolumn{1}{c|}{1} &
  4 \\ \hline
Cross Cutting &
  \multicolumn{1}{c|}{{ \textbf{3}}} &
  \multicolumn{1}{c|}{2} &
  \multicolumn{1}{c|}{0} &
  \multicolumn{1}{c|}{0} &
  \multicolumn{1}{c|}{2} &
  0 &
  \multicolumn{1}{c|}{{\underline{\smash{{2}}}}} &
  \multicolumn{1}{c|}{{\textbf{3}}} &
  \multicolumn{1}{c|}{\textbf{3}} &
  \multicolumn{1}{c|}{{2}} &
  \textbf{3} &
  \multicolumn{1}{c|}{{1}} &
  \multicolumn{1}{c|}{0} &
  \multicolumn{1}{c|}{0} &
  \multicolumn{1}{c|}{1} &
  0 \\ \hline
Encrypt Data &
  \multicolumn{1}{c|}{8} &
  \multicolumn{1}{c|}{\textbf{18}} &
  \multicolumn{1}{c|}{2} &
  \multicolumn{1}{c|}{5} &
  \multicolumn{1}{c|}{8} &
  \textbf{10} &
  \multicolumn{1}{c|}{3} &
  \multicolumn{1}{c|}{{8}} &
  \multicolumn{1}{c|}{{8}} &
  \multicolumn{1}{c|}{{7}} &
  4 &
  \multicolumn{1}{c|}{\textbf{13}} &
  \multicolumn{1}{c|}{2} &
  \multicolumn{1}{c|}{5} &
  \multicolumn{1}{c|}{7} &
  {6} \\ \hline
Identify Actors &
  \multicolumn{1}{c|}{3} &
  \multicolumn{1}{c|}{\textbf{7}} &
  \multicolumn{1}{c|}{1} &
  \multicolumn{1}{c|}{1} &
  \multicolumn{1}{c|}{1} &
  \textbf{7} &
  \multicolumn{1}{c|}{1} &
  \multicolumn{1}{c|}{{3}} &
  \multicolumn{1}{c|}{{3}} &
  \multicolumn{1}{c|}{{3}} &
  {1} &
  \multicolumn{1}{c|}{\textbf{5}} &
  \multicolumn{1}{c|}{1} &
  \multicolumn{1}{c|}{1} &
  \multicolumn{1}{c|}{1} &
  \textbf{5} \\ \hline
Limit access &
  \multicolumn{1}{c|}{3} &
  \multicolumn{1}{c|}{\textbf{5}} &
  \multicolumn{1}{c|}{0} &
  \multicolumn{1}{c|}{1} &
  \multicolumn{1}{c|}{1} &
  \textbf{5} &
  \multicolumn{1}{c|}{0} &
  \multicolumn{1}{c|}{\textbf{3}} &
  \multicolumn{1}{c|}{\textbf{3}} &
  \multicolumn{1}{c|}{{2}} &
  {0} &
  \multicolumn{1}{c|}{{2}} &
  \multicolumn{1}{c|}{{0}} &
  \multicolumn{1}{c|}{{1}} &
  \multicolumn{1}{c|}{0} &
  {2} \\ \hline
Limit exposure &
  \multicolumn{1}{c|}{0} &
  \multicolumn{1}{c|}{1} &
  \multicolumn{1}{c|}{1} &
  \multicolumn{1}{c|}{0} &
  \multicolumn{1}{c|}{0} &
  1 &
  \multicolumn{1}{c|}{0} &
  \multicolumn{1}{c|}{0} &
  \multicolumn{1}{c|}{0} &
  \multicolumn{1}{c|}{0} &
  0 &
  \multicolumn{1}{c|}{\textbf{1}} &
  \multicolumn{1}{c|}{\textbf{1}} &
  \multicolumn{1}{c|}{{0}} &
  \multicolumn{1}{c|}{{0}} &
  \textbf{1} \\ \hline
Lock computer &
  \multicolumn{1}{c|}{0} &
  \multicolumn{1}{c|}{0} &
  \multicolumn{1}{c|}{0} &
  \multicolumn{1}{c|}{0} &
  \multicolumn{1}{c|}{0} &
  0 &
  \multicolumn{1}{c|}{0} &
  \multicolumn{1}{c|}{0} &
  \multicolumn{1}{c|}{0} &
  \multicolumn{1}{c|}{0} &
  0 &
  \multicolumn{1}{c|}{0} &
  \multicolumn{1}{c|}{0} &
  \multicolumn{1}{c|}{0} &
  \multicolumn{1}{c|}{0} &
  0 \\ \hline
Manage User Session &
  \multicolumn{1}{c|}{{\underline{\smash{\textbf{4}}}}} &
  \multicolumn{1}{c|}{2} &
  \multicolumn{1}{c|}{0} &
  \multicolumn{1}{c|}{0} &
  \multicolumn{1}{c|}{0} &
  2 &
  \multicolumn{1}{c|}{{\underline{\smash{{2}}}}} &
  \multicolumn{1}{c|}{\textbf{4}} &
  \multicolumn{1}{c|}{\textbf{4}} &
  \multicolumn{1}{c|}{\textbf{4}} &
  {2} &
  \multicolumn{1}{c|}{0} &
  \multicolumn{1}{c|}{0} &
  \multicolumn{1}{c|}{0} &
  \multicolumn{1}{c|}{0} &
  0 \\ \hline
Validate Inputs &
  \multicolumn{1}{c|}{{\underline{\smash{\textbf{31}}}}} &
  \multicolumn{1}{c|}{20} &
  \multicolumn{1}{c|}{10} &
  \multicolumn{1}{c|}{7} &
  \multicolumn{1}{c|}{2} &
  14 &
  \multicolumn{1}{c|}{{\underline{\smash{{15}}}}} &
  \multicolumn{1}{c|}{{24}} &
  \multicolumn{1}{c|}{{25}} &
  \multicolumn{1}{c|}{\textbf{30}} &
  {19} &
  \multicolumn{1}{c|}{{4}} &
  \multicolumn{1}{c|}{3} &
  \multicolumn{1}{c|}{1} &
  \multicolumn{1}{c|}{1} &
  2 \\ \hline
Verify Message Integrity &
  \multicolumn{1}{c|}{2} &
  \multicolumn{1}{c|}{\textbf{3}} &
  \multicolumn{1}{c|}{1} &
  \multicolumn{1}{c|}{0} &
  \multicolumn{1}{c|}{0} &
  \textbf{3} &
  \multicolumn{1}{c|}{1} &
  \multicolumn{1}{c|}{\textbf{2}} &
  \multicolumn{1}{c|}{\textbf{2}} &
  \multicolumn{1}{c|}{\textbf{2}} &
  {1} &
  \multicolumn{1}{c|}{\textbf{2}} &
  \multicolumn{1}{c|}{{1}} &
  \multicolumn{1}{c|}{{0}} &
  \multicolumn{1}{c|}{0} &
  \textbf{2} \\ \hline
TOTAL &
  \multicolumn{1}{c|}{{\underline{\smash{\textbf{101}}}}} &
  \multicolumn{1}{c|}{84} &
  \multicolumn{1}{c|}{17} &
  \multicolumn{1}{c|}{16} &
  \multicolumn{1}{c|}{16} &
  67 &
  \multicolumn{1}{c|}{{\underline{\smash{{56}}}}} &
  \multicolumn{1}{c|}{{92}} &
  \multicolumn{1}{c|}{{94}} &
  \multicolumn{1}{c|}{\textbf{96}} &
  \textbf{62} &
  \multicolumn{1}{c|}{39} &
  \multicolumn{1}{c|}{8} &
  \multicolumn{1}{c|}{9} &
  \multicolumn{1}{c|}{11} &
  28 \\ \hline
\end{tabular}%
}
\end{table*}

%% file: tables/tableRQ2_OWASP.tex
\begin{table*}[]
\centering

\caption{Comparison of \MST with SOTA approaches for the verification of security design principles in the OWASP Top 10 list. Highlight and underlining follow the same conventions as Table~\ref{table:RQ2:comparison:design}.}
\label{table:RQ2:comparison:owasp}

\resizebox{\textwidth}{!}{%
\begin{tabular}{|l|cccccc||ccccc|ccccc|}
\hline
\multicolumn{1}{|c|}{\multirow{2}{*}{\textbf{OWASP Security Risk}}} &
  \multicolumn{6}{c||}{\textbf{Weaknesses addressed by}} &
  \multicolumn{5}{c|}{\textbf{\begin{tabular}[c]{@{}c@{}}Weaknesses addressed by MST\\  but not by\end{tabular}}} &
  \multicolumn{5}{c|}{\textbf{\begin{tabular}[c]{@{}c@{}}Weaknesses not addressed by MST \\ but addressed by\end{tabular}}} \\ \cline{2-17} 
\multicolumn{1}{|c|}{} &
  \multicolumn{1}{c|}{\textbf{MST}} &
  \multicolumn{1}{c|}{\textbf{Any other}} &
  \multicolumn{1}{c|}{\textbf{Zap}} &
  \multicolumn{1}{c|}{\textbf{DA2}} &
  \multicolumn{1}{c|}{\textbf{Sonar}} &
  \textbf{SA2} &
  \multicolumn{1}{c|}{\textbf{Any other}} &
  \multicolumn{1}{c|}{\textbf{Zap}} &
  \multicolumn{1}{c|}{\textbf{DA2}} &
  \multicolumn{1}{c|}{\textbf{Sonar}} &
  \textbf{SA2} &
  \multicolumn{1}{c|}{\textbf{Any other}} &
  \multicolumn{1}{c|}{\textbf{Zap}} &
  \multicolumn{1}{c|}{\textbf{DA2}} &
  \multicolumn{1}{c|}{\textbf{Sonar}} &
  \textbf{SA2} \\ \hline
\begin{tabular}[c]{@{}l@{}}Broken Access Control\\  \color{white}n\end{tabular}
 &
  \multicolumn{1}{c|}{{\ul \textbf{15}}} &
  \multicolumn{1}{c|}{11} &
  \multicolumn{1}{c|}{4} &
  \multicolumn{1}{c|}{3} &
  \multicolumn{1}{c|}{1} &
  11 &
  \multicolumn{1}{c|}{{ \underline{\smash{ 5 }}}} &
  \multicolumn{1}{c|}{{11}} &
  \multicolumn{1}{c|}{{13}} &
  \multicolumn{1}{c|}{\textbf{14}} &
  {5} &
  \multicolumn{1}{c|}{1} &
  \multicolumn{1}{c|}{0} &
  \multicolumn{1}{c|}{1} &
  \multicolumn{1}{c|}{0} &
  1 \\ \hline
\begin{tabular}[c]{@{}l@{}}Cryptographic Failures\\  \color{white}n\end{tabular}
 &
  \multicolumn{1}{c|}{3} &
  \multicolumn{1}{c|}{\textbf{14}} &
  \multicolumn{1}{c|}{1} &
  \multicolumn{1}{c|}{6} &
  \multicolumn{1}{c|}{6} &
  {8} &
  \multicolumn{1}{c|}{2} &
  \multicolumn{1}{c|}{{3}} &
  \multicolumn{1}{c|}{3} &
  \multicolumn{1}{c|}{3} &
  2 &
  \multicolumn{1}{c|}{\textbf{13}} &
  \multicolumn{1}{c|}{1} &
  \multicolumn{1}{c|}{{6}} &
  \multicolumn{1}{c|}{{6}} &
  \textbf{7} \\ \hline
\begin{tabular}[c]{@{}l@{}} Injection\\  \color{white}n\end{tabular}
 &
  \multicolumn{1}{c|}{{\ul \textbf{18}}} &
  \multicolumn{1}{c|}{16} &
  \multicolumn{1}{c|}{7} &
  \multicolumn{1}{c|}{5} &
  \multicolumn{1}{c|}{0} &
  11 &
  \multicolumn{1}{c|}{{\underline{\smash{ 5 }}}} &
  \multicolumn{1}{c|}{{14}} &
  \multicolumn{1}{c|}{{14}} &
  \multicolumn{1}{c|}{\textbf{18}} &
  {8} &
  \multicolumn{1}{c|}{3} &
  \multicolumn{1}{c|}{3} &
  \multicolumn{1}{c|}{1} &
  \multicolumn{1}{c|}{0} &
  1 \\ \hline
\begin{tabular}[c]{@{}l@{}} Insecure Design\\  \color{white}n\end{tabular}
 &
  \multicolumn{1}{c|}{{\ul \textbf{12}}} &
  \multicolumn{1}{c|}{8} &
  \multicolumn{1}{c|}{3} &
  \multicolumn{1}{c|}{1} &
  \multicolumn{1}{c|}{2} &
  4 &
  \multicolumn{1}{c|}{{\underline{\smash{ 6 }}}} &
  \multicolumn{1}{c|}{{10}} &
  \multicolumn{1}{c|}{\textbf{12}} &
  \multicolumn{1}{c|}{{10}} &
  {9} &
  \multicolumn{1}{c|}{2} &
  \multicolumn{1}{c|}{1} &
  \multicolumn{1}{c|}{1} &
  \multicolumn{1}{c|}{0} &
  1 \\ \hline
\begin{tabular}[c]{@{}l@{}} Security Misconfiguration\\  \color{white}n\end{tabular}
 &
  \multicolumn{1}{c|}{3} &
  \multicolumn{1}{c|}{3} &
  \multicolumn{1}{c|}{0} &
  \multicolumn{1}{c|}{0} &
  \multicolumn{1}{c|}{1} &
  3 &
  \multicolumn{1}{c|}{1} &
  \multicolumn{1}{c|}{\textbf{3}} &
  \multicolumn{1}{c|}{\textbf{3}} &
  \multicolumn{1}{c|}{{2}} &
  1 &
  \multicolumn{1}{c|}{1} &
  \multicolumn{1}{c|}{0} &
  \multicolumn{1}{c|}{0} &
  \multicolumn{1}{c|}{0} &
  1 \\ \hline
\begin{tabular}[c]{@{}l@{}}Vulnerable and Outdated \\ Component\end{tabular} &
  \multicolumn{1}{c|}{0} &
  \multicolumn{1}{c|}{0} &
  \multicolumn{1}{c|}{0} &
  \multicolumn{1}{c|}{0} &
  \multicolumn{1}{c|}{0} &
  0 &
  \multicolumn{1}{c|}{0} &
  \multicolumn{1}{c|}{0} &
  \multicolumn{1}{c|}{0} &
  \multicolumn{1}{c|}{0} &
  0 &
  \multicolumn{1}{c|}{0} &
  \multicolumn{1}{c|}{0} &
  \multicolumn{1}{c|}{0} &
  \multicolumn{1}{c|}{0} &
  0 \\ \hline
\begin{tabular}[c]{@{}l@{}}Identification and Authentication \\ Failures\end{tabular} &
  \multicolumn{1}{c|}{11} &
  \multicolumn{1}{c|}{\textbf{12}} &
  \multicolumn{1}{c|}{0} &
  \multicolumn{1}{c|}{2} &
  \multicolumn{1}{c|}{2} &
  \textbf{12} &
  \multicolumn{1}{c|}{4} &
  \multicolumn{1}{c|}{\textbf{11}} &
  \multicolumn{1}{c|}{{10}} &
  \multicolumn{1}{c|}{{10}} &
  4 &
  \multicolumn{1}{c|}{{5}} &
  \multicolumn{1}{c|}{0} &
  \multicolumn{1}{c|}{1} &
  \multicolumn{1}{c|}{1} &
  \textbf{5} \\ \hline
\begin{tabular}[c]{@{}l@{}}Software and Data Integrity\\  Failures\end{tabular} &
  \multicolumn{1}{c|}{1} &
  \multicolumn{1}{c|}{\textbf{3}} &
  \multicolumn{1}{c|}{2} &
  \multicolumn{1}{c|}{0} &
  \multicolumn{1}{c|}{1} &
  \textbf{3} &
  \multicolumn{1}{c|}{1} &
  \multicolumn{1}{c|}{1} &
  \multicolumn{1}{c|}{{1}} &
  \multicolumn{1}{c|}{1} &
  1 &
  \multicolumn{1}{c|}{\textbf{3}} &
  \multicolumn{1}{c|}{2} &
  \multicolumn{1}{c|}{0} &
  \multicolumn{1}{c|}{1} &
  \textbf{3} \\ \hline
\begin{tabular}[c]{@{}l@{}}Security Logging and Monitoring\\  Failures\end{tabular} &
  \multicolumn{1}{c|}{1} &
  \multicolumn{1}{c|}{\textbf{2}} &
  \multicolumn{1}{c|}{0} &
  \multicolumn{1}{c|}{0} &
  \multicolumn{1}{c|}{0} &
  \textbf{2} &
  \multicolumn{1}{c|}{0} &
  \multicolumn{1}{c|}{\textbf{1}} &
  \multicolumn{1}{c|}{\textbf{1}} &
  \multicolumn{1}{c|}{\textbf{1}} &
  0 &
  \multicolumn{1}{c|}{\textbf{1}} &
  \multicolumn{1}{c|}{0} &
  \multicolumn{1}{c|}{0} &
  \multicolumn{1}{c|}{0} &
 \textbf{1}  \\ \hline
\begin{tabular}[c]{@{}l@{}}Server-side Request Forgery\\  (SSRF)\end{tabular} &
  \multicolumn{1}{c|}{0} &
  \multicolumn{1}{c|}{0} &
  \multicolumn{1}{c|}{0} &
  \multicolumn{1}{c|}{0} &
  \multicolumn{1}{c|}{0} &
  0 &
  \multicolumn{1}{c|}{0} &
  \multicolumn{1}{c|}{0} &
  \multicolumn{1}{c|}{0} &
  \multicolumn{1}{c|}{0} &
  0 &
  \multicolumn{1}{c|}{0} &
  \multicolumn{1}{c|}{0} &
  \multicolumn{1}{c|}{0} &
  \multicolumn{1}{c|}{0} &
  0 \\ \hline
\begin{tabular}[c]{@{}l@{}}TOTAL\\  \color{white}n\end{tabular}
 &
  \multicolumn{1}{c|}{{64}} &
  \multicolumn{1}{c|}{\textbf{69}} &
  \multicolumn{1}{c|}{17} &
  \multicolumn{1}{c|}{17} &
  \multicolumn{1}{c|}{13} &
  54 &
  \multicolumn{1}{c|}{24} &
  \multicolumn{1}{c|}{{54}} &
  \multicolumn{1}{c|}{{57}} &
  \multicolumn{1}{c|}{\textbf{59}} &
  {30} &
  \multicolumn{1}{c|}{{29}} &
  \multicolumn{1}{c|}{7} &
  \multicolumn{1}{c|}{10} &
  \multicolumn{1}{c|}{8} &
  20 \\ \hline
\end{tabular}%
}
\end{table*}

%% file: tables/tableRQ2_top25.tex
\begin{table*}[]
\centering

\caption{Comparison of \MST with SOTA approaches for the verification of weaknesses in the CWE Top 25 list. Highlights follow the same convention ofTable~\ref{table:RQ2:comparison:design}.}
\label{table:RQ2:comparison:top}
\resizebox{\textwidth}{!}{%
\begin{tabular}{|cccccc||ccccc|ccccc|}
\hline
\multicolumn{6}{|c||}{\textbf{Weaknesses addressed by}} &
  \multicolumn{5}{c|}{\textbf{Weaknesses addressed by MST but not by}} &
   \multicolumn{5}{c|}{\textbf{\begin{tabular}[c]{@{}c@{}}Weaknesses not addressed by MST \\ but addressed by\end{tabular}}} \\ \hline
\multicolumn{1}{|c|}{\textbf{MST}} &
  \multicolumn{1}{c|}{\textbf{Any other}} &
  \multicolumn{1}{c|}{\textbf{Zap}} &
  \multicolumn{1}{c|}{\textbf{DA2}} &
  \multicolumn{1}{c|}{\textbf{Sonar}} &
  \textbf{SA2} &
  \multicolumn{1}{c|}{\textbf{Any other}} &
  \multicolumn{1}{c|}{\textbf{Zap}} &
  \multicolumn{1}{c|}{\textbf{DA2}} &
  \multicolumn{1}{c|}{\textbf{Sonar}} &
  \textbf{SA2} &
  \multicolumn{1}{c|}{\textbf{Any other}} &
  \multicolumn{1}{c|}{\textbf{Zap}} &
  \multicolumn{1}{c|}{\textbf{DA2}} &
  \multicolumn{1}{c|}{\textbf{Sonar}} &
  \textbf{SA2} \\ \hline
\multicolumn{1}{|c|}{\textbf{15}} &
  \multicolumn{1}{c|}{\textbf{15}} &
  \multicolumn{1}{c|}{6} &
  \multicolumn{1}{c|}{6} &
  \multicolumn{1}{c|}{4} &
  14 &
  \multicolumn{1}{c|}{5} &
  \multicolumn{1}{c|}{{10}} &
  \multicolumn{1}{c|}{{10}} &
  \multicolumn{1}{c|}{\textbf{12}} &
  {6} &
  \multicolumn{1}{c|}{5} &
  \multicolumn{1}{c|}{1} &
  \multicolumn{1}{c|}{1} &
  \multicolumn{1}{c|}{1} &
  5 \\ \hline
\end{tabular}%
}
\end{table*}

%% file: RQ3.tex
\subsection{RQ3: testability guidelines}
\label{subsec:RQ3}


\subsubsection{Analysis Procedure}

This research question investigates the possibility to define testability guidelines that support engineers in automatically testing software systems with \MST. 
More precisely, we aim to identify a set of features (hereafter \emph{testability features}) that should be provided either by the software under test or by the test framework and environment. Testability guidelines should indicate which testability features are required to detect specific categories of weaknesses, e.g., targeting a security design principle or entailing a high risk.

To identify testability features, we study the weaknesses that can be discovered by \MST, 
in the CWE view for common security architectural tactics.
We identify, for each weakness, a set of features necessary to enable automated testing with our approach.
\TSE{2.3}{For each Web-specific function used in our \MRs, we determine the inputs sent to the SUT and the outputs retrieved from the SUT. Further, we determine what SUT's interfaces should receive the inputs selected by the \MR and produce the outputs retrieved by the \MR; based on them, we derive our testability features. We expect our set of features to be complete because we developed \MST's Web-specific functions.}
\input{tables/tableDistributionCondition_MST}

\input{tables/tableDistributionConditionOWASPTop10_MST}

The identification of features or, more generally, factors that affect or influence software testability is the subject of active research on testability. 
A recent survey~\cite{testability2019survey} lists 21 testability factors: 
\textit{observability}, \textit{controllability}, \textit{complexity}, 
\textit{cohesion}, \textit{understandability}, \textit{inheritance},
\textit{reliability}, \textit{availability}, \textit{flexibility}, \textit{test suite reusability},
\textit{maintainability},
\textit{unit size}, 
\textit{statefulness},
\textit{isolateability},
\textit{software process capability},
\textit{modularity},
\textit{test support environment},
\textit{fault-proneness},
\textit{manageability}, 
\textit{quality of the test suite}, and
\textit{self-documentation}. 
To guide engineers towards the inspection of the proposed testability features for \MST, we match each testability feature to the testability factors in the literature. This allows us to group related testability features.

\input{tables/tableDistributionConditionCWETop25_MST}

Finally, we analyze the distribution of the testability features across the security design principles of the CWE view for common security architectural tactics,
the security risks in the OWASP Top 10 CWE view,
and the weaknesses in the CWE Top 25 view.
This analysis should assist engineers in prioritizing the implementation of testability features for the system under test, based on the targeted weaknesses and security design principles.

\subsubsection{Results}

Table~\ref{table:ApplicationCondition} presents the testability features for \MST and the corresponding testability factors.
In total, we have 10 testability features. Six features concern the system under test, while the other four features concern the test environment (see testability factor \emph{Test support environment}).

In the case of \MST, three out of the 21 testability factors in the literature are required; these are (i) \emph{Controllability} (i.e., the degree to which it is possible to control the state of the component under test~\cite{testability2019survey}), 
(ii) \emph{Observability} (i.e., how easy it is to observe the behavior of a program in terms of its outputs, effects on the environment, and other hardware and software components~\cite{testability2019survey}), and (iii) \emph{Test Support Environment}. In our context, testability factor \emph{Test Support Environment} refers to the capability of the testing environment or framework to provide features to analyze system outputs or to alter the inputs transmitted to the system under test. The required testability factors are mostly determined by the type of testing performed by \MST: security vulnerability testing at the system level by mimicking the actions performed by a malicious user. Therefore, to determine if the output of the system is correct, \MST may require improved \emph{Observability}. To test the system under specific configurations, it needs high \emph{Controllability}, and to automate activities typically performed by malicious users manually, the \emph{Test Support Environment} requires a high degree of automation.

Before discussing the distribution of testability features across design principles, we explain some of the testability features for the weaknesses in Table~\ref{table:AC-Weaknesses}. For instance, weakness \textit{Improper Authentication} is a generic weakness associated with design principle \textit{Authenticate Actors}. It indicates that the system under test does not properly verify the identity claimed by an actor~\cite{CWE_ImproperAuthentication}.
\MST can be applied to identify this weakness when the feature under test is accessible through a URL/path (see TF1 in Table~\ref{table:ApplicationCondition}). We match this testability feature to testability factor \textit{Controllability}.
In weakness \textit{Insufficient Session Expiration} in Table~\ref{table:AC-Weaknesses}, a Web system permits malicious users to reuse old session credentials or session IDs for authorization~\cite{CWE_InsufficientSessionExpiration}. \MST automatically identifies this weakness only when it is possible to modify the values of the parameters passed in HTTP requests (TF2 concerning \emph{Test Support Environment}), which is supported by our \MST toolset, and when the system under test provides a feature to configure the system time (TF8 concerning \emph{Controllability}), which is usually feasible through  secure shell connection, a feature leveraged by our toolset). For instance, an \MR in \SMRL can modify the HTTP-request (e.g., session IDs and cookie values) to reuse old session credentials. 

Table~\ref{table:distributionCondition_MST} presents the distribution of the testability features across the security design principles in the CWE view for common security architectural tactics. Please note that there are more than one testability feature for some of the weaknesses associated with the security design principles. An example is weakness \textit{Insufficient Session Expiration}, which is associated with testability features \textit{TF2} and \textit{TF8} (see Table~\ref{table:AC-Weaknesses}). TF2 supports the
test engineer to reuse an old session (e.g., credentials or ID) by modifying the corresponding request parameters; TF8 helps to change the system time in order to invalidate this session (i.e., make it expired).

In total, we identify {\NAZANIN{102}} testability features for 101 weaknesses concerning the 12 security design principles in 
Table~\ref{table:distributionCondition_MST}. Security design principles \emph{Authorize Actors} and \emph{Validate Inputs} are the ones that require the most testability features; 
this mostly depends on the fact that they are related to the largest subset of weaknesses (see Table~\ref{table:ACSUM_MST}). In Table~\ref{table:distributionCondition_MST}, rows \emph{Total} and \emph{\% of weaknesses} show that the two testability features with the largest number of associated weaknesses are TF1 and TF2 {\NAZANIN{(22\% and 40\%, respectively). These two features are respectively related to testability factors \textit{Controllability} and \emph{test support environment}. 
}}

{\NAZANIN{In our analysis, we observe that \emph{controllability}, \emph{test support environment} and \emph{observability} are required to address 48\%, 48\% and 4\% of the weaknesses, respectively.}} 
These numbers are not fully in line with the literature on the topic, where the two most popular factors are observability (mentioned in 101 papers) and controllability (82 papers)~\cite{testability2019survey}.
We believe that this difference is due to the fact that \MST automatically simulates actions performed by a user on a Web system under specific conditions (e.g., after performing a login). It thus requires a high degree of controllability to exercise the features under test or control the state of the system, instead of a high degree of observability. On the other hand, the literature on testability mostly concerns functional and robustness testing, which requires a high degree of observability. 
The high relevance of the testability factor \emph{Test Support Environment} for \MST is due to the fact that, to mimic a malicious user, it is necessary to automate all the actions typically performed manually by malicious users.

Tables~\ref{table:distributionConditionOWASPTop10_MST} and \ref{table:distributionConditionCWETop25_MST} present the testability features that enable testing for the weaknesses in the OWASP Top 10 and CWE Top 25 views, respectively.
In Table~\ref{table:distributionConditionOWASPTop10_MST}, numbers are in line with the ones in Table~\ref{table:distributionCondition_MST}.
{\NAZANIN{
Indeed, the testability features that are required for testing a higher subset of weaknesses are also TF1 and TF2 in Tables~\ref{table:distributionConditionOWASPTop10_MST}. 
In Table~\ref{table:distributionConditionCWETop25_MST}, TF2 and TF3 have a significant role in testing the features.}}

Tables~\ref{table:distributionCondition_MST},~\ref{table:distributionConditionOWASPTop10_MST}, and \ref{table:distributionConditionCWETop25_MST} provide testability guidelines for engineers. They enable engineers to determine, based on the security requirements of the system under test, which testability features need to be enabled. For example, if the system provides authorization and authentication features, engineers need to implement security design principles \textit{Authenticate Actors} and \textit{Authorize Actors}. Consequently, it might be useful to ensure that these two features under test are accessible through a URL/path (TF1), that the testing framework supports both modifying parameter values (TF2), and that it is possible to log-in with a predefined list of credentials (TF3).  
Moreover, TF1, TF2, and TF3 enable test automation for highly critical weaknesses, which concern code injection, authorization, and authentication. 
In addition, by looking at the testability features addressing more than 10\% of the vulnerabilities in
Tables~\ref{table:distributionCondition_MST},~\ref{table:distributionConditionOWASPTop10_MST}, and \ref{table:distributionConditionCWETop25_MST}
(i.e., at least four weaknesses associated with the OWASP Top 10 risks, two weakness in the CWE Top 25 list, and ten weaknesses concerning the security design principles), 
it is possible to identify a minimal set of testability features (i.e., TF1, TF2, TF3, and TF4) that should be prioritized to automatically verify both security design principles and top security risks. Since TF2 is provided by our \MST implementation and available in manual Web testing frameworks, and, further, TF1, TF3, and TF4 are common in Web systems, we conclude that \MT is
likely applicable in most software projects without the need for adapting existing design or testing frameworks.

\TSE{2.10}{To summarize, controllability and test support environment are the most required testability factors for \MST. Our results show that the most required testability features (i.e., TF1, TF2, TF3, and TF4) are either provided by \MST or available in most Web systems, thus suggesting that \MST can be applied to a wide variety of Web systems.}

%% file: tables/tableDistributionCondition_MST.tex
\begin{table*}[tb]

\scriptsize
\caption{Distribution of testability features for \MST.}
\label{table:distributionCondition_MST}
\begin{center}
\begin{tabular}{|p{3.55cm}@{}|p{0.65cm}@{}|
p{0.65cm}@{}|p{0.65cm}@{}|p{0.65cm}@{}|p{0.65cm}@{}|
p{0.65cm}@{}|
p{0.65cm}@{}|
p{0.65cm}@{}|
p{0.65cm}@{}|
p{0.8cm}@{}|
p{0.8cm}@{}|
p{0.8cm}@{}|
p{0.8cm}@{}|
p{0.7cm}|}
\hline
\multirow{2}{3.55cm}{\textbf{Security Design Principle}}&\multicolumn{10}{c|}{\textbf{Testability Feature}}\\
\cline{2-11}
&
  \textbf{TF1} &
  \textbf{TF2} &
  \textbf{TF3} &
  \textbf{TF4} &
  \textbf{TF5} &
  \textbf{TF6} &
  \textbf{TF7} &
  \textbf{TF8} &
  \textbf{TF9} &
  \textbf{TF10} \\ \hline
   Audit			    		& - &    - & - & - & - &- & 1 & - & - & -  \\ \hline
   Authenticate Actors	    	&    2&	3&	2&	3&	-&	-&	-&	2&	-&	-  \\ \hline
   Authorize Actors    		&   14&	4&	9&	2&	1&	2&	1&	-	&-&	1 \\ \hline
   Cross Cutting	    		&    - &    3 & - & - & - & - & - & - & - & -  \\ \hline
    Encrypt Data		    	& - &1&	1&	3&	-&	-&	2&	1&	-&	- \\ \hline
   Identify Actors			    &    1 & - & - & - & - & - & - & - & - &    2 \\ \hline
   Limit Access    			&    1 &    1 & - & - & - & - & - & - & 1 & -  \\ \hline
   Limit Exposure  			& - & - & - & - & - & - & - & - & - & -  \\ \hline
   Lock Computer   			& - & - & - & - & - & - & - & - & - & -  \\ \hline
   Manage User Sessions     	& - &   1 & - & - & 1 & - & - &    1 &    2 & -  \\ \hline
   Validate Inputs 			&    4 &    27 & - & - & - & - & - & - & - & -  \\ \hline
    Verify Message Integrity    & 1&    1 & - & - & - & - & - & - & - & -  \\ \hline
   \textbf{Total}   	        &    23 &    41 &    12 &    8 &    2 &    2 &    4 &    4 &    3 &   3  \\ \hline
\textbf{\% of weaknesses *} 	&    22\tiny{\%} &    40\tiny{\%} &    12\tiny{\%} &    8\tiny{\%} &    2\tiny{\%} &    2\tiny{\%} &    4\tiny{\%} &    4\tiny{\%} &    3\tiny{\%} &    3\tiny{\%}  \\ 
\hline
\end{tabular}
\end{center}
\footnotesize{* Percentage of weaknesses that can be discovered thanks to a testability feature.}\\
\end{table*}

%% file: tables/tableDistributionConditionOWASPTop10_MST.tex
\begin{table*}[tb]

\scriptsize
\caption{Distribution of testability features of \MST for the weaknesses associated with the OWASP Top 10 security risks.
}
\label{table:distributionConditionOWASPTop10_MST}
\begin{center}
  
  \begin{tabular}{|p{6.95cm}@{}|p{0.65cm}@{}|p{0.65cm}@{}|p{0.65cm}@{}|p{0.65cm}@{}|p{0.65cm}@{}|p{0.65cm}@{}|p{0.65cm}@{}|p{0.65cm}@{}|p{0.65cm}@{}|p{0.8cm}@{}|p{0.8cm}@{}|p{0.8cm}@{}|p{0.8cm}@{}|p{0.17cm}|}
\hline
\multirow{2}{3.55cm}{\textbf{OWASP Security Risk}}&\multicolumn{10}{c|}{\textbf{Testability Feature}}\\
\cline{2-11}
&
  \textbf{TF1} &
  \textbf{TF2} &
  \textbf{TF3} &
  \textbf{TF4} &
  \textbf{TF5} &
  \textbf{TF6} &
  \textbf{TF7} &
  \textbf{TF8} &
  \textbf{TF9} &
  \textbf{TF10} \\ \hline
   Broken Access Control					&    7 &   3 &    4 & - & - & - & 1 & - & - & -  \\ \hline
   Cryptographic Failures		& - & - & - &    2 & - & - &    - &    1 & - & -  \\ \hline
   Injection	&    3 &    14 & - & - & - & - & - & - & 1 & -  \\ \hline
   Insecure Design		&    2&	3&	3&	-&	1&	1&	2&	-&	-&	-  \\ \hline
   Security Misconfiguration		&-&   1& - &    2& - & - & - & - & - & -  \\ \hline
   Vulnerable and Outdated Component	& - & - & - & - & - & - &    - & - & - & -  \\ \hline
   Identification and Authentication Failures		&    1&	3&	2&	4&	-&	-&	-&	1&	1&	- \\ \hline
   Software and Data Integrity Failures 	& - &    1 & - & - & - & - & - & - & - &-  \\ \hline
   Security Logging and Monitoring Failures	& - &    - & - & - & - & - & 1 & - & - & -  
\\ \hline
   Server-side Request Forgery (SSRF)  \&	Monitoring    & - & - & - & - & - & - & - & - & - & -  \\ \hline

\textbf{Total} 	&    13 &    25 &   9 &    8&   1 &    1 &    4 &    2 &   2 &    0  \\ \hline
\textbf{\% of weaknesses *} 	&    20\tiny{\%} &    38\tiny{\%} &    14\tiny{\%} &    12\tiny{\%} & 
    2\tiny{\%} &    2\tiny{\%} &    6\tiny{\%} &    3\tiny{\%} &    3\tiny{\%} &    0\tiny{\%}  \\ 
\hline
\end{tabular}
\end{center}
\footnotesize{* Percentage of weaknesses that can be discovered thanks to a testability feature.}\\

\end{table*}

%% file: tables/tableDistributionConditionCWETop25_MST.tex
\begin{table*}[tb]

\scriptsize
\caption{Distribution of testability features of \MST for the CWE Top 25 weaknesses.}
\label{table:distributionConditionCWETop25_MST}
\begin{center}

  \begin{tabular}{|@{\hspace{1mm}}p{9cm}@{\hspace{1mm}}|p{0.65cm}@{}|p{0.65cm}@{}|p{0.65cm}@{}|p{0.65cm}@{}|p{0.65cm}@{}|p{0.65cm}@{}|p{0.65cm}@{}|p{0.65cm}@{}|p{0.65cm}@{}|p{0.65cm}|}
\hline
\multirow{2}{3.55cm}{\textbf{CWE Top 25 Weakness}}&\multicolumn{10}{c|}{\textbf{Testability Feature}}\\
\cline{2-11}
&
  \textbf{TF1} &
  \textbf{TF2} &
  \textbf{TF3} &
  \textbf{TF4} &
  \textbf{TF5} &
  \textbf{TF6} &
  \textbf{TF7} &
  \textbf{TF8} &
  \textbf{TF9} &
  \textbf{TF10} \\ \hline

   Improper Neutralization of Input During Web Page Generation ('Cross-site Scripting')  		& - &    1 & - & - & - & - & - & - & - & -  \\ \hline
    Improper Input Validation			        & - &    1 & - & - & - & - & - & - & - & -  \\ \hline
   Improper Neutralization of Special Elements used in an SQL Command ('SQL Injection')			        & - &    1 & - & - & - & - & - & - & - & -  \\ \hline
   Improper Limitation of a Pathname to a Restricted Directory ('Path Traversal')			&    1 & - & - & - & - & - & - & - & - & -  \\ \hline
    Cross-Site Request Forgery (CSRF)	    & - &    1 & - & - & - & - & - & - & - & -  \\ \hline
    Unrestricted Upload of File with Dangerous Type       & - & - & - & - & - &    1 & - & - & - & - \\ \hline
    Missing Authentication for Critical Function 	& - & - &    1 & - & - & - & - & - & - & -  \\
 \hline
    Improper Authentication  & - & - &    1 & - & - & - & - & - & - & -  \\ \hline
   Missing Authorization    	    &    - & - & 1 & - & - & - & - & - & - & - \\ \hline
   Incorrect Default Permissions   	    & - & - &    1 & - & - & - & - & - & - & - \\ \hline
   Exposure of Sensitive Information to an Unauthorized Actor           & 1 & - &   - & - & - & - & - & - & - & -   \\ \hline
   Insufficiently Protected Credentials   	    & - & - & 1 & - & - & - &   - & - & - & - \\ \hline
    Incorrect Permission Assignment for Critical Resource   	    & 1 &    -& - & - & - & - & - & - & - & - \\ \hline
   Improper Restriction of XML External Entity Reference    	    & - &    1 & - & - & - & - & - & - & - & - \\ \hline
   Improper Neutralization of Special Elements used in a Command ('Command Injection')    	    & - &    1 & - & - & - & - & - & - & - & - \\ \hline

\textbf{Total} 	                &  3&	6&	5&	0&	0&	1&	0&	0&	0&	0  \\ \hline
\textbf{\% of weaknesses *} 	&    20\tiny{\%} &    40\tiny{\%}  &    33\tiny{\%} &    0\tiny{\%} &    0\tiny{\%} &    7\tiny{\%} &    0\tiny{\%} &    0\tiny{\%} &    0\tiny{\%} &    0\tiny{\%} \\ 
\hline
\end{tabular}
\end{center}

\footnotesize{* The percentage of weaknesses that can be discovered thanks to the testability feature.}\\
\end{table*}

%% file: evaluation.tex

\section{Empirical Evaluation}
\label{sec:evaluation}

In this section, we investigate, based on two open-source case studies, the following RQs:

\begin{itemize}


\item \textit{\textbf{RQ4.}} \textit{\textbf{Is \MST effective?}} The goal of this research question is to assess whether \MST enables, in a reliable manner, the automated detection of security vulnerabilities.

\item \textit{\textbf{RQ5.}} \textit{\textbf{Is \MST 
\TSE{1.3}{efficient}?}} The goal of this research question is to analyze whether the execution time entailed by \MST is acceptable in practice.

\end{itemize}

To perform our empirical evaluation we relied on our implementation of \MST, a toolset that extends the Eclipse IDE~\cite{EclipseIDE}.
Additional information about the toolset, including executable files, download instructions, and a screencast, is available on the tool's website at \url{https://sntsvv.github.io/SMRL/}. A replicability package is provided as well~\cite{Replicability}.

\input{subjects}


\input{RQ4}

\input{RQ5}


\input{validity}

%% file: subjects.tex
\subsection{Subjects of the Evaluation}
\label{sec:subjects}

We applied \MST to Jenkins~\cite{Jenkins}, a leading open source automation server, and Joomla~\cite{Joomla}, a popular open source content management system.
The two case study subjects differ in programming languages and underlying frameworks. Jenkins is a Java Web application that we can execute within any servlet container~\cite{Jetty}; Joomla is a PHP application that relies on the MySQL RDBMS~\cite{Mysql} and the Apache HTTP server~\cite{Apache}. We chose them because they represent modern Web systems having a plug-in architecture and Web interfaces with advanced features such as Javascript-based login and AJAX interfaces. 
Their differences in programming languages and types of inputs may lead to different vulnerabilities and contribute to the generalizability of our empirical results.
We used Jenkins version 2.121.1 and Joomla version 3.8.7. We selected the Jenkins and Joomla versions affected by all the vulnerabilities triggerable from the Web interface, discovered in 2018 and reported in the Common Vulnerabilities and Exposures (CVE) database~\cite{CVE} after June 1st, 2018. Jenkins 2.121.1 and Joomla 3.8.7 are affected by 20 and 16 vulnerabilities, respectively.
Our catalog of \MRs addresses 40\% (8 out of 20) and 31\% (5 out of 16) of the vulnerabilities affecting Jenkins and Joomla, respectively. This outcome is consistent with our analysis in RQ1.
Among these vulnerabilities, we selected only the ones (eight vulnerabilities for Jenkins and two for Joomla) whose effects could be manually reproduced in our environment (i.e., we could observe a security failure). 
For instance, we could replicate only two vulnerabilities for Joomla due to the lack of a detailed description of the attack scenarios. 
 (one is a software fault that received a CVE identifier~\cite{NewCVE}, the other two are related to Jenkins' default configuration). In addition, we considered, for Jenkins, three new vulnerabilities we discovered with \MST.
We have, for Joomla, one new additional configuration-related vulnerability discovered by \MST. 
To summarize, we considered 11 vulnerabilities for Jenkins and 3 for Joomla (see Table~\ref{table:subjectVulnrabilities}). In the table, we provide two CWE IDs when the CVE report refers to a generic vulnerability type (e.g., CWE\_863 for incorrect authorization in reference~\cite{JV6}), but a more specific one fits better (e.g., CWE\_280 about improper handling of privileges that may lead to incorrect authorization).

\input{tables/tableVulnerabilities}

We configured, for each subject, our data collection framework with multiple users having different roles. We used four credentials for Jenkins and six credentials for Joomla. We executed, for each role, the data collection framework to crawl the system under test for a maximum of 300 minutes. The data collection took 1000 minutes for Jenkins and 2280 minutes for Joomla. Crawljax completed the crawling in less than 300 minutes for the anonymous role in Jenkins and Joomla because it visited all states. The data collection time for Joomla was long because it has two different user interfaces (i.e., user and administrative interfaces). The crawling led to 156 and 147 input sequences for Jenkins and Joomla, respectively. Also, we implemented Selenium-based test scripts 
\TSE{2.4}{(four for Jenkins and one for Joomla)} to exercise use cases not covered by Crawljax. These scripts entail a small overhead but address limitations in the crawler (cost-benefit trade-off).
\TSE{2.4}{Based on the URLs in the input sequences collected by Crawljax, we determined which features in the Jenkins and Joomla documentation had been tested. We then identified the features not being tested (i.e., configuring site settings in Joomla, creating new jobs, and canceling jobs in Jenkins).}

%% file: tables/tableVulnerabilities.tex
\begin{table}[t]
\center
\scriptsize
\caption{Vulnerabilities considered in our empirical evaluation.}
\label{table:subjectVulnrabilities}
\begin{tabular}{|r|r|p{2cm}|p{4cm}|}
\hline

  \textbf{Subject} &
  \textbf{Ref.} & \textbf{Vuln. Type} &
    \textbf{Description} 
 \\ 
 \hline
\multirow{9}{*}{Jenkins} &
  \cite{JV6} 
  &
  CWE\_863,
  CWE\_280 &
  Jenkins does not perform a permission check for URLs handling cancellation of queued builds, allowing users with Overall/Read permission to cancel queued builds. \\
 &
 \cite{JV5}&
 CWE\_863, CWE\_285 &
  Jenkins does not perform a permission check for the URL that initiates agent launches,
  allowing users with Overall/Read permission to initiate agent launches. \\
 &
  \cite{JV4}&
 CWE\_200, CWE\_668 &
  Files indicating when a plugin file was last extracted into the Jenkins \texttt{plugins/} directory are accessible via HTTP by users having Overall/Read permissions. This allows unauthorized users to determine the likely install date of a given plugin. \\
 &
 \cite{JV9}&
  CWE\_22 &
  In the file name parameter of a Job configuration, users with Job/Configure permissions can specify a relative path escaping the base directory. Such path can be used to upload a file on the Jenkins host, resulting in an arbitrary file write vulnerability. \\
 &
 \cite{JV2}&
  CWE\_200 &
  Users with Overall/Read permission ar able to access the URL serving agent logs on the UI due to a lack of permission checks. \\
 &
 \cite{JV8}&
  CWE\_384 &
  Jenkins does not invalidate the existing session when a user signs up for a new user account. This allows session fixation. \\
 &
  
  \cite{JV15}&
  CWE\_521 &
  Jenkins does not require users to have strong passwords, which makes it easier for attackers to compromise user accounts. \\
 &
  \cite{JV16}&
  CWE\_262 &
  Jenkins does not integrate any mechanism for managing password aging; consequently, users aren't incentivized to update passwords periodically. \\
 &
  \cite{JV12} &
  CWE\_79 &
  Jenkins does not set Content-Security-Policy headers for files uploaded as file parameters to a build, resulting in a stored XSS vulnerability.\\
  &
     \cite{JV1}&
  CWE\_863 &
  Users with Overall/Read permission can access the URL used to cancel scheduled restart jobs initiated through the update center due to a lack of permission checks. \\
 &
 
   \cite{JV3}&
  CWE\_287 &
  Users with a valid cookie can remain logged in even if the Remember me feature has been disabled in the Jenkins configuration. \\

  \hline
\multirow{3}{*}{Joomla} &
 
 \cite{JV13}&
  CWE\_863 &
  Inadequate checks on the tags search fields can lead to an access level violation. \\
 &
 
 \cite{JV14}&
  CWE\_200 &
  Inadequate checks allow users to see the names of tags that were either unpublished or published with restricted view permission . \\
 &
  \cite{JV16} &
  CWE\_262 &
  Joomla does not integrate any mechanism for managing password aging; consequently, users aren't incentivized to update passwords periodically.\\
\hline
\end{tabular}
\end{table}

%% file: RQ4.tex


\subsection{RQ4: effectiveness}
\label{subsec:empirical:RQ5}

\subsubsection{Experiment design}
An automated testing approach is effective if it helps detect a large proportion of the faults affecting the software under test and generates a limited number of false alarms. Ideally, \MST should identify all the vulnerabilities affecting our case study subjects that can be reproduced in our environment.
We executed \MST with the two case study subjects, considering all the \MRs in our catalog.
For each MT failure reported by \MST, we manually verified if the test input actually triggered any vulnerability (true positive).

\input{tables/tableResultsNew.tex}

We measured specificity and sensitivity~\cite{Lane2003}. 
\TSE{2.11}{For specificity, we consider the set of follow-up inputs $F$ as follows:}
\TSEstart{}
\begin{equation*}
\begin{aligned}
F=&F_{TP}+F_{FP}+F_{TN}+F_{FN}
\end{aligned}
\end{equation*}

with $TP$, $FP$, $TN$, and $FN$ standing for true positives (the follow-up input leads to a failure because the vulnerability has been exercised), false positives (the follow-up input leads to a failure for the wrong reason), true negatives (the follow-up input does not lead to any failure because no vulnerability has been exercised), and false negatives (the follow-up input does not lead to any failure although the vulnerability has been exercised), respectively. In our experiments, we did not observe any false negative ($F_{FN}=0$). Indeed, we investigated the URLs known to be vulnerable and verified that they always lead to failures when they are the target of HTTP requests with parameters that exercise their vulnerability. Further, we assume that our case study subjects are unlikely to be affected by unknown vulnerabilities (i.e., false negatives) because the selected software versions have been used for years. Therefore, we refine our definition of $F$ as follows:
\begin{equation*}
\begin{split}
F=&F_{TP}+F_{FP}+F_{TN}\\
F_{TN}=&F-F_{TP}-F_{FP}\\
F_{TN}+F_{FP}=&F-F_{TP}\\
\end{split}
\end{equation*}

Based on the above, specificity can be computed as follows:
\begin{equation*}
\begin{split}
\mathit{Specificity}=&\frac{F_{TN}}{F_{TN}+F_{FP}}\\
=&\frac{F-F_{TP}-F_{FP}}{F-F_{TP}}\\
\end{split}
\end{equation*}

In other words, specificity (true negative rate) is the ratio of follow-up inputs not triggering any vulnerability that (correctly) do not lead to any MT failure.

Further, the standard definition of false positive rate leads to:
\begin{equation*}
\begin{split}
\mathit{FPR}=&\frac{F_{FP}}{F_{TN}+F_{FP}}\\
=&1-\mathit{Specificity}
\end{split}
\end{equation*}

$(1-\mathit{Specificity})$ measures the proportion of unwarranted \MT failures, which enables us to discuss the effort wasted by engineers with \MST.
We report both the overall specificity of \MST (across all the follow-up inputs generated in our experiment) and the specificity distribution across \MRs.

\emph{Sensitivity} (true positive rate) can be computed as
\begin{equation*}
\begin{split}
\mathit{Sensitivity}=&\frac{F_{TP}}{F_{TP}+F_{FN}}
\end{split}
\end{equation*}

However, for reasons provided below, we measure \textit{sensitivity} as the ratio of vulnerabilities discovered:

\begin{equation*}
\begin{split}
\mathit{Sensitivity}=&\frac{V_{TP}}{V_{TP}+V_{FN}}
\end{split}
\end{equation*}

with $V_{TP}$ being the number of vulnerabilities discovered and $V_{FN}$ being the number of vulnerabilities not discovered.
We employ a coarse granularity (vulnerabilities instead of follow-up inputs) because, by relying on vulnerabilities, sensitivity matches the fault detection rate, which is a standard metric to assess software testing approaches  since engineers would like to know if a vulnerability is discovered, not how many time \MST reports it.

\TSEstop{}

%

\begin{figure}[tb]
\begin{center}
\includegraphics[width=8.4cm]{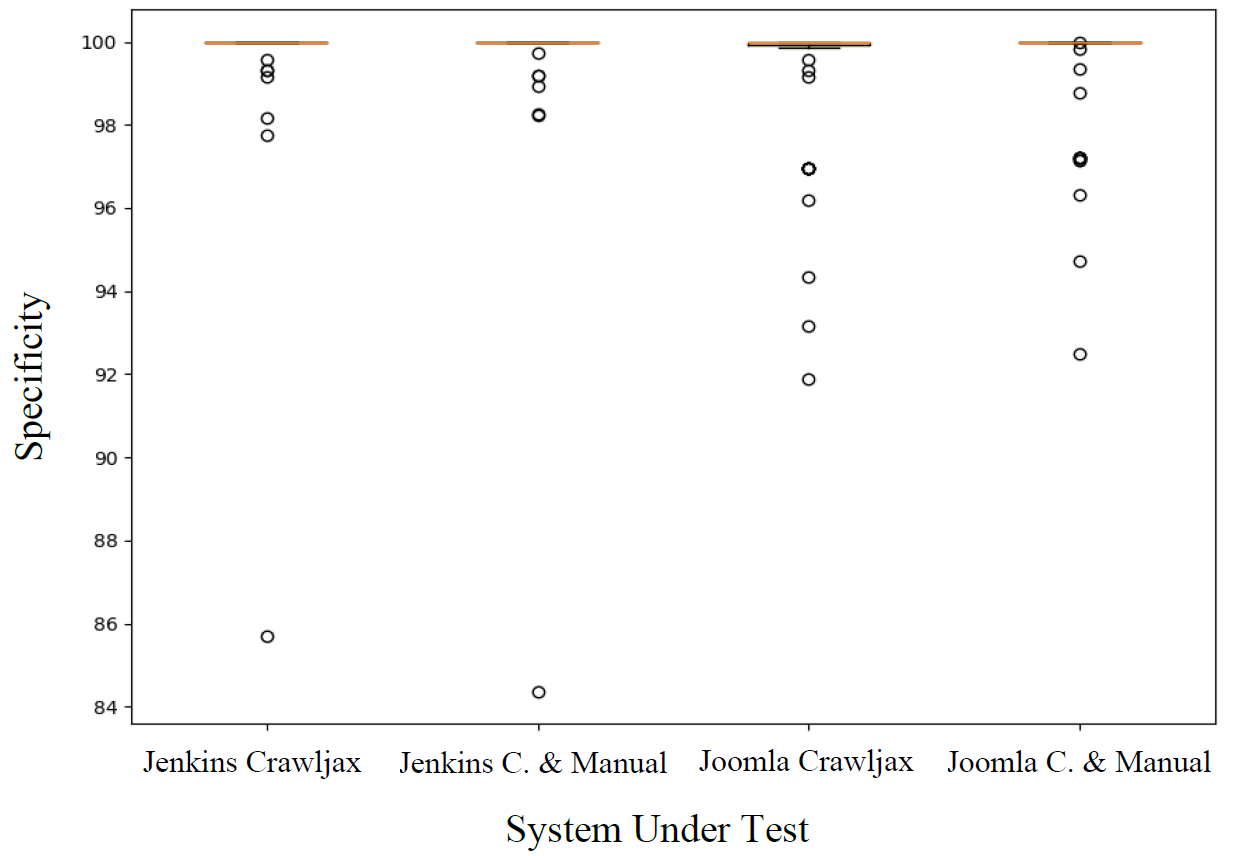}
 \caption{RQ4: Specificity distribution (values in Table~\ref{tab:specificity})} 
\label{fig:specificity}
\end{center}
\end{figure}

\input{tables/tableSpecificity}

\subsubsection{Results}

Table~\ref{table:resultsRQ2} summarizes the results obtained with the two data collection methods supported by \MST, i.e., based on Crawljax only or integrating Crawljax and manual scripts.

We observe that the approach has \textbf{extremely high specificity} when relying on the crawler (99.87\%) and when combining the crawler and manual inputs  (99.81\%). The high specificity indicates that 
only a negligible fraction of follow-up inputs  leads to false alarms (121 out of 93359, 0.13\%, and 103 out of 55174, 
0.19\%).
False alarms are due to limitations in Crawljax, which, for Jenkins, did not traverse all the URLs provided by the GUI for all users. 
Consequently, \MRs concerning authorization vulnerabilities fail.
However, 
it is easy to determine that the URLs causing the false alarms should be accessible to all users.

Fig.~\ref{fig:specificity} shows boxplots presenting the distribution of specificity across \MRs. Specificity is high for every \MR, with the median being $100\%$. The lowest whisker\footnote{Computed as $\textit{first quartile} - 1.5 * \textit{Inter Quartile Range}$} is 99.94\% for Joomla with Crawljax only, which indicates that, without outliers, the minimum specificity is above 99\%. In practice, for all our \MRs, only a very small proportion of follow-up inputs leads to false alarms.

The worst specificity outlier is 84.38\% for \texttt{OTG\_INPVAL\_004} with Jenkins (five false positives out of 32 follow-up inputs tested). The false positives are mainly due to asynchronous actions in the source inputs, which remain to be completed when the follow-up input is executed and thus lead to different outputs for the source and follow-up inputs, making the \MR fail. However, the five false positives lead to a limited waste of developers' time. Indeed, to discover these false positives, it is sufficient to manually test the URL of the original and the follow-up action, which does not take more than ten minutes in total.



\textbf{Sensitivity is high} when data collection relies on both Crawljax and manual test scripts: 81.81\% for Jenkins and 100\% for Joomla.
Since sensitivity reflects the fault detection rate (i.e.,  the proportion of vulnerabilities discovered), we conclude that our approach is \textbf{highly effective}. 
Overall, \MST detects 85.71\% of the vulnerabilities targeted in our evaluation.
It misses two of the eight targeted vulnerabilities in Jenkins. We can reveal one missing vulnerability only if the server configuration is modified during test execution~\cite{JV3}. Unfortunately, our toolset does not support the server configuration during test execution.
We cannot reproduce the other missing vulnerability since it concerns the termination of Jenkins' reboot~\cite{JV1}, which is not interruptible when Jenkins is not overloaded 
(our case).


When the data collection relies on Crawljax only, sensitivity drops below 70\% for both Jenkins and Joomla.
The low sensitivity occurs because of the incapability of our crawler to exercise some particular interactions. For example, Jenkins requires quick system interactions 
to exercise some features (e.g., writing a valid Unix command in a textbox to enqueue a batch job and then quickly pressing a button to delete it from the queue). Joomla, instead, requires interactions with a widget showing all the available tags (in the presence of multiple widgets, Crawlajx may fail to systematically exercise all the widgets).
However, even when the data collection is based on Crawljax only, with 9 out of 14 (64.29\%) vulnerabilities detected, \TSE{3.15.12}{we nevertheless consider the overall fault detection rate satisfactory.}  
Indeed, automatically detecting 64\% of the vulnerabilities not targeted by SOTA approaches, without the need for any manual test script, is beneficial. 


The benefits of \MST mostly stem from the \MRs in our catalog being reusable to test any Web system.
Furthermore, the required manual test scripts are few and inexpensive to implement.
For the Web systems above, we manually wrote 5 test scripts which only contain 31 actions in total.
This manual effort is negligible compared to 93359 input sequences (544806 actions) \TSE{3.15.13}{automatically generated by our approach to test the two systems when Crawljax and manual inputs are combined.} 
A traditional way to verify the same scenarios would require \CHANGEDLAST{93359} manually implemented test scripts, each providing a distinct input sequence and a dedicated oracle (e.g., an assertion statement).
Therefore, we conclude that \MST provides an advantageous cost-effectiveness trade-off compared to current practice.
%

%% file: tables/tableResultsNew.tex

\begin{table}[tb]
\scriptsize
\caption{Summary of RQ4 results grouped by data collection method.}
\begin{tabular}{|@{\hspace{0.1cm}}p{1.6cm} @{\hspace{0.05cm}}| @{\hspace{0.05cm}}p{1.5cm} |  @{\hspace{0.08cm}}r@{\hspace{0.08cm}} | @{\hspace{0.08cm}}r@{\hspace{0.08cm}}    |  @{\hspace{0.08cm}}r@{\hspace{0.08cm}} | @{\hspace{0.08cm}}r@{\hspace{0.08cm}} |  }
\hline
\multirow{2}{*}{\textbf{Case study}}&\textbf{Discovered}&\multicolumn{2}{c}{\textbf{Crawljax}}&\multicolumn{2}{c|}{\textbf{Crawljax \& Manual}}\\
&\textbf{Vulnerabilities}&\textbf{Specificity}&\textbf{Sensitivity}&\textbf{Specificity}&\textbf{Sensitivity}\\
\hline
Jenkins& \cite{JV6,JV5,JV4,JV2,JV9,JV8,JV15,JV16,JV12}& \CHANGEDLAST{99.94}\% & 63.64\% &   \CHANGEDLAST{99.90}\% & 81.81\%  \\
\hline
Joomla&  \cite{JV13,JV14,JV16}& 99.79\% & 66.67\% &   99.71\% & 100.00\% \\
\hline
\textbf{Overall} & 12 & 99.87\% & 64.29\% &   99.81\% & 85.71\%  \\
\hline
\end{tabular}
\label{table:resultsRQ2}
\end{table}%


%% file: tables/tableSpecificity.tex
\begin{table}[]
\centering
\scriptsize
\caption{RQ4: Specificity distribution}
\label{tab:specificity}
\begin{tabular}{|l|c|c|c|c|}
\hline
&
\multicolumn{2}{c|}{Jenkins} &
  \multicolumn{2}{c|}{Joomla} \\ 
  \cline{2-5}
&
  Crawljax &
  C. \& Manual &
  Crawljax &
  C. \& Manual\\ 
 \hline
First Quarltile  & 100        & 100         & 99.94     & 100      \\ \hline
Second Quarltile & 100        & 100         & 100       & 100        \\ \hline
Third Quarltile  & 100        & 100         & 100       & 100        \\ \hline
Lower whisker    & 100        & 100         & 99.84     & 100      \\ \hline
Upper whisker    & 100        & 100         & 100       & 100        \\ \hline
\end{tabular}%


\end{table}

%% file: RQ5.tex

\subsection{RQ5: \TSE{1.3}{Efficiency}}
\label{subsec:empirical:RQ5}
\subsubsection{Experiment design}

The execution time of \MST depends on the time required to run the crawler and the time required to execute the \MRs.

\begin{figure}[tb]
\begin{center}
\includegraphics[width=8.4cm]{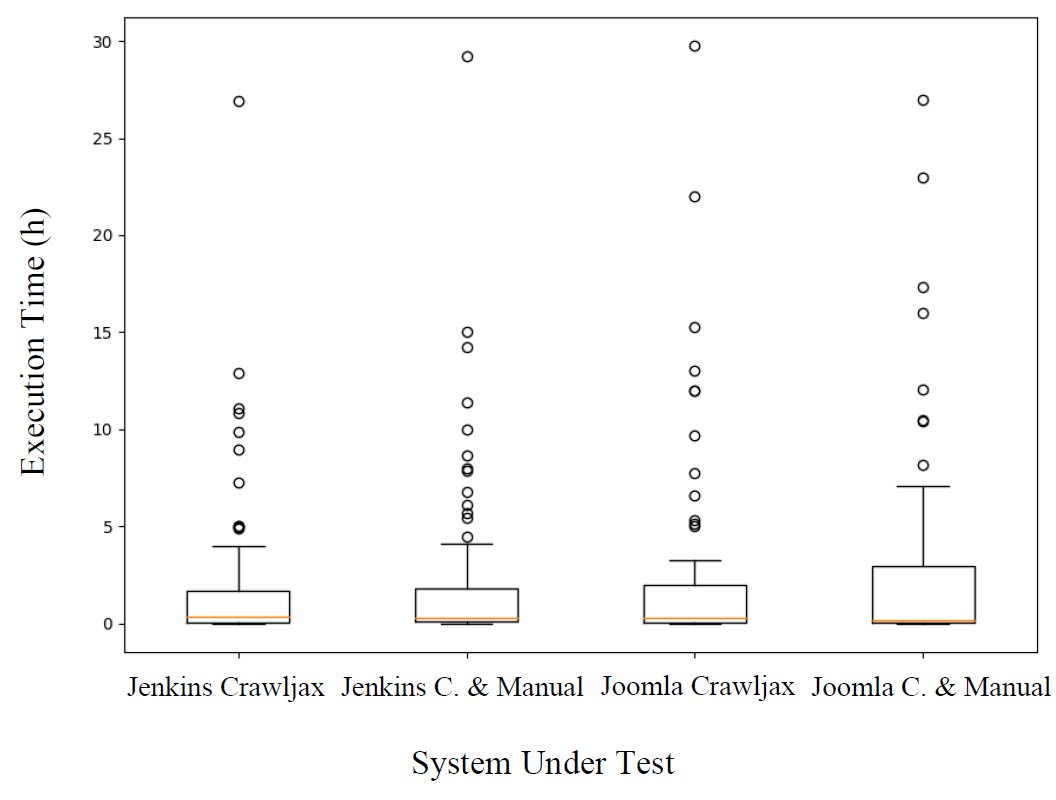}
 \caption{RQ5. Box-plot reporting execution time  in hours (values in Table~\ref{tab:executionTime}). Some outliers have been removed to ease visualization.} 
\label{fig:executionTime}
\end{center}
\end{figure}

\input{tables/tableExecutionTime}

The execution time of crawling depends on the number of user roles to be tested and the number of actions (i.e., inputs that can be provided through links and Web forms on different Web pages) for the SUT. 
However, in our context, the number of user roles has a limited impact because crawling can be parallelized.
Since the development of an efficient crawler is out of the scope of this paper, we did not perform an empirical evaluation of the \TSE{1.3}{efficiency} of our crawler (i.e., the extended Crawljax). The interested reader is referred to the original Crawljax publication for further discussions~\cite{crawljax:tweb12}. In this RQ, we discuss the time required to execute the \MRs in our catalog.




We aimed to determine if \MST is efficient enough to be used in practice and what are the main factors that influence \MRs' execution time. To do so, we kept track of the time required to execute each \MR considered for RQ4 and discussed the execution time distribution across \MRs.

By design, the time required to execute an \MR is driven by the number of source inputs and follow-up inputs executed, the number of actions belonging to source and follow-up inputs, and the complexity of the \MR executed (e.g., the number of instructions in the \MR and computational complexity of the functions invoked by the \MR). Finally, the hardware components used to run the Web server can have different response times and, consequently, affect the execution time of the \MR. 

Based on the above, we studied the correlation between execution time and the number of source inputs, follow-up inputs, and actions in follow-up inputs by relying on the non-parametric Spearman's 
correlation coefficient.
As for hardware, we did not study the effect of different hardware configurations on \MST \TSE{1.3}{efficiency} but executed our experiments in scenarios that we consider feasible for testing software with \MST:
a virtual machine  installed on professional desktop PCs (Dell G7 7500, RAM 16Gb, Intel(R) Core(TM) i9-10885H CPU @ 2.40GHz) and terminal access to a shared remote server with Intel(R) Xeon(R) Gold 6234 CPU (3.30GHz) and 8 CPU cores.


\subsubsection{Results}

Fig.~\ref{fig:executionTime} provides boxplots reporting the execution time distribution, in hours, for the \MRs executed to address RQ4. The highest third quartile is 3.03 hours and indicates that 75\% of the \MRs can be executed for both systems overnight or in half a working day (i.e., less than four hours) if we parallelize the executions of the \MRs (e.g., by running each on a dedicated virtual environment). Further, when excluding outliers, the maximum execution time is about seven hours and a half for Joomla tested with manual and Crawljax-based inputs (see \emph{Upper whisker} in Table~\ref{tab:executionTime}), which indicates that most \MRs can be executed overnight. Considering that security testing is currently performed manually and is error-prone, we believe that the cost of setting up parallelized execution environments is justified.


From Table~\ref{tab:executionTime}, we can see that, across configurations (i.e., executing \MST with inputs derived by relying on Crawljax only or by combining Crawljax and manual inputs), between four and six \MRs  could not be executed overnight because they took more than 14 hours. 
These \MRs look for injection, authorization, and authentication problems by relying on a catalog of crafted values passed to URL parameters or form inputs. In practice, they test the same URL multiple times with different users and inputs, leading to a combinatorial explosion.
However, in our framework, it is possible to parallelize the execution of a single \MR across multiple nodes (e.g., virtual machines running on a Cloud or grid infrastructure); each node iterates over a subset of source inputs. With ten execution nodes for each of these six \MRs, it should be feasible to execute them overnight. For example, we parallelized the execution of the \MR leading to the worst-case execution time (i.e., \texttt{CWE\_138...\_OTG\_AUTHZ\_001b} in Jenkins) and observed a maximum execution time of 14 hours, thus confirming  our assumption. 


\input{tables/tableSpearmanCorrelation}

Table~\ref{tab:spearmanCorr} reports, for our experiments, the 
Spearman's coefficients capturing the correlation between execution time and the number of  (a) source inputs, (b) follow-up inputs, and (c) actions in follow-up inputs.
We observe that, as expected, the execution time significantly correlates (i.e., coefficient above 0.5 and p-value below 0.05) with the number of actions in follow-up inputs; indeed, each action leads to a request execution in the browser.
We also note a correlation with the number of follow-up inputs, \TSE{3.15.14}{which is lower in the case of Jenkins.} Such low correlation for Jenkins is observed when testing with source inputs derived from both Crawljax and manual test cases. Indeed, manual test cases include a smaller set of actions than the ones collected by the crawler, \TSE{3.15.15}{but they test URLs that are also covered by the source inputs derived by the crawler.} Consequently, since several \MRs do not test the same URL twice, \MST, when data collection is based on Crawljax and manual test cases, tends to generate follow-up inputs that are shorter and, therefore, quicker to execute (the long follow-up inputs derived from the Crawljax source inputs are not executed because they include URLs already tested by relying on the short manual source inputs). We do not observe the same trend in Joomla, likely because the crawled source inputs contain fewer actions (the average number of actions for each source input is 5.9 for Joomla and 6.7 for Jenkins).
Finally, the correlation between the execution time and the number of source inputs is more limited for Jenkins and not significant for Joomla. This limited correlation is mainly due to source inputs not leading to follow-up inputs (e.g., because a precondition does not hold) and, consequently, causing execution times that vary a lot across \MRs even if the number of source inputs processed by these \MRs is the same.
\TSE{3.7}{Concluding, efficiency does not depend on the number of input interfaces but on the complexity of the features of the Web-system under test. The former drives the number of source inputs generated (low correlation with execution time), and the latter drives the length of the source and follow-up inputs (higher correlation with execution time). We selected well-known case study subjects representative of a broad set of systems; engineers testing systems not similar to Jenkins and Joomla may observe different efficiency results.}

%% file: tables/tableExecutionTime.tex
\begin{table}[]
\centering
\scriptsize
\caption{RQ5: Distribution of execution time (hours)}
\label{tab:executionTime}
\begin{tabular}{|p{1.7cm}|p{1cm}|p{1.5cm}|p{1cm}|p{1.5cm}|}
\hline
&
\multicolumn{2}{c|}{Jenkins} &
  \multicolumn{2}{c|}{Joomla} \\ 
  \cline{2-5}
&
  Crawljax &
  C. \& Manual &
  Crawljax &
  C. \& Manual\\ 
 \hline
First Quarltile  & 0.06       & 0.11        & 0.05      & 0.084      \\ \hline
Second Quarltile & 0.38       & 0.34        & 0.28       & 0.22       \\ \hline
Third Quarltile  & 2.63       & 2.38        & 2.01       & 3.03        \\ \hline
Lower whisker    & 0.017      & 0.017       & 0.017     & 0.017      \\ \hline
Upper whisker    & 6.48       & 5.80        & 4.95     & 7.44       \\ \hline
\# of \MRs taking more than 14 hours    & 6       & 5       & 5     & 3       \\ \hline
\end{tabular}%


\end{table}

%% file: tables/tableSpearmanCorrelation.tex
\begin{table*}[]
\centering
\caption{RQ5: Spearman correlation}
\label{tab:spearmanCorr}
\resizebox{\textwidth}{!}{%
\begin{tabular}{|l|cc|cc|cc|cc|}
\hline
\multicolumn{1}{|c|}{\multirow{3}{*}{Set of features\textbackslash System under test}} & 
\multicolumn{4}{c|}{Jenkins} &
  \multicolumn{4}{c|}{Joomla} \\ \cline{2-9}
&
  \multicolumn{2}{c|}{Crawljax} &
  \multicolumn{2}{c|}{Crawljax \& Manual} &
  \multicolumn{2}{c|}{Crawljax} &
  \multicolumn{2}{c|}{Crawljax \& Manual} \\ \cline{2-9} 
\multicolumn{1}{|c|}{} &
  \multicolumn{1}{l|}{Spearman} &
  P-value &
  \multicolumn{1}{l|}{Spearman} &
  P-value &
  \multicolumn{1}{l|}{Spearman} &
  P-value &
  \multicolumn{1}{l|}{Spearman} &
  P-value \\ \hline

Execution time and number of source inputs &
  \multicolumn{1}{c|}{0.39} &
  0.00076 &
  \multicolumn{1}{c|}{0.38} &
  0.00071 &
  \multicolumn{1}{c|}{0.16} &
  0.20 &
  \multicolumn{1}{c|}{0.17} &
  0.18 \\ \hline

Execution time and number of follow-up inputs &
  \multicolumn{1}{c|}{0.52} &
  1.91e-06 &
  \multicolumn{1}{c|}{0.36} &
  0.0017 &
  \multicolumn{1}{c|}{0.56} &
  4.15e-07 &
  \multicolumn{1}{c|}{0.55} &
  7.15e-07 \\ \hline

Execution time and number of actions in follow-up inputs &
  \multicolumn{1}{c|}{0.51} &
  4.19e-06 &
  \multicolumn{1}{c|}{0.43} &
  0.00013 &
  \multicolumn{1}{c|}{0.52} &
  3.41e-06 &
  \multicolumn{1}{c|}{0.52} &
  5.72e-06 \\ \hline
\end{tabular}%
}
\end{table*}

%% file: validity.tex
\section{Threats to Validity}
\label{sec:empirical:validity}


\subsubsection{Internal validity}

\TSE{2.1, 3.5}{Two co-authors inspected all the results to identify applicable SOTA oracle automation strategies for security testing (RQ1), 
to determine what \MR can address a vulnerability type (RQ2), and to analyze the causes undermining the applicability (RQ2) and testability factors enabling \MST (RQ3). For RQ1, one researcher proposed oracle types always confirmed by the second researcher. For RQ2, one researcher provided an \MR or indicated the reason for not being able to apply \MST. The second researcher verified the implemented \MR and proposed alternative implementations (when necessary). 
Regarding the reasons for not applying \MST, the second researcher provided alternatives when in disagreement. 
By considering the category assigned by each researcher to each weakness considered for RQ2, we computed the Cohen's Kappa coefficient to measure inter-rater agreement, leading to a Kappa equal to $0.707$, with a 95\% confidence interval between $0.611$ and $0.804$, which indicates \emph{substantial agreement}.
Disagreement is impossible in RQ3 since the identification of testability factors is based on the utility functions used by the \MR. Once the first researcher assigned testability factors to each \MR, the second researcher verified the correctness of the outcome.}

To minimize \emph{implementation errors} in RQ4 and RQ5, we carefully inspected and tested the \MST toolset before running our experiments. Also, we executed our \MRs on DVWA~\cite{DVWA}, a security benchmark, thus ensuring that our \MRs can discover some of the targeted vulnerabilities (injection vulnerabilities of different kinds, in this case).


\subsubsection{Conclusion validity}

\TSE{3.5}{For RQ1, RQ2, and RQ3, we reported the proportion of items in a given catalog that belong to a certain class: types of oracles (RQ1), feasible \MRs (RQ2), inapplicability causes and testability features (RQ3). For RQ4, we discussed specificity and sensitivity. Therefore, statistical tests were not required for RQ1, RQ2, RQ3, and RQ4.} 

For RQ5, the underlying distribution of the data (i.e., execution time and the number of executed source inputs, follow-up inputs, and actions in follow-up inputs) is not known in our context. Therefore, we relied on Spearman's rank correlation, which is non-parametric, to avoid violating the assumptions of parametric tests for correlation analysis.

For RQ4 and RQ5, the sources of \emph{randomness affecting results} might be (i) the workload of the machines used to run the experiments (slowing down the performance of both \MST and the case study subjects) and (ii) the presence of other users interacting with the software under test (affecting both execution time and system outputs). To mitigate these effects, we run the experiments in dedicated environments with the study subjects used only by our framework.

\subsubsection{Construct validity}

We discuss \emph{construct validity} in terms of 
face, content, convergent, and predictive validity~\cite{Ralph2018}.

For RQ4 and RQ5, the constructs we considered in our work are effectiveness and efficiency. Effectiveness is measured through two reflective indicators, which are sensitivity and specificity. Efficiency is measured in terms of execution time.
Concerning \emph{face validity}, we believe our indicators are appropriate. Indeed, sensitivity is the ratio of vulnerabilities discovered and corresponds to the fault detection rate, commonly used in software testing papers to evaluate testing effectiveness.
Specificity captures the proportion of follow-up inputs not leading to failures (i.e., a large majority for stable systems like Jenkins and Joomla) that, correctly, do not trigger any failure report in \MST. 
Specificity captures developers' time wasted on investigating false alarms. Indeed, this is proportional to the number of follow-up inputs checked in vain.
Execution time is a direct measure that enables us to assess if, for systems similar to our case study subjects, \MST is efficient enough \TSE{3.15.16}{to be integrated into software development.}



\emph{Content validity} concerns the breadth of the construct. High sensitivity (i.e., fault detection rate) is a condition for a software testing technique to be useful. However, it does not imply that engineers can understand the root cause of a failure. Successful root cause analysis depends on other factors, such as the software engineer's experience and knowledge about the software under test and the availability of appropriate debugging and logging tools, which are out of the scope of this paper.
The false positive rate, a measure of the effect of false alarms, is equal to \emph{1 - specificity} and is complementary to sensitivity; therefore, we believe our choice of measurements to be complete. 
To discuss efficiency, 
we rely on the number of source inputs, follow-up inputs, and actions in follow-up inputs; they capture the complexity of the SUT since they reflect the actions that might be performed by an attacker and are automatically derived through a crawler exercising the system under test. 
Other metrics for the size of the SUT (e.g., lines of code, number of Web pages) may not capture its complexity (e.g., knowing the number of Web pages is insufficient to understand how many form inputs might be submitted to the SUT). Therefore, we believe that studying the correlation between our selected complexity metrics and execution time is adequate to discuss the efficiency of the approach.


About \emph{convergence}, the number of source inputs is weakly correlated to the number of follow-up inputs 
(i.e., $0.14 \le \text{Spearman's} \rho \ge 0.23$ for all the subjects). 
The source inputs filtered by \MRs' preconditions may explain this weak correlation. The number of follow-up inputs is strongly correlated with the number of actions in follow-up inputs 
(i.e., Spearman's $\rho \ge 0.85$ for all the subjects), 
which is expected since we derive follow-up inputs by traversing the crawler graph with DFS. 

As for \emph{predictive validity}, we reported statistics for RQ5 based on a non-parametric correlation coefficient (Spearman’s) because the collected data (i.e., number of  follow-up inputs, number of actions in follow-up inputs, and execution time) does not appear to be normally distributed.


\subsubsection{External validity}

\TSE{3.5}{In RQ1 and RQ2, we considered specific catalogs (the \emph{OWASP testing book}, the \emph{CWE view for common security architectural tactics}, the CWE view for \emph{OWASP Top 10}, and the CWE view for \emph{CWE Top 25}) that include a set of vulnerability types considered broad and complete by security researchers~\cite{Elder2022}. In RQ3, we derived our results from the characteristics of the implemented \MRs (i.e., what enables their execution). 
Since our \MRs are not system-specific, our testability features apply to any Web system.}

To strengthen the generalizability of our conclusions in RQ4, we selected systems that are representative of modern Web systems but different in terms of technical and process aspects. For RQ5, we executed our experiments by relying on hardware commonly available to Web and security engineers (i.e., laptops, virtual machines, and web servers).
\TSE{3.6}{Contrary to our initial \MST study~\cite{Mai2020a}, we did not analyze the Web system developed in the context of the EDLAH2~\cite{EDLAH} project because the project ended and our license to access the system expired. However, the interfaces exposed by Joomla are more complex than the ones of the EDLAH2 system (e.g., a larger set of accessible Web pages).}

%% file: related.tex
\section{Related Work}
\label{sec:related}

This section covers the related work across three categories: (i) \textit{software security testing}, (ii) \textit{model-based security testing}, and (iii) \textit{metamorphic testing}. In the first category, we present existing security testing strategies and discuss how \MST complements them. Model-based security testing is one of the standard security testing methods~\cite{Felderer-MBST-2016,TianYang-ResearchOnSoftwareSecurityTesting-2010}. Although it is a relatively new research field, many model-based approaches have been published lately, and we, therefore, present and compare them with \MST. Finally, in the last category, we discuss \MT techniques since they represent the foundation on which \MST has been defined.   

\subsection{Software Security Testing} 
Security testing approaches can be categorized~\cite{Felderer2016} as follows:
(i) security functional testing validating whether the specified security properties are implemented correctly and 
(ii) security vulnerability testing mimicking attacks that target typical system vulnerabilities. \MST can be applied to both security functional and vulnerability testing since \MRs can capture security properties (e.g., a login screen should always be shown after a session timeout) and the properties of the inputs and outputs involved in discovering a vulnerability (e.g., the output generated when requesting an admin resource without authentication should differ from the one obtained with authentication).

Many security vulnerability testing approaches rely on an implicit test oracle, i.e., one that relies on tacit knowledge to distinguish between correct and incorrect system behavior~\cite{Haley2008}. 
It is the case for approaches targeting buffer overflows, memory leaks, unhandled exceptions, and denial of service~\cite{Ognawala-SymbExecutionLowLevelVuln-ASE-2016,Bekrar2011,Takanen2018fuzzing}, which mostly rely on mutational fuzzing~\cite{fuzzingbook2019}, i.e., the generation of new inputs through the random modification of existing ones.  
For instance, Bekrar et al.~\cite{Bekrar2011} present a technique combining mutational fuzzing with data tainting and coverage analysis to identify security vulnerabilities in file processors and network protocols.
Implicit oracles deal with simple abnormal system behavior, such as unexpected system termination, and are system-agnostic. 
The literature, however, lacks explicit oracles for vulnerabilities leading to invalid outputs (e.g., providing data to a user who is not supposed to access it); indeed, such outputs are system-specific and difficult to capture with the mechanisms used for implicit oracles (e.g., runtime exceptions).

\input{tables/tableRelatedWork.tex}

Vulnerability testing approaches for code injections also suffer from the oracle problem~\cite{Raghavan2000,Kals2006,Martin2008,Bau2010,Appelt-SLQmutation-ISSTA-2014,Salas2014,Tripp-XSS-ISSTA-2013,Appelt2013}. 
For instance, test cases targeting SQL injections are in the form of HTTP requests that trigger responses from a Web application while a crawler receives responses in which some predefined keywords (e.g., ``invalid'') are searched~\cite{Raghavan2000}. When no keywords are detected, the crawler cannot determine whether the injection is filtered by the system under test. 
To resolve this problem, Huang et al.~\cite{Huang2003web} proposed an MT-like technique that sends multiple HTTP requests, i.e., one request with an injection, an intentionally invalid request, and a valid request. They compare the responses to determine if the request with the injection is filtered. 
Unfortunately, MT-like approaches that address a broader set of security vulnerabilities are missing.
In contrast, \MST targets a wide variety of security vulnerabilities (including code injection vulnerabilities); a detailed analysis of security vulnerabilities that can and cannot be addressed by our approach was provided in Section~\ref{sec:usage}.

\subsection{Model-based Security Testing} 
Model-based approaches~\cite{Felderer-MBST-2016,Felderer2011} mostly target security vulnerability testing (e.g., References~\cite{Bertolino-PolPa-AST-2012,blome2013vera,he2008attack,Marback2013,jurjens2002umlsec,jurjens2005sound,jurjens2005secure,Jurjens2008,xu2006threat,Martin-PolicyCoverage-ICICS-2006,Wimmel2002,Whittle2008,xu2012automated,xu2015automated}), whereas some solutions address security functional testing (e.g., References~\cite{Le2007,Mouelhi2008,Mouelhi2009,Xu2012,Masood2010,Martin-PoliciesMutationTestingFramework-WWW-2007,Martin-TestingPoliciesChangeImpact-IWSESS-2007}). 
Most of them only generate test sequences from security models and do not address the oracle problem. For instance, Marback et al.~\cite{Marback2013} propose a model-based security testing approach that automatically generates security test sequences from threat trees. 

Model-based approaches that generate test cases with oracles~\cite{xu2012automated,xu2015automated,Xu2012} 
rely on mappings between model-level abstractions (i.e., tokens in markings of PrT networks) and executable code implementing the oracle logic (e.g., searching for error messages in system output). 
For instance, Xu et al.~\cite{xu2012automated,xu2015automated} automatically generate executable vulnerability test cases from formal threat models. Further, model-level test oracles (tokens in markings of attack paths) are directly mapped to implementation-level code.
But such mapping is not feasible when it is difficult to specify precise test oracles and automatically compare expected values to the actual results. 
The same problem also affects another approach that targets access control policies~\cite{Xu2012}.
Overall, these approaches do not free engineers from significant implementation effort since they require the manual implementation of the executable oracle code. 
\subsection{Metamorphic Testing} 
With \MT, we aim to address the limitations of security testing approaches described above. Indeed, \MT supports oracle automation thanks to \MRs 
that can precisely capture the relations between test inputs and outputs.
In practice, \MT frees engineers from implementing a specific oracle (e.g., test case assertions) for each test case.
Considerable research has been devoted to developing \MT approaches for various domains such as computer graphics (e.g.,~\cite{Mayer2006,Guderlei2007,Just2009,Kuo2011}), simulation (e.g.,~\cite{Chen2009,Ding2011,Murphy2011}), Web services (e.g.,~\cite{Chan2007b,Sun2011,Zhou2012}), embedded systems (e.g.,~\cite{Tse2004,Chan2007,Kuo2011b,Jiang2013}), compilers (e.g.,~\cite{Tao2010,Le2014}), variability and decision support (e.g.,\cite{segura2011automated,segura2010automated,segura2015automated,kuo2010metamorphic}), bioinformatics (e.g.,~\cite{chen2009innovative,pullum2012early,ramanathan2012verification}), numerical programs (e.g.,~\cite{chen2002metamorphic,aruna2014metamorphic}), and machine learning (e.g.,~\cite{Xie2009,Murphy2008}). However, very little attention has been paid to its application in security testing~\cite{Segura2016}.

Preliminary applications of \MT to security testing focus on the functional testing of security components (i.e., testing encryption programs in the absence of oracles~\cite{sun2014property}, verifying the output of code obfuscators and the rendering of login interfaces~\cite{ChenMTSecurity2016}), and the verification of low-level properties broken by specific security bugs 
(e.g., the heartbleed bug~\cite{Heartbleed} which affects the relation between the size of the payload data field of an SSL message and the length declared in the same message). 
Although these works demonstrate the feasibility of \MT for security testing, they focus on a narrow set of vulnerabilities and do not automate the generation of executable metamorphic test cases, which are manually implemented based on the identified \MRs. In contrast, \MST supports the specification of \MRs for many vulnerabilities and automates the generation of executable metamorphic test cases from the \MRs.

Although \MT is highly automatable, \MT research has mostly focused on the application of \MT to specific testing problems~\cite{Segura2016}. For instance, Kuo et al.~\cite{Kuo2011b} report on a case study using metamorphic testing to detect faults in a wireless metering system. Ding et al.~\cite{Ding2011} present a case study for fault detection in a Monte Carlo modeling program simulating photon propagation. Chen et al.~\cite{chen2002metamorphic} focus on applying \MT to programs implementing partial differential equations. These works do not provide any systematic method to specify \MRs and proper tool support. In general, few approaches provide tool support enabling engineers to write system-level \MRs~\cite{Segura2016}. 
Those which require that \MRs be defined either as Java methods~\cite{ToolZhu} or method pre-/post-conditions~\cite{MurphyJML}  limit the adoption of \MT to verify system-level security properties; indeed, they target single methods and not the output provided by the system as a whole.
Since \MRs often employ a declarative notation,  
engineers need significant, additional effort to translate abstract, declarative \MRs into an imperative programming language. To avoid such overhead, we propose a DSL as part of \MST (see Section~\ref{sec:dsl}). Segura et al.~\cite{segura2017template,segura2017metamorphictemplate} provide a template-based approach for describing \MRs. The proposed template aims to ease communication among practitioners but does not support security-related language constructs.
Further, there is no automated support to transform template-based \MRs into executable test cases.

The notion of \emph{metamorphic property} (e.g., the \textit{permutative} property that specifies that the order of inputs should not affect the output) introduced by Murphy et al.~\cite{murphy2008properties} forms the basis for \emph{general metamorphic relations}, which are analogous to metamorphic relation  patterns. Zhou et al.~\cite{zhou2018metamorphic} define the notion of metamorphic relation pattern (MRP) as \textit{an abstraction that characterizes a set of (possibly infinitely many) metamorphic relations}. Subclasses of MRPs are proposed in the literature: \textit{metamorphic relation input pattern (MRIP)}~\cite{zhou2018metamorphic} and \textit{metamorphic relation output pattern (MROP)}~\cite{segura2017metamorphic}. An MRIP is an abstraction characterizing the relations among the source and follow-up inputs of a set of metamorphic relations; an MROP describes an abstract relation among the source and follow-up outputs. 

Segura et al.~\cite{segura2017metamorphic} propose six MROPs (equivalence, equality, subset, disjoint, complete, and difference) for testing RESTful web APIs (implementing create, read, update, or delete operations over a resource). 
In our \MR catalog, we leverage some of these patterns to define output conditions; in particular, our \MRs verify equality, difference, and subset (i.e., what we achieve with \TEXTTT{userCanRtrieveContent}, which checks if the output is a subset of what was already observed in previous executions).
Segura et al.~\cite{segura2019metamorphic} also define a catalog of MRPs for query-based systems, which focus on the properties of inputs being conditions (e.g., the keywords used when executing a search engine, which can be joined) and outputs being data sets (e.g., the results returned by the search engine, which can be disjoint, subsets, or shuffled).
Zhou et al.~\cite{zhou2018metamorphic} propose a \textit{symmetry} MRP and a \textit{change direction} MRIP to test the system from "different viewpoints", e.g., checking if an object recognition system recognizes the same object regardless of whether it is played forward or backward (changing direction). 
The works described above assume that source inputs are either single items or item sets, which simplifies the definition of patterns capturing mathematical properties (e.g., symmetry or equality). 
In our work, we focus instead on source inputs and outputs that are action sequences and corresponding output sequences; action sequences are necessary to describe interactions with complex systems. Consequently, our patterns capture the different operations to be performed on these input/output sequences. The patterns provided in the literature are part of our \MRs, but they capture only output conditions, as described above.


\subsection{Summary}
In Table~\ref{table:RelatedWork}, based on a set of features necessary for security testing, we summarize the differences between \MST and related work.
For each approach, the symbol '+' indicates that the approach provides the feature, '-' indicates that it does not,
and 'NA' indicates that the feature is not applicable because it can be implemented only by approaches in other categories (i.e., Metamorphic Testing, Model-based Security Testing, or other Software Security Testing approaches).
For instance, Ognawala et al.~\cite{Ognawala-SymbExecutionLowLevelVuln-ASE-2016} employ symbolic execution to detect memory out-of-bounds/buffer overflow vulnerabilities caused by unhandled memory operations. Therefore, none of the features related to \MT, such as the support for specifying \MRs, are considered for Ognawala et al.~\cite{Ognawala-SymbExecutionLowLevelVuln-ASE-2016}, as depicted in Table~\ref{table:RelatedWork}.
Most automated security testing approaches do not address the oracle problem~\cite{Raghavan2000,Kals2006,Martin2008,Bau2010,Appelt-SLQmutation-ISSTA-2014,Salas2014,Tripp-XSS-ISSTA-2013,Appelt2013} or rely on an implicit test oracle~\cite{Ognawala-SymbExecutionLowLevelVuln-ASE-2016,Bekrar2011,Takanen2018fuzzing}. The few approaches that do address the oracle problem focus on a limited number of security vulnerabilities~\cite{Huang2003web}. Also, most model-based security testing approaches do not address the oracle problem since they only generate test sequences from security models~\cite{Marback2013,Wimmel2002,Lebeau-ICST-2013,Whittle2008}. Some model-based approaches derive test cases with oracles~\cite{xu2012automated,xu2015automated,Xu2012} but require mappings between model-level abstractions and executable code implementing the oracle logic.  
\MT can overcome these limitations,
but existing \MT solutions target a few specific security vulnerabilities and do not support automated testing based on \MRs capturing general security properties. To overcome these limitations, we need a dedicated DSL and algorithms that automate the execution of \MT. To the best of our knowledge, \MST is the only approach that supports, with a DSL, the specification of \MRs capturing a wide range of security properties, automates the generation of executable metamorphic test cases from the \MRs, and automatically detects various vulnerabilities based on those relations.

%
%

%% file: tables/tableRelatedWork.tex
\begin{table*}[tb]
\scriptsize
\caption{Summary and comparison of related work.}
\label{table:RelatedWork}
\begin{tabular}{
|@{\hspace{0.05cm}}p{0.4cm}
|@{\hspace{0.02cm}}p{2.85cm}
|@{\hspace{0.05cm}}p{1.35cm} 
|@{\hspace{0.05cm}}p{1.80cm} 
|@{\hspace{0.05cm}}p{2.10cm} 
|@{\hspace{0.05cm}}p{2.20cm} 
|@{\hspace{0.05cm}}p{2.00cm}
|@{\hspace{0.05cm}}p{1.60cm}
|@{\hspace{0.05cm}}p{1.40cm}|}
\hline

& &\textbf{Support for various vulnerabilities}&\textbf{No need for implicit oracles}&\textbf{Model-based test generation including oracle}&\textbf{No need for model-based mappings or manual oracle implementation}&\textbf{Application of MT to security testing}
&\textbf{DSL support to specify \MRs}&\textbf{Publicly Available Tool support for \MT}\\
\hline

& \textbf{\MST}& $+$ & $+$ & $\mathit{NA}$ & $+$ & $+$ & $+$ & $+$\\
\hline

\multirow{11}{*}{\rotatebox{90}{\textbf{Software Security}} \rotatebox{90}{\textbf{\,\,\,\,\,\,\,\,\,\,\,\, Testing}}} & Ognawala et al.~\cite{Ognawala-SymbExecutionLowLevelVuln-ASE-2016}& $-$ & $-$ & $\mathit{NA}$ & $\mathit{NA}$ & $\mathit{NA}$ & $\mathit{NA}$ & $\mathit{NA}$ \\
\cline{2-9}

& Bekrar et al~\cite{Bekrar2011}& $+$ & $-$ & $\mathit{NA}$ & $\mathit{NA}$ & $\mathit{NA}$ & $\mathit{NA}$ & $\mathit{NA}$\\
\cline{2-9}

& Takanen et al.~\cite{Takanen2018fuzzing}& $+$ & $-$ & $\mathit{NA}$ & $\mathit{NA}$ & $\mathit{NA}$ & $\mathit{NA}$ &$\mathit{NA}$ \\
\cline{2-9}

& Kals et al.~\cite{Kals2006}& &  & $\mathit{NA}$ & $\mathit{NA}$ & $\mathit{NA}$ & $\mathit{NA}$ & $\mathit{NA}$\\
\cline{2-9}

& Martin et al.~\cite{Martin2008}& $-$ & $+$ & $\mathit{NA}$ & $\mathit{NA}$ & $\mathit{NA}$ & $\mathit{NA}$ & $\mathit{NA}$\\
\cline{2-9}

& Bau et al.~\cite{Bau2010}& $-$ & $+$ & $\mathit{NA}$ & $\mathit{NA}$ & $\mathit{NA}$ & $\mathit{NA}$ & $\mathit{NA}$\\
\cline{2-9}

& Appelt et al.~\cite{Appelt-SLQmutation-ISSTA-2014}& $-$ & $+$ & $\mathit{NA}$ & $\mathit{NA}$ & $\mathit{NA}$ & $\mathit{NA}$ & $\mathit{NA}$\\
\cline{2-9}

& Salas et al.~\cite{Salas2014}& $-$ & $+$ & $\mathit{NA}$ & $\mathit{NA}$ & $\mathit{NA}$ & $\mathit{NA}$ & $\mathit{NA}$\\
\cline{2-9}

& Tripp et al.~\cite{Tripp-XSS-ISSTA-2013}& $-$ & $+$ & $\mathit{NA}$ & $\mathit{NA}$ & $\mathit{NA}$ & $\mathit{NA}$ &$\mathit{NA}$ \\
\cline{2-9}

& Appelt et al.~\cite{Appelt2013}& $-$ & $+$ & $\mathit{NA}$ & $\mathit{NA}$ & $\mathit{NA}$ & $\mathit{NA}$ & $\mathit{NA}$\\
\cline{2-9}

& Huang et al.~\cite{Huang2003web}& $-$ & $+$ & $\mathit{NA}$ & $\mathit{NA}$ & $\mathit{NA}$ & $\mathit{NA}$ &$\mathit{NA}$ \\
\hline

\multirow{16}{*}{\rotatebox{90}{\textbf{\,\,\,\,Model-based}} \rotatebox{90}{\textbf{Security Testing}}} & Le Traon et al.~\cite{Le2007}& $\mathit{NA}$ & $+$ & $-$ & $-$ & $\mathit{NA}$ & $\mathit{NA}$ & $\mathit{NA}$ \\
\cline{2-9}

& Mouelhi et al.~\cite{Mouelhi2008,Mouelhi2009}&  $\mathit{NA}$ & $+$ & $-$ & $-$ & $\mathit{NA}$ & $\mathit{NA}$ & $\mathit{NA}$ \\
\cline{2-9}

& Martin et al.~\cite{Martin-TestingPoliciesChangeImpact-IWSESS-2007}& $\mathit{NA}$ & $+$ & $-$ & $-$ & $\mathit{NA}$ & $\mathit{NA}$ & $\mathit{NA}$ \\
\cline{2-9}

& Martin et al.~\cite{Martin-PoliciesMutationTestingFramework-WWW-2007}& $\mathit{NA}$ & $+$ & $-$ & $-$ & $\mathit{NA}$ & $\mathit{NA}$ & $\mathit{NA}$ \\
\cline{2-9}

& Wimmel and J{\"u}rjens~\cite{Wimmel2002}& $\mathit{NA}$ & $+$ & $-$ & $-$ & $\mathit{NA}$ & $\mathit{NA}$ & $\mathit{NA}$ \\
\cline{2-9}

& Masood et al.~\cite{Masood2010}& $\mathit{NA}$ & $+$ & $-$ & $-$ & $\mathit{NA}$ & $\mathit{NA}$ &$\mathit{NA}$ \\
\cline{2-9}

& Bertolino et al.~\cite{Bertolino-PolPa-AST-2012}& $+$ & $+$ & $-$ & $-$ & $\mathit{NA}$ & $\mathit{NA}$ & $\mathit{NA}$ \\
\cline{2-9}

& Blome et al.~\cite{blome2013vera}& $+$ & $-$ & $-$ & $+$ & $\mathit{NA}$ & $\mathit{NA}$ & $\mathit{NA}$ \\
\cline{2-9}

& He et al.~\cite{he2008attack}& $+$ & $+$ & $-$ & $-$ & $\mathit{NA}$ & $\mathit{NA}$ & $\mathit{NA}$\\
\cline{2-9}

& Marback et al.~\cite{Marback2013}& $+$ & $+$ & $-$ & $-$ & $\mathit{NA}$ & $\mathit{NA}$ & $\mathit{NA}$\\
\cline{2-9}

& J{\"u}rjens et al.~\cite{Jurjens2008}& $+$ & $+$ & $-$ & $-$ & $\mathit{NA}$ & $\mathit{NA}$ & $\mathit{NA}$ \\
\cline{2-9}

& Xu et al.~\cite{xu2006threat}& $+$ & $+$ & $-$ & $-$ & $\mathit{NA}$ & $\mathit{NA}$ &$\mathit{NA}$ \\
\cline{2-9}


& Lebeau et al.~\cite{Lebeau-ICST-2013}& $+$ & $+$ & $-$ & $-$ & $\mathit{NA}$ & $\mathit{NA}$ & $\mathit{NA}$\\
\cline{2-9}

& Whittle et al.~\cite{Whittle2008}& $+$ & $+$ & $-$ & $-$ & $\mathit{NA}$ & $\mathit{NA}$ &$\mathit{NA}$ \\
\cline{2-9}

& Xu et al.~\cite{xu2012automated,xu2015automated}& $+$ & $+$ & $+$ & $-$ & $\mathit{NA}$ & $\mathit{NA}$ & $\mathit{NA}$ \\
\cline{2-9}

& Xu et al.~\cite{Xu2012}& $\mathit{NA}$ & $+$ & $+$ & $-$ & $\mathit{NA}$ & $\mathit{NA}$ &$\mathit{NA}$ \\
\hline

\multirow{32}{*}{\rotatebox{90}{\textbf{\,\,\,Metamorphic}} \rotatebox{90}{\textbf{\,\,\,\,\,\,\,\,\,Testing}}} & Mayer and Guderlei~\cite{Mayer2006}& $\mathit{NA}$ & $+$ & $\mathit{NA}$ & $+$ & $-$ & $-$ & $-$ \\
\cline{2-9}

&  Just and Schweiggert~\cite{Just2009}& $\mathit{NA}$ & $+$ & $\mathit{NA}$ & $+$ & $-$ & $-$ & $-$ \\
\cline{2-9}

& Kuo et al.~\cite{Kuo2011}& $\mathit{NA}$ & $+$ & $\mathit{NA}$ & $+$ & $-$ & $-$ & $-$ \\
\cline{2-9}

& Jameel et al.~\cite{jameel2015test}& $\mathit{NA}$ & $+$ & $\mathit{NA}$ & $+$ & $-$ & $-$ & $-$ \\
\cline{2-9}

& Chan et al.~\cite{chan2007piping,chan2010finding}& $\mathit{NA}$ & $+$ & $\mathit{NA}$ & $+$ & $-$ & $-$ & $-$ \\
\cline{2-9}

& Chen et al.~\cite{Chen2009}& $\mathit{NA}$ & $+$ & $\mathit{NA}$ & $+$ & $-$ & $-$ & $-$ \\
\cline{2-9}

& Ding et al.~\cite{Ding2011}& $\mathit{NA}$ & $+$ & $\mathit{NA}$ & $+$ & $-$ & $-$ & $-$ \\
\cline{2-9}

& Murphy et al.~\cite{Murphy2011}& $\mathit{NA}$ & $+$ & $\mathit{NA}$ & $+$ & $-$ & $-$ & $-$ \\
\cline{2-9}

& Sim et al.~\cite{sim2005metamorphic}& $\mathit{NA}$ & $+$ & $\mathit{NA}$ & $+$ & $-$ & $-$ & $-$ \\
\cline{2-9}

& Chan et al.~\cite{Chan2007b}& $\mathit{NA}$ & $+$ & $\mathit{NA}$ & $+$ & $-$ & $-$ & $-$ \\
\cline{2-9}

& Sun et al.~\cite{Sun2011}& $\mathit{NA}$ & $+$ & $\mathit{NA}$ & $+$ & $-$ & $-$ & $+$ \\
\cline{2-9}


& Zhou et al.~\cite{Zhou2012}& $\mathit{NA}$ & $+$ & $\mathit{NA}$ & $+$ & $-$ & $-$ & $+$ \\
\cline{2-9}

& Tse et al.~\cite{Tse2004}& $\mathit{NA}$ & $+$ & $\mathit{NA}$ & $+$ & $-$ & $-$ & $-$ \\
\cline{2-9}

& Chan et al.~\cite{Chan2007}& $\mathit{NA}$ & $+$ & $\mathit{NA}$ & $+$ & $-$ & $-$ & $-$ \\
\cline{2-9}

& Kuo et al.~\cite{Kuo2011b}& $\mathit{NA}$ & $+$ & $\mathit{NA}$ & $+$ & $-$ & $-$ & $-$ \\
\cline{2-9}

& Jiang et al.~\cite{Jiang2013}& $\mathit{NA}$ & $+$ & $\mathit{NA}$ & $+$ & $-$ & $-$ & $-$ \\
\cline{2-9}

& Tao et al.~\cite{Tao2010}& $\mathit{NA}$ & $+$ & $\mathit{NA}$ & $+$ & $-$ & $-$ & $+$ \\
\cline{2-9}

& Yao et al.~\cite{yao2012research}& $\mathit{NA}$ & $+$ & $\mathit{NA}$ & $+$ & $-$ & $-$ & $-$ \\
\cline{2-9}


& Segura et al.~\cite{segura2011automated,segura2010automated}& $\mathit{NA}$ & $+$ & $\mathit{NA}$ & $+$ & $-$ & $-$ & $+$ \\
\cline{2-9}

& Kuo et al.~\cite{kuo2010metamorphic}& $\mathit{NA}$ & $+$ & $\mathit{NA}$ & $+$ & $-$ & $-$ & $-$ \\
\cline{2-9}

& Chen et al.~\cite{chen2009innovative}& $\mathit{NA}$ & $+$ & $\mathit{NA}$ & $+$ & $-$ & $-$ & $-$ \\
\cline{2-9}

& Pullum et al.~\cite{pullum2012early}& $\mathit{NA}$ & $+$ & $\mathit{NA}$ & $+$ & $-$ & $-$ & $-$ \\
\cline{2-9}

& Chen et al.~\cite{chen2002metamorphic}& $\mathit{NA}$ & $+$ & $\mathit{NA}$ & $+$ & $-$ & $-$ & $-$ \\
\cline{2-9}

& Aruna
and Prasad~\cite{aruna2014metamorphic}& $\mathit{NA}$ & $+$ & $\mathit{NA}$ & $+$ & $-$ & $-$ & $-$ \\
\cline{2-9}

& Xie et al.~\cite{Xie2009}& $\mathit{NA}$ & $+$ & $\mathit{NA}$ & $+$ & $-$ & $-$ & $-$ \\
\cline{2-9}

& Murphy et al.~\cite{Murphy2008}& $\mathit{NA}$ & $+$ & $\mathit{NA}$ & $+$ & $-$ & $-$ & $-$ \\
\cline{2-9}

& Segura et al.~\cite{segura2017metamorphic}& $\mathit{NA}$ & $+$ & $\mathit{NA}$ & $+$ & $-$ & $-$ & $+$ \\
\cline{2-9}

& Segura et al.~\cite{segura2017template,segura2017metamorphictemplate}& $\mathit{NA}$ & $+$ & $\mathit{NA}$ & $+$ & $-$ & $+$ & $-$ \\
\cline{2-9}

& Chen et al.~\cite{ChenMTSecurity2016}& $-$ & $+$ & $\mathit{NA}$ & $+$ & $+$ & $-$ & $-$ \\
\cline{2-9}

& Sun et al.~\cite{sun2014property}& $-$ & $+$ & $\mathit{NA}$ & $+$ & $+$ & $-$ & $-$ \\
\cline{2-9}


& Luu et al.~\cite{luu2021testing}& $\mathit{NA}$ & $+$ & $\mathit{NA}$ & $+$ & $-$ & $-$ & $-$ \\
\cline{2-9}

& Lascu et al.~\cite{lascumetamorphic}& $\mathit{NA}$ & $+$ & $\mathit{NA}$ & $+$ & $-$ & $-$ & $+$ \\
\cline{2-9}

& Ayerdi et al.~\cite{ayerdi2021generating}& $\mathit{NA}$ & $+$ & $\mathit{NA}$ & $+$ & $-$ & $-$ & $+$ \\
\cline{2-9}


& Xu et al.~\cite{xu2021using}& $\mathit{NA}$ & $+$ & $\mathit{NA}$ & $+$ & $-$ & $-$ & $-$ \\
\hline

\end{tabular}
\\
\\

\vspace{-6.7mm}
\end{table*}%

%% file: conclusion.tex
\section{Conclusion}
\label{sec:conclusion}

In this paper, we presented an approach, \MST, that enables engineers to specify metamorphic relations (\MRs) capturing security properties of Web systems, and that automatically detects security vulnerabilities based on those relations. Our approach aims to alleviate the oracle problem in security testing. 

Our contributions include (1) a DSL and supporting tools for specifying \MRs for security testing, (2) a catalog of \MRs inspired by OWASP guidelines and vulnerability descriptions in the CWE~\cite{CWE}, (3) a data collection framework crawling the system under test to derive input data automatically, 
and (4) a testing framework automatically performing security testing based on the \MRs and the input data~\cite{WebSMRL}.

Our analysis of the OWASP guidelines shows that our approach can automate 39\% of the security testing activities not currently targeted by SOTA techniques, indicating that it significantly contributes to addressing the oracle problem in security testing. Further, our catalog of \MRs can detect 101 vulnerability types in the CWE view for security design principles (45\% of the total), thus highlighting the broad applicability of \MST in the security testing context.

Our empirical results with two open-source Web systems show that the approach requires limited manual effort and detects 85.71\% of the targeted vulnerabilities, thus suggesting it is highly effective. Moreover, since at most 0.19\% of the executed follow-up inputs led to a false alarm, the impact of false alarms is minimal. Finally, the execution of our \MRs can be parallelized, thus enabling metamorphic security testing to be automatically performed overnight.
